\documentclass{amsart}
\usepackage{pgfplots}
\usepackage{color}

\newtheorem{theorem}{Theorem}[section]
\newtheorem{lemma}{Lemma}
\newtheorem{corollary}{Corollary}
\newtheorem{proposition}{Proposition}
\theoremstyle{conjecture}
\newtheorem{conjecture}{Conjecture}
\theoremstyle{definition}
\newtheorem{definition}{Definition}
\theoremstyle{question}
\newtheorem{question}{Question}
\theoremstyle{questions}

\theoremstyle{remark}
\newtheorem*{remark}{Remark}
\theoremstyle{remarks}
\newtheorem*{remarks}{Remarks}
\theoremstyle{example}
\newtheorem*{example}{Examples}
\numberwithin{equation}{section}

\begin{document}
\title{Notes on Invariant Measures for Loop Groups}

\author{Doug Pickrell}
\email{pickrell@arizona.edu}

\begin{abstract}Let $K$ denote a simply connected compact
Lie group and let $G=K^{\mathbb C}$, the complexification.  It is known that
there exists an $LK$ bi-invariant probability measure on
a natural hyperfunction completion of the complex loop group $LG$. There are various generalizations,
involving positive line bundle valued measures on the hyperfunction completion, replacing $K$ with a symmetric space, 
replacing $LK$ (the configuration space of the principal chiral model) with (the homotopy equivalent space of )
gauge equivalence classes of $K$-connections on the 2-sphere (the configuration space of $YM_3$), and so on. 
The purpose of these notes is to publicize a number
of conjectures and questions concerning how
these measures are characterized, how they are explicitly represented, and how they are
potentially relevant to quantum sigma models and $YM_3$.

\end{abstract}

\maketitle



\setcounter{section}{-1}

\section{Introduction}

Given a locally compact second countable (lcsc) topological group, there exists a unique 
translation bi-invariant measure class, and this class is represented by an essentially
unique left invariant measure. Conversely, given a standard Borel group with a left translation invariant measure class, 
there is a unique lcsc topology such that the invariant measure class is the Haar measure class.

In quantum field theory it is necessary to do analysis on infinite dimensional spaces, which are not locally compact. Whereas a finite dimensional smooth manifold has
a unique Lebesgue measure class, for an infinite dimensional space of fields, there is often a variety of `interesting' measure classes, because
a finite measure is `almost' supported on a compact set. Identifying the relevant measure class (or classes)
for an infinite dimensional problem is a nontrivial issue.

A central example is Wiener measure on the continuous loop group $C^0(S^1,K)$ (or a path group), where $K$ is a compact Lie group. 
This probability measure is heuristically of the form
$$d\nu_{\frac 1T E}=\frac{1}{\mathfrak Z_T} exp(-\frac 1T E(g))\prod_{\theta\in S^1}d\lambda_K(g(\theta))$$
where $d\lambda_K$ denotes the unique invariant probability measure on $K$, $E(g)$ is the energy of a loop
$$E(g)=\int_{S^1}|g^{-1}\frac{dg}{d\theta}|^2d\theta$$
and the background product measure is the Haar measure for the compact product group $\prod_{\theta\in S^1} K$ (the $T>0$ has many interpretations, depending on context, and we have implicitly fixed an $Ad$ invariant inner product on the Lie algebra of $K$). Wiener measure is the Feynman measure for a free particle on $K$, and the corresponding ground state is (a square root of) the Haar measure on $K$. As suggested by the heuristic formula, Wiener measure has a local character, and for each $T$ the measure class of $\nu_{\frac 1T E}$ is translation bi-invariant with respect to the dense subgroup 
$W^1(S^1,K)$, loops having one derivative in the $L^2$ Sobolev sense. However a Wiener loop is non-differentiable at all points, almost surely,
and for $T\ne T'$, $\nu_{\frac 1T E}$ and $\nu_{\frac{1}{T'} E}$ represent disjoint measure classes. Moreover there
are many other measure classes on $C^0(S^1,K)$ which are invariant with respect to some dense subgroup, e.g. heat kernel measure classes (which are inversion and bi-invariant
with respect to groups of the form $W^s(S^1,K)$, for some degree of smoothness $s>1/2$, the critical degree of smoothness for $S^1$). 

The simplest object of these notes is a limit of Wiener measure which (I claim) is relevant to quantum field theory, where $S^1$ is interpreted as space.
  
\begin{theorem}\label{theorem1}Suppose that $K$ is a simply connected
compact Lie group, and let $G$ denote the complexification. There
exists a Borel probability measure $\mu_0$ on the formal completion,
$\mathbf LG$, of the complex loop group $C^{0}(S^1,G)$ which is bi-invariant
with respect to $L_{fin}K$, the group of loops in $K$ with finite Fourier series
(in some matrix representation of $K$). 
\end{theorem}

This theorem is proven in \cite{Pi1}, and more directly in \cite{Pi2}. The basic strategy is to show
that Wiener measure  on $C^0(S^1,K)\subset\mathbf LG$ has weak
limits in Riemann-Hilbert coordinates for $\mathbf LG$ as $T\uparrow\infty$. More grandly we expect that 
\begin{equation}\label{limits1}d\lambda_K \stackrel{T\downarrow 0}{\leftarrow} \nu_{\frac 1T E}\stackrel{T\uparrow \infty}{\rightarrow} \{ \begin{matrix} \prod_{S^1}d\lambda_K& \text{ weakly relative to } C^0(\prod_{S^1} K)\\ \mu_0& \text{ weakly relative to } BC(\mathbf LG) \end{matrix}\end{equation}
where $BC(\cdot)$ denotes bounded continouous functions on a space. A crucial ingredient in the proof of the
bi-invariance of the limit $\mu_0$ is the asymptotic invariance of
Wiener measure, established earlier by Marie and Paul Malliavin
(see \cite{MM2} and section 4.1 of Part III of \cite{Pi1}). The assumption that $K$ is semi-simple in the theorem is essential; the statement is simply false
if $K=\mathbb T$, the circle (one must introduce a level $l>0$ to obtain a true statement in non-semi-simple cases, when the dual Coxeter number vanishes).

Whereas it makes sense to evaluate a random Wiener measure distributed loop $g\in \mathbf LG$ at a point of $S^1$ (because Wiener measure 
is supported on $C^0(S^1,K)$), it does not make sense to evaluate a random $\mu_0$ distributed `loop' at a point of $S^1$. Wiener measure (for closed loops and
paths) is of central importance in quantum mechanics, where the path parameter is interpreted as (imaginary) time. The measure $\mu_0$ is of significance in
2d quantum field theory, where $S^1$ is interpreted as space. Conjecturally, analogous to $\lambda_K$ for a free particle on $K$, (a square root of) $\mu_0$ is the groundstate for a free closed string on $K$ (i.e. the sigma model with target $K$), and analogous to $\nu_{\frac 1T E}$ for a free particle, there is a Markov process on (distributional) fields living on (Euclidean) space-time $S^1 \times \mathbb R$ (or a more general Riemannian surface) for the evolution of the free string. 

The purpose of these notes is to publicize a number of conjectures and questions about the measure $\mu_0$, its deformations (in particular a probability measure $\mu_l$ parameterized by level $l$), further generalizations, and their potential uses, such as the speculation about free strings above. I have thought about 
these matters for a long time, and I have a lot of opinions. It is painful to acknowledge how little I can prove. 

\subsection{ Outline of the Notes}

The first half of Section \ref{background} is a review of the definitions of the formal and hyperfunction completions of $C^{\infty}(S^1,G)$, denoted by $\mathbf L G$ and $Hyp(S^1,G)$, respectively. The natural inclusions
$$L_{fin}K \subset C^{\infty}(S^1,K) \subset C^{\infty}(S^1,G)\subset Hyp(S^1,G)\subset \mathbf L G$$
are homotopy equivalences. The natural observables on the completions (see Section \ref{calculate1} below) are vastly different from the observables on continuous loop spaces, e.g. evaluating a  loop at a point. The completions are not groups (hence convolution of measures is not defined), and the anti-holomorphic involution corresponding to the unitary form $C^{\infty}(S^1,K)\to C^{\infty}(S^1,G)$ does not continuously extend to the completions. A basic theme of these notes is that the support of the measure $\mu_0$ is a kind of surrogate for the missing unitary form of these completions (To clarify: The anti-involution $C^{\infty}(S^1,G)\to C^{\infty}(S^1,G):g \to g^*$ does extend to the completions, hence it makes sense to speak of the positive generalized loops of the completions. Inversion $g\to g^{-1}$ does not extend, hence $g\to g^{-*}$ does not extend). 

The second half of Section \ref{background} involves more sophisticated Lie theoretic structure which is useful for example to explain root subgroup factorization (an analogue of Fourier series for group valued periodic functions, and a key tool for understanding Toeplitz determinants and the measures $\mu_l$) for higher rank groups. We will frequently specialize to $K=SU(2)$, to minimize references to the second half of Section \ref{background}.

In Section \ref{existence} there is an outline of the proof of Theorem \ref{theorem1}, and a discussion of the support properties of $\mu_0$. The expectation is that a generalized loop distributed according to $\mu_0$ is, in various senses, as close to being a function on $S^1$ as one can hope, without actually being a function (or even an equivalence class of functions, as in measure theory). This is the typical behavior for the spatial trace of a scalar quantum field in two dimensions.

Section \ref{uniqueness} concerns uniqueness questions about $L_{fin}K$ bi-invariant measures and measure classes on the (very thick) formal completion $\mathbf L G$. An unsurprising but important takeaway: our expectation is that the theory of $L_{fin}K$ bi-invariant measures and measure classes on $\mathbf L G$ is markedly different from the theory of $K$ bi-invariant measures and measure classes on $G$. Nonetheless this is a useful analogy to bear in mind (This is actually more than an analogy; these two examples fit into a common framework involving symmetrizable Kac-Moody groups, as we will describe at the end of Section \ref{Generalizations}).

In Sections \ref{calculate1} and \ref{calculate2} we discuss how to calculate the measure $\mu_l$. This is the point where the subject
comes alive. There are currently two main conjectures involving two completely different perspectives. The first concerns what we can observe for (deterministic) generalized loops.
Given a real analytic embedding of $S^1$ into a closed Riemann surface, there is an induced map from the hyperfunction completion to the space of holomorphic $G$ bundles on the surface. This `observable' should be viewed as an analogue of mapping a continuous loop to its value at a point. The first conjecture (which may require slight qualification) is that $\mu_0$ pushes forward to a natural normalized (Goldman) symplectic volume element on the subset of stable bundles. Proving this first conjecture, for a general $LK$ bi-invariant measure on the hyperfunction completion, is possibly the most efficient way to prove the uniqueness conjectures of Section \ref{uniqueness}. This generalizes in a natural way
to the deformations of $\mu_0$. 

The second conjecture is that so called root subgroup coordinates are independent random variables with respect to $\mu_l$; see the product formulas (\ref{productmeasure1}) and (\ref{productmeasure5}) in the $SU(2)$ and general cases, respectively. This conjecture largely reduces to a deceptively simple task: proving that these product expressions
(are well-defined and) are invariant with respect to the anti-involution
$g \to g^*$ on the completions.  There are other interesting pushforwards and coordinates, and the relevant formulas are simply missing.
One of special importance is the $\theta_+$ distribution for $\mu_l$, where $\theta_+\in H^1(\Delta,\mathfrak g)$ is the linear Riemann-Hilbert coordinate for $g\in Hyp(S^1,G)$.
A $YM_3$ analogue of this problem is mentioned below, which surprisingly turns out to be tractable.

Section \ref{representations} concerns representation theory. Given a fixed positive integral level $l$, an important problem is to correctly formulate an analytic version of the (holomorphic) Peter-Weyl theorem. There is a well-known algebro-geometric version of the Peter-Weyl theorem for arbitrary symmetrizable Kac-Moody groups due to Kac and Peterson. But this is inadequate for an application we have in mind to sewing for the (chiral) Wess-Zumino-Witten model of conformal field theory, see below. A key ingredient is a diagonal distribution conjecture, which is a corollary of the root subgroup conjectures of Section \ref{calculate2}. It appears that the decomposition of $L^2(\mu_0)$ with respect to the bi-invariant action of $L_{fin}K$ is rather dull: the action on the orthogonal complement of the constant function $1$ may be irreducible. On the other hand a corollary of the uniqueness conjectures of Section \ref{uniqueness} is that $C^{\omega}Homeo(S^1)$ fixes $\mu_0$ and hence acts unitarily on the Hilbert space of half densities. This is highly decomposable, in a potentially interesting way, as a consequence of the first perspective in Section \ref{calculate1}. But how to decompose this action is presently unclear.

In Section \ref{Generalizations} there are some comments on generalizations. The basic existence result for invariant measures generalizes to (formal and hyperfunction completions of) loops into symmetric spaces. It is a challenge to formulate a generalization for loops into more general compact homogeneous targets (this seems doable) or Einstein targets (which will undoubtedly require new ideas). The vague general question is, given a Riemannian manifold $X$, not necessarily compact but admitting a Wiener type measure $\nu_{\frac 1T E}$ on
$C^0(S^1,X)$, does there exist some analogue of the hyperfunction completion such that there is an interesting large $T$ limit analogous to $\mu_0$? In another direction, I originally thought the theory might extend to symmetrizable Kac-Moody algebras. This now seems doubtful, and it is enlightening to understand why. 

The measures $\mu_l$, and various generalizations, are interesting because of their potential applications to quantum field theory. Ultimately these measures should be understood as emerging from this qft perspective as fixed time distributions for various 2d Feynman measures.

In Section \ref{sigmamodel}, I discuss the two dimensional sigma model with target (compact simply connected) $K$, i.e. the principal chiral model, for which $LK$ is the configuration space.  This is a field theory which is classically conformally invariant and `integrable', in the sense that there exists a zero curvature representation of the classical equations. In the plane there
is additionally a classical Yangian symmetry. At the quantum level it is a canon of physics that on the one hand conformal invariance is broken, while on the other hand in the plane a deformed Yangian symmetry survives. This is manifested by the emergence of massive particles, and there exist explicit conjectures for the masses and scattering matrices in terms of evaluation representations for Yangian quantum groups. This Yangian symmetry is `not compatible with periodic boundary conditions'. Consequently it is a longstanding puzzle to rigorously formulate the model in $RS^1\times \mathbb R$ (where $R$ is the radius of the circle), or more generally against a possibly curved Riemannian background, e.g. to construct a model satisfying Segal's axioms for qft. In this section I posit that the measure $\mu_0$ is essentially the vacuum, and try to draw some conclusions, based on properties of $\mu_0$ outlined in previous sections. This point of view suggests the possibility that, just as a free quantum field (i.e. the sigma model with target $\mathbb R$) is an assembly of harmonic oscillators, when $R$ is finite, the $K$-valued chiral model might be an assembly of harmonic oscillators and `spherical harmonic oscillators'. This is an alluring possibility, not a formal conjecture. I have long struggled and failed to devise a test for this speculation.

In Section \ref{sewing} we discuss sewing, from a global point of view, for the chiral $WZW_l$ model (a holomorphic half of the full $WZW_l$ model for closed strings). It is well-known that heuristically the basic sewing mechanism is the conjectural holomorphic Peter-Weyl theorem alluded to above. We spell this out. It would be a dream if the 
analytic point of view of these notes was useful for the full WZW model (involving left and right movers), or even better, the corresponding boundary conformal field theory (for which, to my knowledge, sewing rules have not been proven). Unfortunately I do not see this.

There is a well-known correspondence between the mathematical technology associated with loop groups and the chiral and WZW models on the one hand, and 3D Yang-Mills and Chern-Simons on the other hand (e.g. the configuration spaces are essentially homotopic). In Section \ref{YM} I comment on some mathematical aspects of this correspondence, which have been uncovered by Karabali, Nair, and other physicists. In particular they assert the existence of a (finite) measure that corresponds to $\mu_0$. In an earlier version of these notes
I was unsure whether this was real or merely heuristic. Recently Guillarmou, Kupiainen and Rhodes (\cite{GKR}) have shown that the physicists' assertion is very likely correct. 
At a (fairly convincing) heuristic level, it appears that the analogue of (\ref{limits1}) is 
\begin{equation}\label{limits4}d\lambda_{H^1(S^{(2)},K)} \stackrel{T\downarrow 0}{\leftarrow} \nu_{\frac 1T YM_2}\stackrel{T\uparrow \infty}{\rightarrow} \{ \begin{matrix} \text{Ashtekar measure}& \text{ wrt } BC(\{\text{holonomy functors}\}) \\ \widetilde{\mu_0}& \text{ wrt } BC(\bigsqcup_{\rho\in H^1(S^{(2)},K)}G\backslash Map(\widetilde{S^{(2)}},G/K)^{\rho})  \end{matrix}\end{equation}
where the first measure is the normalized Goldman volume form on moduli space, and the measure $\widetilde {\mu_0}$ is (the finite part of) a measure (or more precisely a family of `conformally invariant' measures parameterized by $Bun_G^0(S^{(2)})$) studied in (\cite{GKR}). Assuming its truth, my hypothesis that $L^2(\mu_0)$ is a natural model for the chiral model Hilbert space corresponds exactly to Karabali and Nair's claim that $L^2(\widetilde{\mu_0})$ is a natural model for the Hilbert space of $YM_3$. Whereas I hypothesize that $\mu_0$ is essentially the vacuum for the chiral model, Karabali and Nair claim that the $YM_3$ vacuum is a density times $\widetilde{\mu_0}$. This is the point where my exposition simply trails off. Incidentally, the $YM_2$ measure (like the GKR measure) has a relatively explicit
expression in the coordinate system which is cryptically referenced in (\ref{limits4}). 

It is irresistible to wonder if there is a three dimensional analogue of (\ref{limits1}) and (\ref{limits4}) involving a $YM_3$ measure (which remains to be discovered).  

\subsubsection{Notation} In reference to the Kontsevich/Segal perspective on qft: $S^{(d)}$ will denote a space of dimension $d$ (with geometric properties which we will specify), and $\Sigma^{(d)}$ will denote a Euclidean space time of dimension $d$ (with geometric properties which we will specify).

\section{Loop Groups and Completions}\label{background}

The polynomial loop groups $L_{fin}K$ and $L_{fin}G:=G(\mathbb
C[z,z^{-1}])$ consist of maps from $S^1$ into $K$ ($G$,
respectively) which have finite Fourier series, with pointwise
multiplication (relative to a fixed faithful matrix representation).
Recall $\mathbb C[z]$ is the algebra over $\mathbb
C$ generated by $z$, and $\mathbb C[z,z^{-1}]$ is the algebra of
finite Laurent series. Neither of these (skeletal, or minimal) loop groups is
a Lie group. They are well-suited for algebra (as in \cite{KP}), but for some purposes of analysis, it is essential
to consider completions (as in \cite{PS}).

The analytic loop group, $H^0(S^1,G)$ (loops which are holomorphic in a collar containing $S^1$),
is a complex Lie
group. Let $D$ ($D^*$) denote the $\underline{closed}$ unit disk
($\{z:|z|\ge 1\}\cup\{\infty\}$, respectively).
A dense open neighborhood of the identity in $H^0(S^1,G)$ consists of those loops
which have a unique Riemann-Hilbert factorization
\begin{equation}\label{0.2}g=g_{-}\cdot g_0\cdot g_{+}\end{equation}
where $g_{-}\in H^0(D^{*},\infty ;G,1)$, $g_0\in G$, and $g_{+}\in
H^ 0(D,0;G,1)$ (thus (\ref{0.2}) is an equality of holomorphic
functions which holds on a thin collar of $S^1$, the collar
depending upon $g$). A model for this neighborhood is
$$H^1(D^{*},\mathfrak g)\times G\times H^1(D,\mathfrak g)$$
where the linear coordinates are determined by
$\theta_{+}=g_{+}^{-1}(\partial g_{+})$, $\theta_{-}=(\partial g_{
-})g_{-}^{-1}$.  The (left or right) translates of this
neighborhood by elements of $L_{fin}K$ cover
$H^0(S^1,G)$.

In general there are decompositions
\begin{equation}\label{Birkhoff1}G(\mathbb C[z,z^{-1}])=\bigsqcup_{\lambda \in
Hom(S^1,T)}\Sigma_{\lambda}^{G(\mathbb C[z,z^{-1}])},\quad
\Sigma_{\lambda}^{G(\mathbb C[z,z^{-1}])}=G(\mathbb
C[z^{-1}])\cdot\lambda\cdot G(\mathbb C[z])\end{equation} and
\begin{equation}\label{Birkhoff2}H^0(S^1,G)=\bigsqcup_{\lambda\in
Hom(S^1,T)}\Sigma_{\lambda}^{H^0(S^1,G)},\quad
\Sigma_{\lambda}^{H^0(S^1,G)}= H^0(D^*,G)\cdot\lambda\cdot
H^0(D,G)\end{equation}
where $T$ is a maximal torus of $K$. Note the `top stratum' $\Sigma_{\lambda=1}^{H^0(S^1,G)}$
is the dense open neighborhood of the previous paragraph. There are finer decompositions
over the affine Weyl group $W\propto Hom(S^1,T)$, where for example the `top stratum' consists of
$g$ as in (\ref{0.2}) such that $g_0\in G$ has a standard LDU factorization. We will loosely refer to all
of these decompositions, and generalizations below, as `Birkhoff decompositions'.

The hyperfunction completion, $Hyp(S^1,G)$, is modeled on the
space
$$H^1(\Delta^{*},\mathfrak g)\times G\times H^1(\Delta ,\mathfrak g)$$
where $\Delta$ and $\Delta^{*}$ denote the
$\underline {open}$ disks centered at $ 0$ and $\infty$,
respectively, and the transition functions are obtained by
continuously extending the transition functions for the analytic
loop space of the preceding paragraph. In the case of $SU(2)$, transition
functions are written down explicitly in Section 3 of \cite{Pi2}.  The global definition is
$$Hyp(S^1,G)=G(H^0(S^1_{-}))\times_{H^0(S^1,G)}G(H^0(S^1_{+}))$$
where $H^0(S^1_{\pm})$ denote the direct limits of the spaces
$H^0(\{r<\vert z\vert <1\})$ and $H^0(\{1<\vert z\vert <r\})$, as
$ r\uparrow 1$ and $r\downarrow 1$, respectively.

From the global
definition it is clear that $H^0(\Delta,G)$ ($H^0(\Delta^*,G)$) acts naturally
from the right (left, respectively) of $Hyp(S^1,G)$, hence the analytic loop group
$H^0(S^1,G)$ acts naturally
from both the left and right of $Hyp(S^1,G)$. There is a natural action
of $C^{\omega}Homeo(S^1)$ on $Hyp(S^1,G)$:
given a real analytic (orientation-preserving) homeomorphism $\sigma$ and
an equivalence class $[g_1,g_2]\in Hyp(S^1,G)$ (thus
$g_1:\{1-\epsilon<|z|<1\}\to G$ and $g_2:\{1<|z|<1+\epsilon\}\to G$ are defined and
holomorphic for sufficiently small $\epsilon$),
\begin{equation}\label{viraction}\sigma_*([g_1,g_2])=[g_1\circ \sigma^{-1},g_2\circ \sigma^{-1}]\end{equation}

The formal completion is defined in a similar way, where
$H^1(\Delta ,\mathfrak g)$ is replaced by the corresponding space
of formal power series
$$H^1_{formal}(\Delta ,\mathfrak g)=\{\theta_{+}=(\theta_1+\theta_2z+
..)dz,\quad\theta_i\in \mathfrak
g\}\simeq\prod_1^{\infty}\mathfrak g$$
The global definition of
the formal completion is
$$\mathbf LG=G(\mathbb C((z^{-1})))\times_{G(\mathbb C[z,z^{-1}])}G(
\mathbb C((z)))$$ where $\mathbb C((z))$ is the field of formal
Laurent series $\sum a_nz^n$, $a_n=0$ for $n<<0$. From the global
definition it is clear that $G(\mathbb C((z)))$ ($G(\mathbb C((z^{-1})))$)
acts naturally
from right (left, respectively) of $\mathbb L G$, and hence $L_{fin}G$ acts naturally
from both the left and right of $\mathbb L G$.
Mokler has clarified the nature of the formal completion from the point of
view of algebraic geometry, see \cite{Mokler}. A disadvantage of the formal completion
is that only the subgroup $PSU(1,1)$ (of linear fractional transformations) of $C^{\omega}Homeo(S^1)$
acts as in (\ref{viraction}).

These completions have generalized `Birkhoff decompositions'
$$Hyp(S^1,G)=\bigsqcup_{\lambda\in Hom(S^1,T)}\Sigma^{hyp}_{\lambda},\quad\Sigma_{
\lambda}^{hyp}=H^0(\Delta^{*},G)\cdot\lambda\cdot H^0(\Delta
,G),$$ and
$$\mathbf LG=\bigsqcup_{\lambda\in Hom(S^1,T)}\Sigma_{\lambda}^{\mathbf LG},\quad
\Sigma_{\lambda}^{\mathbf LG}=G(\mathbb
C[[z^{-1}]])\cdot\lambda\cdot G(\mathbb C[[z]])$$ where $\mathbb
C[[\zeta ]]$ denotes formal power series in $\zeta$. These decompositions are compatible with the Birkhoff factorizations
(\ref{Birkhoff1}) and (\ref{Birkhoff2}). It is the existence of
these coherent decompositions, corresponding to different
smoothness conditions, which imply that the natural inclusions
$$L_{fin}K\subset L_{fin}G\subset H^0(S^1,G) \subset C^0(S^1,G)\subset Hyp(S^1,G)\subset \mathbf L G$$
(and various other analytic completions) are all homotopy equivalent (see 8.6.6 in \cite{PS}).

For the measure-theoretic purposes of these notes, we mainly need the top
stratums corresponding to $\lambda=1$.
In both the formal and hyperfunction cases, the top stratum is open and dense,
and for each point $g$ in the top stratum, there is a unique factorization as
in (\ref{0.2}), where in the hyperfunction case $g_{\pm}$ are
$G$-valued holomorphic functions in the open disks $\Delta$
and $\Delta^{*}$, respectively, and in the formal case $g_{\pm}$
are simply formal power series satisfying the appropriate
algebraic equations determined by $G$.  We will refer to
$g_{-},g_0,g_{+}$ ($\theta_{ -},g_0,\theta_{+}$, respectively) as
the Riemann-Hilbert coordinates (linear Riemann-Hilbert
coordinates, respectively) of $g$.

\begin{remark} $C^0(S^1,G)$ is a complex Lie group. To obtain a coordinate neighborhood of the identity,
choose a coordinate neighborhood $1\in U\subset G$ of $1$. Then $C^0(S^1,U)$ is a coordinate neighborhood
of $1\in C^0(S^1,G)$. We cannot evaluation a generic $g\in Hyp(S^1,G)$ at a point in $S^1$. A
coordinate neighborhood
of $1\in Hyp(S^1,G)$ is the top stratum with coordinates $\theta_{ -},g_0,\theta_{+}$, as in the previous paragraph;
the Taylor coefficients of $\theta_{\pm}$ are smeared values of $g$, as in quantum field theory.

Given a finite set of points $V\subset S^1$, there is an evaluation map
$$eval_V:C^0(S^1,G)\subset \prod_{S^1}G \to \prod_V G: g \to (g(v))|_{v\in V}$$
In Subsection \ref{modulispace} we will discuss the analogue of these projections for $Hyp(S^1,G)$ - instead
of points in $S^1$, we consider analytic loops in Riemann surfaces.
\end{remark}

Of central importance, there exists a holomorphic line bundle
$\mathcal L \to \mathbf L G$ such that the corresponding $\mathbb C^{\times}$ bundle
is a completion of the universal central extension
$$0\to \mathbb C^{\times} \to \widehat{C^{\infty}(S^1,G)} \to C^{\infty}(S^1,G)\to 0$$
This and the unitary form
$$0\to \mathbb T \to \widehat{C^{\infty}(S^1,K)} \to C^{\infty}(S^1,K)\to 0$$
are described in chapter 4 of \cite{PS} (here we are using the assumption that $\mathfrak g$ is simple).
The restrictions of these extensions to $L_{fin}G$ and $L_{fin}K$, respectively, are
the `minimal untwisted affine Kac-Moody groups', the group structures of which
are elucidated in \cite{KP} in terms of generators and relations. The natural maximal unitary
universal central extension is
$$0\to \mathbb T \to \widehat{W^{1/2}(S^1,K)} \to W^{1/2}(S^1,K)\to 0$$
where $W^{1/2}(S^1,K)$ is the ($C^0$ Lie) group of equivalence classes of loops with half a derivative
in the $L^2$ sense (with the natural Polish topology).

\begin{remark} It is problematic to define a corresponding maximal complex universal central extension.
The difficulty is that $W^{1/2}(S^1,G)$ is not a group, because $W^{1/2}$ does not imply boundedness. If one
imposes boundedness, then for example the Birkhoff factorization fails - the factors are not necessarily
bounded. I do not know how to resolve this tension, which recurs in a number of slightly different ways.

Given the measure/quantum field theoretic preoccupations of these notes, this fussing about lack of boundedness for $W^{1/2}$
loops may be completely irrelevant (this is a classical preoccupation).
\end{remark}

\subsubsection{Appendix: The Abelian Case}\label{abeliancase}

In these notes we are assuming that $\mathfrak k \subset \mathfrak g$ are simple. But it is occasionally useful to contemplate `the abelian case' $i\mathbb R \subset \mathbb C$. In this abelian case $Hyp(S^1, \mathbb C^{\times})$
is a group; in fact it factors as a product of $Hom(S^1,S^1)$ and the identity component, which is a quotient of its Lie algebra
$$0 \to 2\pi i\mathbb Z \to Hyp(S^1,\mathbb C)\to Hyp(S^1,\mathbb C^{\times})\to 0 $$
where an ordinary hyperfunction on $S^1$ with linear triangular factorization $f=f_-+f_0+f_+$ maps to the multiplicative hyperfunction with Riemann-Hilbert factorization $exp(f)=exp(f_-)exp(f_0)exp(f_+)$. Moreover there is a unitary form, because it makes sense to say that $f$ is real, i.e. $f_0\in\mathbb R$ and $f_+^*=f_-$.

\subsection{Supplementary Notation}\label{supplement}

For the most part these notes should be intelligible using the notation
that we have established to this point. For this to be possible, at several points
we specialize to $K=SU(2)$. However at some points, e.g. when we discuss
root subgroup factorization for higher rank groups, it will be necessary to
use more structure.

A venerable source for notational conventions for Kac-Moody algebras is naturally \cite{Kac}.
However, at least in my view, \cite{Carter} has made a few improvements. One difference between
 these two sources: Kac employs the convention
of denoting a finite dimensional algebra, and the associated structure, using dots,
and denoting the associated (untwisted) affine extension, and the associated structure, with the absence of dots,
whereas Carter uses $(\cdot)^0$ in place of a dot. I have found it convenient to stick with Kac's convention.
One other note: I like to use $r$ for rank and $l$ for level. This will no doubt offend some people. Sorry.

\subsubsection{Finite Type Structure}

Following Kac's convention, $\dot K$ is a simply connected compact Lie group with simple
Lie algebra $\dot{\mathfrak k}$, $\dot{\mathfrak g}$ is the complexification,
 and $\dot{\mathfrak g} \to \dot{\mathfrak g}:x
\to -x^*$ is the anticomplex involution fixing $\dot{\mathfrak
k}$.

Fix a triangular decomposition
\begin{equation}\label{6.1}\dot {\mathfrak g}=\dot {\mathfrak n}^{-}\oplus\dot
{\mathfrak h}\oplus\dot {\mathfrak n}^{ +}\end{equation} such that
$\dot{\mathfrak t}=\dot{\mathfrak k} \cap \dot{\mathfrak h}$ is
maximal abelian in $\dot{\mathfrak k}$; this implies
$(\dot{\mathfrak n}^+)^*=\dot{\mathfrak n}^-$.

We introduce the
following standard notations: for each positive root $\dot \alpha$, $\dot h_{\dot\alpha}$
denotes the corresponding coroot; $\{\dot{\alpha_j}: 1\le j\le r\}$ is
the set of simple positive roots; $\{\dot{h_j}:=\dot h_{\dot\alpha_j}\}$ is the set of
simple coroots; $\{\dot{\Lambda}_j\}$ is the set of fundamental
dominant integral weights; $\dot{\theta}$ is the highest root;  $\dot W$ is
the Weyl group; $\langle\cdot ,\cdot\rangle$ is the `standard
invariant symmetric bilinear form', uniquely determined by the condition (for the dual form)
$\langle\dot{\theta} ,\dot{\theta}\rangle = 2$; and
\begin{equation}\label{dotdfn}
\dot{\rho}:=\frac 12\sum_{\dot\alpha>0}\dot{\alpha}=\sum_{i=1}^r\dot{\Lambda_i}:=\sum_{i=1}^r \dot a_j\dot \alpha_j
\qquad \dot h_{\dot\theta}=\sum_{i=1}^r c_j\dot h_j\end{equation}

\begin{remark} The integers $\dot a_j$ and $c_j$, among other things, are conveniently tabulated in the summary at the end of
\cite{Carter}. However there is a silly error in the summary regarding the relation of the standard form and the Killing form: the correct expression should be
$$\langle\cdot ,\cdot\rangle=\frac{1}{2\dot g}\kappa(\cdot,\cdot)$$
where $\dot g:=1+\sum_{i=1}^r\dot a_j$ is the dual Coxeter number and $\kappa$ is the Killing form.
\end{remark}

For each simple
root $\gamma$, fix a root homomorphism $i_{\gamma}:sl(2,\mathbb C)
\to \dot{\mathfrak g}$ (we denote the corresponding group
homomorphism by the same symbol), and let
$$f_{\gamma}=i_{\gamma}(\left (
\begin{matrix} 0&0\\1&0
\end{matrix} \right)),\quad e_{\gamma}=i_{\gamma}(\left (
\begin{matrix} 0&1\\0&0
\end{matrix} \right)), \quad \text{and} \quad \mathbf{r}_{\gamma}=i_{\gamma}(\left (
\begin{matrix} 0&i\\i&0
\end{matrix} \right)) \in \dot T=exp(\dot{\mathfrak t});$$
$\mathbf{r}_{\gamma}$ is a representative for the simple
reflection $r_{\gamma} \in \dot W$ corresponding to $\gamma$ (we
will adhere to the convention that representatives for Weyl group
elements will be denoted by bold letters).

Introduce the lattices
$$\widehat{\dot T}=\bigoplus_{1\le i\le r} \mathbb Z \dot{\Lambda}_i \quad \text{(weight
lattice)}, \quad \text{and} \quad \check{\dot T}=\bigoplus_{1\le
i\le r} \mathbb Z \dot{h}_i \qquad \text{(coroot lattice)}$$
These lattices and bases are in duality. Recall that the kernel of
$exp:\dot{\mathfrak t} \to \dot T$ is $2 \pi i$ times the coroot
lattice. Consequently there are natural identifications
$$\widehat{\dot T} \to Hom(\dot T,\mathbb T)$$ where a weight $\dot{\Lambda}$
corresponds to the character $exp(2\pi ix) \to exp(2\pi
i\dot{\Lambda} (x))$, for $x\in \dot{\mathfrak h}_{\mathbb R}$,
and
$$\check{\dot
T} \to Hom(\mathbb T,\dot T),$$ where an element $h$ of the coroot
lattice corresponds to the homomorphism $\mathbb T\to \dot
T:exp(2\pi ix)\to exp(2\pi ixh)$, for $x\in \mathbb R$. Also
$$\widehat{\dot{R}}=\bigoplus_{1\le i\le r} \mathbb Z \dot{\alpha}_i \quad \text{(root lattice)}, \quad
\text{and} \quad \check{\dot R}=\bigoplus_{1\le i\le r} \mathbb Z
\dot{\Theta}_i \quad \text{(coweight lattice)}$$ where these bases
are also in duality. The $\dot{\Theta}_i$ are the fundamental
coweights.

The affine Weyl group is the semidirect product $\dot W \propto
\check{\dot T}$. For the action of $\dot W$ on $\dot{\mathfrak
h}_{\mathbb R}$, a fundamental domain is the positive Weyl chamber
$C=\{x:\dot{\alpha}_i (x)>0, i=1,..,r\}$. For the natural affine
action
\begin{equation}\label{affineaction}\dot W \propto \check{\dot T}\times \dot{\mathfrak
h}_{\mathbb R} \to \dot{\mathfrak h}_{\mathbb R}\end{equation} a
fundamental domain is the convex set
$$C_0=\{x\in C:\dot{\theta}(x)<1\} \qquad
\text{(fundamental alcove)}$$ The set of extreme points for the
closure of $C_0$ is $\{0\}\cup\{\frac{1}{\dot a_i}\dot{\Theta}_i\}$.

The Lie algebra $\dot{\mathfrak g}$ is graded by height, where the height of a simple positive root vector
is one. In general, for a positive root
$\dot{\alpha}$,
$\dot{\alpha}(h_{\dot{\delta}})=height(\dot{\alpha})$. When $\dot {\mathfrak g}$ is simply laced,
 i.e. all roots have the same length, this is equal to $\dot \delta(h_{\dot \alpha})$

Let $\dot N^{\pm}=exp(\dot n^{\pm})$ and $\dot A=exp(\dot
h_{\mathbb R})$. An element $g\in \dot N^-\dot T\dot A\dot N^+$
has a unique triangular decomposition
\begin{equation}\label{finitefactorization}g=\dot l(g) \dot d(g) \dot u(g), \quad \text{where} \quad
\dot d=\dot m \dot a=\prod_{j=1}^r \dot {\sigma}_j(g)^{\dot
h_j},\end{equation} and $\dot
{\sigma}_i(g)=\phi_{\dot{\Lambda}_i}(\pi_{\dot {\Lambda}_i}(g)
v_{\dot{\Lambda}_i})$ is the fundamental matrix coefficient for
the highest weight vector corresponding to $\dot {\Lambda}_i$.

\subsubsection{Affine Algebra Structure}

Let $L\dot{\mathfrak g}=C^{\infty}(S^1,\dot{\mathfrak g})$, viewed
as a Lie algebra with pointwise bracket. There is a universal
central extension
\[0\to \mathbb C c\to\tilde {L}\dot {\mathfrak g}\to L\dot {\mathfrak g}\to 0,\]
where as a vector space $\tilde {L}\dot {\mathfrak g}=L\dot
{\mathfrak g}\oplus \mathbb C c$, and in these coordinates
\begin{equation}\label{bracket}[X+\lambda c,Y+\lambda^{\prime} c]_{\tilde {L}\dot {\mathfrak g}}
=[X,Y]_{L\dot {\mathfrak g}}+\frac i{2\pi}\int_{S^1}\langle
X\wedge dY\rangle c. \end{equation} The smooth completion of the
untwisted affine Kac-Moody Lie algebra corresponding to
$\dot{\mathfrak g}$ is
$$\widehat L\dot{\mathfrak g}=\mathbb C d\propto\tilde {L}\dot {\mathfrak g}
\qquad \text{(the semidirect sum)},$$ where the derivation $d$
acts by $d(X+\lambda c)=\frac 1i\frac d{d\theta}X$, for $X\in
L\dot {\mathfrak g}$, and $[d,c]=0$. The algebra generated by
$\dot {\mathfrak k}$-valued loops induces a central extension
\[0\to i\mathbb R c\to\tilde {L}\dot {\mathfrak k}\to L\dot {\mathfrak k}\to 0 \]
and a real form $\widehat L\dot{\mathfrak k}=i\mathbb R d\propto\tilde
{L}\dot {\mathfrak k}$ for $\widehat L\dot{\mathfrak g}$.

We identify $\dot {\mathfrak g}$ with the constant loops in $L\dot
{\mathfrak g}$. Because the extension is trivial over $\dot
{\mathfrak g}$, there are embeddings of Lie algebras
\[\dot {\mathfrak g}\to\tilde {L}\dot {\mathfrak g}\to \widehat L\dot{\mathfrak g}. \]

There are triangular decompositions
\begin{equation}\label{looptriangulardecomposition}\tilde L\dot{\mathfrak
g} =\mathfrak n^{-}\oplus \mathfrak h \oplus \mathfrak n^{ +}
\quad \text{and} \quad \widehat L\dot{\mathfrak g} =\mathfrak
n^{-}\oplus (\mathbb C d+\mathfrak h) \oplus \mathfrak n^{ +},
\end{equation} where $\mathfrak h=\dot {\mathfrak h}+\mathbb C c$
and $\mathfrak n^{\pm}$ is the smooth completion of
$\dot{\mathfrak n}^{\pm} +\dot{\mathfrak g}(z^{\pm 1}\mathbb
C[z^{\pm 1}])$, respectively. The simple roots for $(\widehat
L_{fin}\dot{\mathfrak g},\mathbb C d+\mathfrak h)$ are
$\{\alpha_j: 0\le j\le r\}$, where
\[\alpha_0=\delta-\dot{\theta} ,\quad\alpha_j=\dot{\alpha}_j,\quad j>0, \]
$\delta(d)=1$, $\delta(c)=0$, $\delta (\dot {\mathfrak h})=0$, and the
$\dot{\alpha}_j$ are extended to $\mathbb C d+\mathfrak h$ by
requiring $\dot{\alpha}_ j(c)=\dot{\alpha}_ j(d)=0$. We also
crucially introduce a linear functional $\gamma$ which vanishes on
$\dot{\mathfrak h}$ and satisfies $\gamma(c)=1$, $\gamma(d)=0$. The simple
coroots are $\{h_j:0\le j\le r\}$, where
\[h_0=c-\dot {h}_{\dot{\theta}},\quad h_j=\dot {h}_j,\quad j>0.\]
Note that $c=h_0+\sum_{j=1}^rc_j \dot h_j$ (which explains Carter's convention
that $\dot h_{\dot\theta}=\sum_{i=1}^rc_j\dot h_j$).

For a dominant integral weight $\Lambda$, $\Lambda(c)=l(\Lambda)$ is the level.
In general
\begin{equation}\label{domfunct}\Lambda=\sum_{i=0}^r\Lambda(h_j)\Lambda_j=l\gamma+\dot \Lambda\end{equation}
where the $\Lambda_j$, $j=0,...,r$, are the fundamental dominant integral weights,
and given the level $l$, $\Lambda$ is uniquely determined by $\dot \Lambda$, a dominant integral weight on $\dot{\mathfrak h}_{\mathbb R}$ (the condition that $\Lambda(h_0)=l-\dot\Lambda(h_{\dot\theta})\ge 0$ is
equivalent to $\dot\Lambda(h_{\dot\theta}\le l$).

For $i>0$, the root homomorphism $i_{\alpha_i}$ is
$i_{\dot{\alpha}_i}$ followed by the inclusion $\dot {\mathfrak
g}\subset \tilde L \dot{\mathfrak g}$. For $i=0$
\begin{equation}\label{roothom}i_{\alpha_0}(\left(\begin{matrix} 0&0\\
1&0\end{matrix} \right))=e_{\dot{\theta}}z^{-1},\quad
i_{\alpha_0}(\left(\begin{matrix}
0&1\\
0&0\end{matrix} \right))=f_{\dot{\theta}}z, \end{equation} where
$\{f_{\dot{\theta}},\dot {h}_{\dot{\theta}},e_{\dot{\theta}}\}$
satisfy the $ sl(2,\mathbb C )$-commutation relations, and
$e_{\dot{\theta}}$ is a highest root vector for $\dot {\mathfrak
g}$. The fundamental dominant integral functionals on $\mathfrak
h$ are $\Lambda_j$, $j=0,..,r$, where $\Lambda_i(h_j)=\delta(i-j)$
and $\Lambda_i(d)=0$.

Also set $\mathfrak t=i\mathbb R c \oplus \dot{\mathfrak t}$ and
$\mathfrak a=\mathfrak h_{\mathbb R}=\mathbb R c \oplus
\dot{\mathfrak h}_{\mathbb R}$. Finally
\begin{equation}
\rho:=\sum_{i=0}^r\Lambda_i=\dot {\delta}+\dot g \gamma
\end{equation}
where (to repeat) the dual Coxeter number $\dot g=1+\sum_{i=1}^r\dot a_j$.

\subsubsection{Loop Groups and Extensions}\label{loopgroups}

Let $\Pi:\tilde L \dot G \to L \dot G$ ($\Pi:\tilde L \dot K \to L
\dot K$) denote the universal central $\mathbb C^*$ ($\mathbb T$)
extension of the smooth loop group $L \dot G$ ($L \dot K$,
respectively), as in \cite{PS}. Let $N^{\pm}$ denote the subgroups
corresponding to $\mathfrak n^{\pm}$. Since the restriction of
$\Pi$ to $N^{\pm}$ is an isomorphism, we will always identify
$N^{\pm}$ with its image, e.g. $l\in N^+$ is identified with a
smooth loop having a holomorphic extension to $\Delta$ satisfying
$l(0)\in \dot N^+$. Also set $T=exp(\mathfrak t)$ and
$A=exp(\mathfrak a)$.

As in the finite dimensional case, for $\tilde {g}\in N^{-}\cdot T
A\cdot N^{+}\subset\tilde {L}\dot G$, there is a unique triangular
decomposition
\begin{equation}\label{diagonal}\tilde {g}=l\cdot d \cdot u,\quad \text{where}\quad d
=ma=\prod_{j=0}^r\sigma_j(\tilde { g})^{h_j},\end{equation} and
$\sigma_j=\sigma_{\Lambda_j}$ is the fundamental matrix
coefficient for the highest weight vector corresponding to
$\Lambda_j$. If $\Pi(\tilde {g})=g$, then because
$\sigma_0^{h_0}=\sigma_0^{c-\dot {h}_{\dot{\theta}}}$ projects to
$\sigma_ 0^{-\dot {h}_{\dot{\theta}}}$, $g=l\cdot \Pi(d)\cdot u$,
where
\begin{equation}\label{loopdiagonal}\Pi(d)(g)=
\sigma_0(\tilde {g})^{-\dot
{h}_{\dot{\theta}}}\prod_{j=1}^r\sigma_j(\tilde {g})^{\dot
{h}_j}=\prod_{j=1}^r\left (\frac {\sigma_j(\tilde
{g})}{\sigma_0(\tilde {g})^{\check {a}_j}}\right )^{\dot {h}_j},
\end{equation} and the $\check {a}_j$ are positive integers such
that $\dot {h}_{\dot{\theta}}=\sum\check {a}_j\dot {h}_j$ (these
numbers are also compiled in Section 1.1 of \cite{KW}).

If $\tilde{g} \in \tilde{L}\dot K$, then
$\vert\sigma_j(\tilde{g})\vert$ depends only on $g=\Pi(\tilde g)$.
We will indicate this by writing
\begin{equation}\label{diagonalnotation}\vert\sigma_j(\tilde{g})\vert=\vert\sigma_j\vert(g)
\quad \text{and} \quad a(\tilde g)=a(g),\end{equation} where
$a(\tilde g)$ is defined as in (\ref{diagonal}).

\section{Existence}\label{existence}

In this section we will outline a proof of Theorem \ref{theorem1}, and we will discuss a 
conjecture which implies sharp support properties for $\mu_0$.

\begin{theorem} Suppose (as we do generally in this paper) that $K$ is simply connected and has a simple Lie algebra. Then\

(a) $L_{fin}K$ is dense in $C^{\infty}(S^1,K)$.\

(b) $L_{fin}K$ is generated by $K$ (the constant loops)
and the highest root subgroup $i_{\alpha_0}(SU(2))$, see (\ref{roothom}).
In particular if $K=SU(2)$, then $L_{fin}K$ is generated by the constant subgroup $K$
and the (highest root) subgroup
$$\{\left(\begin{matrix} \bar a&-\bar bz^{-1}\\
bz&a\end{matrix} \right): \qquad \left(\begin{matrix} a&b\\
-\bar {b}&\bar {a}\end{matrix} \right)\in SU(2)\}$$
\end{theorem}

For memorably elegant proofs, see Propositions (3.5.3) and (5.2.5) of \cite{PS}. In the case of $K=SU(2)$, the subgroups
$$K=\{\left(\begin{matrix} a& b\\
-\bar{b}&a\end{matrix} \right)\in SU(2)\} \text{ and } \{\left(\begin{matrix} \bar a&-\bar bz^{-1}\\
bz&a\end{matrix} \right)\} $$
are conjugate via the outer automorphism induced by conjugation by the multi-valued loop
$$\left(\begin{matrix} 0& iz^{-1/2}\\ iz^{1/2}&0 \end{matrix}\right)$$

A corollary of the theorem is that, given the existence of $\mu_0$, to prove its bi-invariance, it suffices
to show it is bi-invariant with respect to $K$ (which will be obvious) and the highest root subgroup (which is isomorphic to $SU(2)$).
For this purpose we need to understand how these subgroups act on $\mathbf LG$. We will now recall how this
can be done in the special case $K=SU(2)$.

The left and right actions of the constants $K$ on $Hyp(S^1,G)$
are completely transparent, in terms of the Riemann-Hilbert
factorization $g=g_{-}g_0g_{+}$ for a generalized loop $g\in \mathbb L G$ (in the top stratum):
$$k_L\cdot g\cdot k_R^{-1}=[k_Lg_{-}k_L^{-1}]\cdot [k_Lg_0k_R^{-1}]\cdot [k_
Rg_{+}k_R^{-1}]$$

We will now explicitly write out the action of the highest root subgroup in the case $K=SU(2)$, in terms of Riemann-Hilbert coordinates
for the top stratum. For
this purpose write
$$g_{-}=\left(\begin{matrix} A(z)&B(z)\\
C(z)&D(z)\end{matrix} \right)=1+\left(\begin{matrix} A_1&B_1\\
C_1&-A_1\end{matrix} \right)z^{-1}+\left(\begin{matrix} A_2&B_2\\
C_2&D_2\end{matrix} \right)z^{-2}+..$$
$$g_0=\left(\begin{matrix}a_0&b_0\\c_0&d_0\end{matrix}\right), \qquad a_0d_0-b_0c_0=1$$
and
$$g_{+}=\left(\begin{matrix} a(z)&b(z)\\
c(z)&d(z)\end{matrix} \right)=1+\left(\begin{matrix} a_1&b_1\\
c_1&-a_1\end{matrix} \right)z+..$$

\begin{lemma} Suppose $h=\left(\begin{matrix} d&cz^{-1}\\
bz&a\end{matrix} \right)\in i_{\alpha_0}(SL(2,\mathbb C))$.\\

(a) If $\alpha:=a+B_1\ne 0$, then $h$ acting on the left of $g=g_{-}g_0g_{+}$ is also in the top stratum
and has factorization
$$[\left(\begin{matrix} d&cz^{-1}\\
bz&a\end{matrix} \right)g_{-}\left(\begin{matrix} \alpha&0\\
\beta(z)&\alpha^{-1}\end{matrix} \right)]
\cdot [\left(\begin{matrix} \alpha^{-1}&0\\
-\gamma_0&\alpha\end{matrix} \right)g_0]
\cdot [g_0^{-1}\left(\begin{matrix} 1&0\\
\frac {b}{\alpha}z&1\end{matrix} \right)g_0g_{+}],$$
where $\gamma_0=\frac {-2abA_1+b^2(B_2-A_1B_1)}{\alpha}$
and $\gamma(z)=\gamma_0-bz$.

(b) Similarly if $\alpha':= d-cC_1\ne 0$, then $g\cdot h^{-1}$ is again in the top stratum and
has factorization
$$[g_{-}g_0\left(\begin{matrix} 1&
-\frac {c}{\alpha'}z^{-1}\\0&1\end{matrix} \right)g_0^{-1}]
\cdot
[g_0\left(\begin{matrix} \alpha'^{-1}&-\beta_0\\
0&\alpha'\end{matrix} \right)]
\cdot [\left(\begin{matrix} \alpha'&\beta(z^{-1})0\\
0&\alpha'^{-1}\end{matrix} \right)g_{+}\left(\begin{matrix} d&cz^{-1}\\
bz&a\end{matrix} \right)^{-1}]$$
where $\beta_0=\frac {2cda_1+c^2(c_2+a_1c_1)}{\alpha'}$
$\beta (z^{-1})=\beta_0+cz^{-1}$.
\end{lemma}

The proof of this is a completely straightforward calculation.

\begin{corollary}\label{corollary1}The space of variables $\{B_1,D_{n-1},B_n\}$ is invariant with respect to the left action of $h$.
Define $B_n'=B_n/D_{n-1}$ (where $D_0=1$). Then
$$B_n'(h\cdot g)=\frac{c+dB_n'}{a+bB_n'}$$
i.e. $B_n'$ is equivariant with respect to the action of $h$ on $\mathbf LG$ and the linear fractional
action on the Riemann sphere.
\end{corollary}

We need two other ingredients, quasi-invariance and asymptotic invariance of Wiener measure:

\begin{theorem}Suppose that $g_L\in W^1(S^1,K)$. 

(a) $\nu_{\frac 1T E}$ is inversion invariant, and left and right quasi-invariant with respect to $g_L$.

(b) (asymptotic invariance) 
$$\int\vert 1-\frac {d\nu_{\frac 1T E}(g_Lg)}{d\nu_{\frac 1T E}
(g)}\vert^pd\nu_{\frac 1T E}(g)\to 0\quad as\quad T \to \infty$$
for every $1\le p<\infty$. \end{theorem}

For this see \cite{MM2} or chapter 4 part II of \cite{Pi1}. It is not necessary to assume that $K$ is simply connected 
(hence semi-simple) in this theorem; we simply need an $Ad$ invariant inner product on $\mathfrak k$.

Modulo technical details (some of which we will fill in the next subsection), the proof of Theorem \ref{theorem1} proceeds as follows. The asymptotic invariance of Wiener measure
and Corollary \ref{corollary1} implies that the $B_n'$ distribution of $\nu_{\frac 1T E}$ converges to the uniform
distribution on the sphere for the action of $SU(2)$, as $T\uparrow \infty$. Taking $n=1$, invariance of Wiener measure with respect to constants implies
that all of the variables $A_1,B_1,C_1$ are tight as $T\uparrow \infty$. Now taking $n=2$ we can claim the same thing about $B_2,C_2,A_2,D_2$. One continues
in this fashion to conclude that all the coefficients of $g_-$ are tight. A similar argument shows that $g_0$ and $g_+$ are tight.
Prohorov's theorem implies that $\nu_{\frac 1T E}$ has limit points as $T\uparrow \infty$, and asymptotic invariance implies that these limit points
are $L_{fin}K$ bi-invariant. This proves the existence of $\mu_0$. 

This argument does not show that $\mu_0$ is unique, and it does not lead to a practical means of calculating the joint distributions for
the random variables of the previous paragraph.

\subsection{A More Quantitative Approach}

In this subsection we will formulate a monotonicity conjecture which clarifies the convergence of $\nu_{\frac 1T E}$ in Riemann-Hilbert coordinates
(In Section \ref{calculate1} we will discuss a coordinate free approach).

The arguments in this section (assuming the monotonicity conjecture) apply more generally to prove the existence of $\mu_l$. For this we
have to consider approximations to $\mu_l$ which heuristically have the form
$$d\nu_{\frac 1T E,l}:=\frac{1}{\mathfrak Z}det(A(g)A(g^{-1}))^{l}d\nu_{\frac 1T E} $$
A regularization is required to give this rigorous meaning, because the determinant is identically zero for $g \notin W^{1/2}(S^1,K)$ and Wiener measure
of $W^{1/2}(S^1,K)$ is zero; see Part II of \cite{Pi1}. In \cite{Pi1} we failed to prove an asymptotic invariance result for this sequence of measures for $l\ne 0$. The monotonicity approach might have a better chance of success. 

The following incidentally fills in a technical detail we omitted in the outline of the proof of Theorem \ref{theorem1}.

\begin{theorem}\label{1.1.2} If $T>0$ then with $\nu_{\frac 1T E}$ probability one, $g\in
C^0(S^1,K)$
has a unique Riemann-Hilbert factorization, $g=g_{-}g_0g_{+}$,
where $g_{\pm}$ extend continuously to $D$ (the closed unit disk)
and $D^*$ (the closed unit disk centered at $\infty$), respectively. The $S^1$ restrictions are almost surely
Holder continuous of order $\mu$, for each $\mu <1/2$.  This
implies that for any $\alpha <1$, almost surely
$$\sum_{n>0}n^{\alpha}\vert\hat {g}_{\pm}(\pm n)\vert^2<\infty $$
\end{theorem}

\begin{proof}  Let $C^s=C^s(S^1)$ denote the Banach space of Holder continuous functions
of order $s$. Gohberg and Krupnik proved that the projections $C^{s}\to C^{s}_{\pm}$
are continuous for each $0<s<1$.  This is the crucial ingredient in proving that for
$g\in C^{s}(S^1,G)$, the factors in the generalized Riemann-Hilbert
factorization are also in $C^{s}$, provided that $0<s <1$ and the factorization exists. This
can be proven in at least two different ways.  One is to
use the theory of decomposing algebras, see Example 2,
Section 5, Chapter II of \cite{CG}, where there is also a reference to the work of Gohberg and Krupnik.
Another is to note that the
geometric approach to factorization in \cite{PS} applies (with
$C^{s}_{\pm}$ in place of $H_{\pm}$), given the continuity of the
projections.

For Wiener measure $\nu_{\frac 1T E}$, it is known that $g\in C^{s}$ for
any $s <1/2$, for almost every $g$ $[\nu_{\frac 1T E}]$.  This proves the
first statement.

The last statement follows from the fact that Holder
continuity of order $s$ implies $L^2$-Sobolev smoothness of order
$\alpha$, provided $\alpha<s$.  \end{proof}

\begin{corollary}\label{1.1.3} For every $l\ge 0$, for fixed
$\alpha <1$, and for each $R>0$,
\begin{equation}\label{1.1.4}\nu_{\frac 1T E ,l}\{n^{\alpha}\vert\hat {g}_{+}(n)\vert^2>R\}\to 0\quad
as\quad n\to\infty \end{equation}
and
\begin{equation}\label{1.1.5}\nu_{\frac 1T E ,l}\{n^{\alpha}\vert\hat {x}_{+}(n)\vert^2>R\}\to 0\quad
as\quad n\to\infty \end{equation}
\end{corollary}

\begin{proof}  Suppose that we are given
nonnegative random variables $X_n$ on a probability space.
We have
$$\{\sum X_n<\infty \}\subset \{\lim_{n\to\infty}X_n=0\}=\bigcap_{
\epsilon >0}\bigcup_N\bigcap_{n\ge N}\{X_n\le\epsilon \}.$$
Thus
$$Prob\{\sum X_n<\infty \}=1\quad\implies\quad\forall\epsilon >0,
\quad\lim_{N\to\infty}Prob\{X_n\le\epsilon ,\forall n\ge N\}=1$$
$$\implies\lim_{N\to\infty}Prob\{\exists n\ge N:X_n>R\}=0\implies\lim_{
n\to\infty}Prob\{X_n>R\}=0.$$
for any $R>0$.

Now recall that $\nu_{\frac 1T E ,l}$ is absolutely continuous with
respect to $\nu_{\frac 1T E}$.  To obtain (\ref{1.1.4}) from Theorem \ref{1.1.2}, take
$X_n=n^{\alpha}\vert\hat {g}_{+}(n)\vert^2$.

To obtain (\ref{1.1.5}) it suffices to show that $x_{+}$ has the same
Besov-space characteristics as $g_{+}$, for $a.e.$ $g$ $[\nu_{\frac 1T E}]$.
Now $x_{+}$ and $g_{+}$ are related via the nonlinear bijective
correspondence
$$H^0(\Delta ,0;\mathfrak g,0)\leftrightarrow H^0(\Delta ,0;G,1):x_{+}
\leftrightarrow g_{+},$$
where $\partial x_{+}=g_{+}\partial g_{+}^{-1}$, essentially a holomorphic version of
the classical Ito map.  We know that $g_{+}$ is bounded with
probability one.  The condition that $g_{+}$ is in $B^{1/p}$ is that
$$\int_{\Delta}\vert\frac {\partial g_{+}}{\partial z}\vert^p(1-\vert
z\vert )^{2-p}<\infty$$
(see \cite{Peller}).  The boundedness of $g_{+}$ is equivalent to
boundedness of $g_{+}^{-1}$, because $detg_{+}=1$.  This clearly
implies that $x_{+}$ also lies in $B^{1/p}$, for any $p$, with
probability one.  This implies $(\ref{1.1.5})$. \end{proof}

\begin{remark}  For $\mathfrak g$-valued, or $\mathfrak k$-valued, functions on
$S^1$, the Ito correspondence preserves Besov or Sobolev or
Holder characteristics only for orders of smoothness
$>1/2$.  It is notable that in the present context, where
we use factorization and holomorphicity to regularize
loops, we can propagate smoothness characteristics for
lower orders from nonlinear to linear variables via our
holomorphic Ito correspondence.  The tradeoff is that
factorization is enormously complicated.\end{remark}

We now turn to the key issue.  The question is for
which $\alpha$ is
$$\nu_{\frac 1T E ,l}\{n^{\alpha}\vert\hat {x}_{+}(n)\vert^2>R\}. \label{1.1.7}$$
a decreasing function of $n$, if any?  The supremum
$\alpha_c=\alpha_c(\frac 1T E ,l)$ over all such $\alpha$ is a kind of critical
exponent.

\begin{conjecture}\label{1.1.8}For $\alpha =0$, (1.1.7) is a
nonincreasing function of $n$, for all $\beta >0$, $l\ge 0$, and
$R\ge 0$, i.e.  $\alpha_c(\beta ,l)\ge 0$.  In particular
$$\nu_{\frac 1T E ,l}\{\vert\hat {x}_{+}(n)\vert >R\}\le\nu_{\frac 1T E ,l}
\{\hat {x}_{+}(1)\vert >R\}$$
\end{conjecture}

In thinking about this conjecture, it is instructive to
compare with the abelian case.  When $K=\mathbb T$, the circle
group, Szego's formula for Toeplitz determinants implies
that
$$d\nu_{\frac 1T E ,l}(x_{+})=\prod_{n>0}\frac {\frac 1T E n^2+2ln}{\pi}e^{
-\frac 12(\beta n^2+2ln)\vert x_n\vert^2}dm(x_n)$$
(see Section 3.1 of Part III of \cite{Pi1}).  Thus (\ref{1.1.7}) equals
$$\int_{n^{\alpha}r^2>R}\frac {\gamma}{\pi}e^{-\frac 12\gamma r^2}
rdr=\int_{u>\frac {\gamma}{2n^{\alpha}}R}e^{-u}du=e^{-\frac {\gamma
R}{2n^{\alpha}}},$$
where $\gamma =\beta n^2+2ln$, $x_n=re^{i\theta}$, and $u=\frac 1
2\gamma r^2$.  To
determine the critical exponent $\alpha_c$, we must determine
those $\alpha$ for which the sequence
$$\frac {\beta n^2+2ln}{n^{\alpha}},$$
is nondecreasing.  If $l=0$, then $\alpha_c=2$.  It is increasing
for $\alpha =1$, for any $l$, so in general the critical exponent
is something between $1$ and $2$, depending upon $l$ (If we
were only interested in large $n$, then in general the
critical exponent would be $2$; if we want something
independent of $\beta$, then we have $\alpha_c=1$).

Suppose first that $l=0$.  For $\alpha =2$, we see that
$$\nu_{\frac 1T E}\{n^{\alpha}\vert x_n\vert^2>R\}=\nu_{\frac 1T E}\{\vert\hat{
\theta}_{+}(n-1)\vert^2>R\}=e^{-\frac {\beta}2R}$$
Thus the $\theta_n$ are independent and identically distributed.

In the nonabelian case it is definitely not the case that
the $\theta_n$ are independent, and there is very firm evidence
indicating that they are not identically distributed.  In
fact in the nonabelian case there
is a kind of shift of the level (in a sense that we will
spell out), so that one definitely expects $\alpha_c$ is more like
$1$ in the nonabelian case.

In the remainder of this subsection, we will explore
some of the consequences of $(\ref{1.1.8})$.

\begin{lemma}\label{1.1.9}There exists a constant $d$
(depending only upon $\mathfrak g$) such that
$$\lim_{\beta\to 0}\nu_{\frac 1T E ,l}\{\vert\hat {x}_{\pm}(1)\vert >R
\}\le\frac d{(1+(R/d)^2)^{l+1}}.$$
\end{lemma}

\begin{proof}  Write $g\in \mathcal N^{-}\cdot\dot {H}\cdot \mathcal N^{
+}$ as $l\cdot diag\cdot u$ and
$u=exp(x)$.  We have
$$w=proj_{\mathfrak g_{\theta}\otimes z^{-1}}(x)=proj_{\mathfrak g_{\theta}}
(\hat {x}_{+}(1)). \label{1.1.10}$$
Now in [Pi] we established that the $\lim\nu_{\frac 1T E ,l}$ distribution
of $w$ has a limit as $\beta\to 0$, and this is given by
$$\frac 1E(1+\vert w\vert^2)^{-2-l}dm(w) \label{1.1.11}$$
(This follows from the asymptotic invariance of the $\nu_{\frac 1T E ,
l}$
and (3.2.3) of Part I of \cite{Pi1}).  Thus
$$\lim_{\beta\to 0}\nu_{\frac 1T E ,l}\{\vert w\vert >R\}=(1+R^2)^{-1-
l}. \label{1.1.12}$$

The measure $\nu_{\frac 1T E ,l}$ is biinvariant with respect to the
action of $K$ on $\bf{L} G$.  We have
$$v=\hat {x}_{+}(1)\in \mathfrak g, \label{1.1.13}$$
the right action of $K$ restricts to the adjoint action on
this variable, and hence the $\nu_{\frac 1T E ,l}$ distribution of $v$ is
invariant with respect to the adjoint action of $K$.  This
action is irreducible, hence the $K$ orbit of $e_{\theta}$ spans all of
$\mathfrak g$.  It follows that
$$\{\vert v\vert >R\}\subset\bigcup_i\{\vert\langle v,g_i\cdot e_{
\theta}\rangle\vert >R/d\} \label{1.1.14}$$
for some constant $d$ and finite set $\{g_i\}\subset K$.  Thus
$$\lim_{\beta\to 0}\nu_{\frac 1T E ,l}\{\vert\hat {x}_{+}(1)\vert >R\}
\le (\#\{g_i\})(1+(R/d)^2)^{-1-l}, \label{1.1.15}$$
by invariance and (\ref{1.1.12}).  \end{proof}

\begin{corollary}\label{1.1.16} The set of measures $\{\nu_{\frac 1T E ,l}
\}$ has
weak limits as $\beta\to 0$.
\end{corollary}

 \begin{proof}  It follows from (1.1.8) and (1.1.9) that
$$\lim_{\beta\to 0}\nu_{\frac 1T E ,l}\{\vert\hat {x}_{+}(n)\vert >R\}
\le\lim_{\beta\to 0}\nu_{\frac 1T E ,l}\{\vert\hat {x}_{+}(1)\vert >R\}
\le\frac d{(1+(R/d)^2)^{1+l}}, \label{1.1.17}$$
for $a.e.R$, for $n>0$.  This implies the existence of weak
limits $\mu_l$.\end{proof}

From (\ref{1.1.17}) it follows that
\begin{equation}\sum_{n>0}\mu_l\{\vert\hat {g}_{+}(n)\vert >n^{\delta}\}<\infty
, \label{1.1.18}\end{equation}
hence by the first Borel-Cantelli lemma
\begin{equation}\mu_l\{\limsup_{n\to\infty}(n^{-\delta}\vert\hat {g}_{+}(n)\vert
)>1\}=0 \label{1.1.19}\end{equation}
provided that $\delta >1/(2+2l)$ (of course we have a similiar
kind of estimate for $g_{-}$).  This easily implies that $g_{+}$ is
actually a holomorphic function in the open unit disk.
Hence we have the following

\begin{corollary} We have $\mu_l(Hyp(S^1,G))=1$, and $\mu_
l$
is quasiinvariant with respect to the action of $L_{an}K$ on
$Hyp(S^1,G)$.
\end{corollary}

\begin{proof}  The first statement follows immediately
from (\ref{1.1.19}).  Since we now know that $\mu_l$ is a weak limit
point of $\nu_{\frac 1T E ,l}$ on $Hyp(S^1,G)$, the result follows from the
asymptotic quasiinvariance of $\nu_{\frac 1T E ,l}$.  \end{proof}

In the abelian case ($l>0$), the factors $g_{\pm}$ very nearly,
but do not quite, have boundary values on $S^1$ (there is a logarithmic divergence).  The
estimates above suggest that the behavior in the
nonabelian case is roughly the same.

\section{Characterizing Measures and Measure Classes}\label{uniqueness}

The first conjecture states that there is a very simple characterization of the measure $\mu_0$
and various deformations:

\begin{conjecture}\label{conj1} There exists a unique Borel probability measure $\mu_0$ on $\mathbf L G$
which is bi-invariant with respect to $L_{fin}K$.

More generally there exists an essentially unique Borel measure (denoted $\mu^{|\mathcal L|^{2l}}$)
with values in the positive line bundle $|\mathcal L|^{2l}:=(\mathcal L \otimes \overline{\mathcal L})^{\otimes l} \to \mathbf L G$ which is bi-invariant with respect to $L_{fin}K$, for $l>-1$.

\end{conjecture}

\begin{remarks}\label{remarksconj1} (a) Fix $l\in \mathbb N$. At a heuristic level, given sections $\sigma_i$ of $\mathcal L^{*\otimes l}$
\begin{equation}\label{heuristic1}\sigma_1\otimes \overline{\sigma_2}d\mu^{|\mathcal L|^{2l}}=\frac{1}{\mathfrak Z}\langle\sigma_1,\sigma_2\rangle d\mu_0\end{equation}
where $\langle \cdot,\cdot \rangle$ is a unitarily invariant Hermitian structure for $\mathcal L^{*\otimes l}$.
This structure is only defined on unitary loops of order $W^{1/2}$, which has $\mu_0$ measure zero.
So we have to think of $d\mu^{|\mathcal L|^{2l}}$ as having values in $|\mathcal L|^{2l}$.

(b) In place of $\mu^{|\mathcal L|^{2l}}$, it is often convenient to consider the coordinate expression
$$d\mu_l:=|\sigma_0|^{2l}\mu^{|\mathcal L|^{2l}}$$
where $\sigma_0$ is the `fundamental matrix coefficient'; $\mu_l$ is an ordinary Borel probability measure - an implicit normalization - which is bi-quasi-invariant with respect to translations by $L_{fin}K$, and having a Radon-Nikodym derivative with a specific form determined by the line bundle. We will clarify this when we write down explicit formulas in the next section.

(c) To put this conjecture into perspective, compare this with the finite dimensional analogue of $K$ acting from the left and right of $G$. For this action the $K$ Haar measure is far from the unique $K\times K$ invariant probability measure on $G$. At one extreme the $K$ Haar measure has minimal support. At the opposite extreme there are $K$ bi-invariant measures which are absolutely continuous with respect to the Haar measure for $G$. In general $K$ bi-invariant probability measures can be classified using the spherical (or Harish-Chandra) transform; see chapter IV of \cite{Helgason}.

In the (in my view unlikely) event that Conjecture \ref{conj1} is false, then there must be an affine analogue of the spherical transform
which classifies the possibilities, see Subsection \ref{Diagdistribution} for a possible analogue.

(d) For some perspective on the centrality of this conjecture, see Subsection \ref{symmetrizable}.

(e) For a related negative result, to put the use of the hyperfunction and formal completions into perspective, see \cite{CM}.
\end{remarks}

One practical reason this conjecture is important is that there are multiple approaches to constructing the
measure $\mu_0$ and its deformations. It is not a priori clear that different approaches yield the same measure.
For example in both \cite{Pi1} and \cite{Pi2}, the basic strategy is to show
that Wiener measure on $C^0(S^1,K)\subset\mathbf LG$ has weak
limits in a coordinate system for $\mathbf LG$ as the inverse temperature
parameter $\beta\to 0$. It is not known that the limit is unique. A crucial ingredient in the proof of the
bi-invariance of these limits is the asymptotic invariance of
Wiener measure (as $\beta \to 0$), established earlier by Marie and Paul Malliavin
(see \cite{MM2} and section 4.1 of Part III of \cite{Pi1}). Conjecturally one
can also use heat kernel measures (denoted
$\nu^{(s)}_t$) which are parameterized by a degree of smoothness $s>1/2$ and temperature (or time) $t$, as $t\uparrow \infty$ (see \cite{Pi3}). Asymptotic invariance has not been proven for this limit (It is possible that one can also use
the limit $s\downarrow 1/2$, but this is a more subtle question which I will return to later).
In the next section we will suggest another possible construction, using explicit formulas; again, it is not clear this will yield the same measure.

Conjecture \ref{conj1} implies a number of statements which will probably have to be resolved first, e.g.

\begin{conjecture}\label{conj10} There exists a unique $L_{fin}K$-invariant Borel probability measure on the fundamental `homogeneous space'
$$\mathbf LG/G(\mathbb C[[z]])=G(\mathbb C((z^{-1})))/G(\mathbb C[z])$$

There is a similar statement involving level $l$ and line bundles.
\end{conjecture}

To clarify the statement, the space in question is an equivariant quotient of $\mathbf L G$, for the left action of
$L_{fin}K$, hence there does exist an invariant measure as in the first part of the statement ($\mu_0$ can be pushed forward). To prove Conjecture \ref{conj1} it is necessary, in linear Riemann-Hilbert coordinates, to show that invariance determines the joint distribution for $\theta_-,g_0$ and $\theta_+$; in Conjecture \ref{conj10} it suffices to show that invariance determines the distribution for $\theta_-$. One can argue that this is more plausible than Conjecture \ref{conj1} because the finite dimensional analogue, $K$ acting on $K/T=G/B^+$, does have a unique invariant measure, whereas this is not true for
$K\times K$ acting on $G$.

I made an attempt to prove uniqueness of $\mu_0$ on the basic homogeneous space (in the case $K=SU(2)$)
in section 2 of \cite{Pi2}, using a series of computable distributions for $\mu_0$. I now think this was wrong-headed. A better approach
is outlined in the next section.

Here is a more technical version of Conjecture \ref{conj10}, involving line bundle valued measures.

 Suppose that $\Lambda$ is a dominant integral functional and consider
the Borel subgroup
$$B^+:=\{\widehat g_+\in \widehat{G(\mathbb C[z])}:g_+(0)\in \dot B^+\}$$
$\Lambda$ defines a character $B^+$,  , and in turn there is a holomorphic line
$$\mathcal L_{\Lambda}:=\widehat{G(\mathbb C((z^{-1})))}\times_{B^+}\mathbb C $$

\begin{conjecture}\label{conj11} (a) There exists a unique $L_{fin}K$-invariant measure $\mu^{|\mathcal L_{\Lambda}|^2}$ having values in the positive line bundle $|\mathcal L_{\Lambda}|^2$ which is normalized such that

$$\mu_{\Lambda}:=\sigma_{\Lambda}\otimes \overline{\sigma_{\Lambda} } \mu^{|\mathcal L_{\Lambda}|^2}$$ is a probability measure.

(b) When $\Lambda=l\Lambda_0=l\gamma$, then $\mu_{\Lambda}=\mu_l$ (this is trivial).

(c) If $\Lambda$ has level $l$, then  $[\mu_{\Lambda}]=[\mu_l]$, i.e. the measure class depends only on the level.
\end{conjecture}

The following refers to the natural action of analytic
homeomorphisms of $S^1$ on the hyperfunction completion of the loop group, (\ref{viraction}).

\begin{conjecture}\label{conj2} The measure $\mu_0$ is supported on the hyperfunction completion $Hyp(S^1,G)$.
It is bi-invariant with respect to $H^0(S^1,K)$ and invariant with respect to $C^{\omega}Homeo(S^1)$.

More generally the measure $\mu^{|\mathcal L|^{2l}}$ is supported on the hyperfunction completion.
It is bi-invariant with respect to the natural actions of $C^{\omega}(S^1,K)$ and $C^{\omega}Homeo(S^1)$.

These measures have the property that the associated unitary representations extend continuously to $W^{1/2}(S^1,K) \times W^{1/2}(S^1,K)$ and $W^{1+1/2}Homeo(S^1)$.
\end{conjecture}

In reference to the first claim, we will gain more insight into the support of the measure $\mu_0$ when we compute in the next section. The use of the hyperfunction completion is akin to asserting that typical Brownian paths are continuous - we can say more, but this is enough for many purposes.

Note that $PSU(1,1)\subset C^{\omega}Homeo(S^1)$, and $PSU(1,1)$ acts in a natural way on $\theta_{\pm}$, in linear Riemann-Hilbert coordinates. Thus we are asserting that the distributions for $\theta_{\pm}$ are conformally invariant, a very strong statement that we can potentially check against specific calculations.

A measure having values in a line bundle determines a measure class on the base.

\begin{conjecture}\label{conj3} The measure classes arising in the previous conjecture exhaust all ergodic $L_{fin}K$ bi-invariant measure classes
on $\mathbf L G$ with the property that the natural associated unitary representation of $L_{fin}K \times L_{fin}K$ (realized naturally, on half-densities for the measure class) extends continuously to a strong operator continuous unitary representation of $W^{1/2}(S,K)\times W^{1/2}(S,K)$.

\end{conjecture}

The continuity requirement is essential. For example Wiener measures and heat kernel measures determine $L_{fin}K$ bi-invariant measure classes which are disjoint from the measure classes in the previous conjectures. $W^{1/2}$ loops are akin to the Cameron-Martin Hilbert space associated to a Gaussian measure on a topological vector space.

This conjecture implies that any such invariant measure class, which a priori has nothing to do with line bundles,
has a representative which is invariant in a natural sense related to the universal central extension of $LK$.

\begin{question} Is $\mu_0$ the unique probability measure on $Hyp(S^1,G)$ which is fixed by $K\times C^{\omega}Homeo(S^1)\times K$?
\end{question}

\section{Computing Invariant Measures I}\label{calculate1}

\subsection{Pushforwards to Moduli Spaces}\label{modulispace}

A first problem: The uniqueness conjectures of the previous section lead to a test. Given any compact manifold $X$ and a strong operator continuous representation
for $Diff(X)$, Shimomura (\cite{Shimomura}) showed that the space of smooth vectors is dense. Assuming
the truth of Conjectures \ref{conj2} and \ref{conj3}, there is a unitary representation of the group of smooth homeomorphisms on $L^2(\mu_0)$ (or more invariantly, on half densities). Hence there should be smooth vectors
(it does not make sense to talk about analytic vectors; diffeomorphisms of a manifold is not an analytic Lie group, see \cite{Milnor}).

The measure $\mu_0$ is a limit of Wiener measures. One way to think about Wiener measure
$\nu_{\frac 1T E}$ is the following.  Suppose that $V$ is a finite set of vertices around $S^1$. There is then a group evaluation homomorphism
$$eval_V:C^0(S^1,K) \to \prod_V K:g \to (g(v))_{v\in V}$$
and
$$(eval_V)_*(\nu_{\frac 1T E})=\frac{1}{p_{2\pi T}(1)}\prod_{e\in E}p_{l(e)T}(g_{\partial e})\prod_{v\in V}d\lambda_K(g_v)$$
where $T=1/\beta$, $l(e)$ denotes the length of an edge $e$, $p_t$ is the heat kernel on $K$, and $d\lambda_K$ is normalized Haar measure on $K$. These projections are coherent (as $V$ varies), and Wiener measure is determined and essentially defined by these projections. The pullbacks of smooth functions are smooth vectors.

A second problem: When we pass to the limit $T \uparrow \infty$, it is no longer true that we can evaluate generalized loops at points, almost surely relative to $\mu_0$. Instead we look at the natural functions which are defined on the hyperfunction completion. This will solve both of our problems.

Suppose that $\Sigma$ is a compact Riemann surface, and let $c:S^1 \to \Sigma$ be an embedded analytic loop in $\Sigma$.  The map $c$ extends
uniquely to a bi-holomorphic embedding
$c:\{1-\epsilon <\vert z\vert <1+\epsilon \}\to\Sigma$ for some $
\epsilon >0$.  Given a pair
$(g,h)$ representing $[g,h]\in Hyp(S^1,G)$, we obtain a holomorphic bundle on $\Sigma$ by using $g$ as
a transition function on an $\epsilon'$-collar to the left of $c$ and
$h$ as a transition function on an $\epsilon'$-collar to the right of
$c$, for some $\epsilon'<\epsilon$, depending upon the pair $(g,h)$.  The
isomorphism class of this bundle is independent of the
choice of $\epsilon'$, and depends only upon $[g,h]\in Hyp(S^1,G)$. This can be summarized as follows:

\begin{proposition} (a) There is an induced map
$$P(c):Hyp(S^1,G) \to Bun_G(\Sigma) $$
where a $G$ valued hyperfunction $g$ maps to the holomorphic $G$ bundle defined by using $g$ for transition functions, and $Bun_G(\Sigma)$ denotes the set (or stack in algebraic geometry) of holomorphic $G$ bundles.

(b) There is a holomorphic action
$$H^0(\Sigma\setminus Im(c),G)\times Hyp(S^1,G)\to Hyp(S^1,G):f,[g,h]\to
[f\vert_{S^1_{-}}\circ c g,h f\vert_{S^1_{+}}^{-1}\circ c],$$
and the mapping $P(c)$ induces an isomorphism of sets
$$Hyp(S^1,G)/c^*H^0(\Sigma\setminus Im(c),G)\to Bun_G(\Sigma)$$

(c) If $\sigma\in C^{\omega}Homeo(S^1)$, then the induced map
$$Hyp(S^1,G)\stackrel{\sigma}{\rightarrow}Hyp(S^1,G)\stackrel{P(c)}{\rightarrow}Bun_G(\Sigma)$$
equals $P(c\circ\sigma^{-1})$.
\end{proposition}

A fundamental theorem of Narasimhan and Seshadri, and Ramanathan, asserts that the open
dense subset $Bun_G^0(\Sigma)$ of stable $G$ bundles can be identified real analytically
with the set of irreducible homomorphisms $\pi_1(\Sigma)\to K$, modulo conjugation by $K$ (see \cite{Donaldson}
for an alternate proof and references). Via this identification, $Bun_G^0(\Sigma)$ supports a unique mapping class group invariant Borel probability measure, the normalized Goldman symplectic volume element (see \cite{Pi1.5}).

\begin{conjecture}\label{conj4} (a) The pushforward $P(c)_*\mu_0$ is equal to
the unique mapping class group invariant probability measure on $Bun_G^0(\Sigma)$.

(b) Given a smooth function $f$ with compact support in $Bun_G^0$, $P(c)^*f$ is a smooth vector for the action
of $C^{\omega}Homeo(S^1)$ on $L^2(\mu_0)$.

(c) There are similar conjectures involving $\mu^{|\mathcal L|^{2l}}$ and the $l$th power of a positive line bundle on the moduli space.
\end{conjecture}

\begin{remark}\label{conj4remarks} (i) One can reformulate (a) in the following way. Given $c$, project $\nu_{\frac 1T E}$ to the moduli space. It seems obvious, but is not so easy to prove, that $P(c)_*\nu_{\frac 1T E}$ is in the Lebesgue class. Now let $\beta \downarrow 0$, i.e. increase the entropy. The limit should be as information free as possible.
It seems clear, but I do not have a proof, that the limit should be in the Lebesgue class and invariant with respect to the subgroup of the mapping class group fixing the homotopy class of $c$. This would imply
that the density is a central function of $(\rho,[c])$, the pairing of $\rho \in H^1(\Sigma,K)$ and the homotopy class of $c$. Thus the crux of the matter seems to be the question, is this density equal to one, as I have conjectured?

(ii) This form of (a) in the first remark makes sense for an abelian group such as the circle $\mathbb T$ in place of $K$. In this abelian case $P(c)$ (restricted to the identity component, see (\ref{abeliancase})) is a group homomorphism
$$Hyp(S^1,\mathbb C^{\times})_0\to H^1(\mathcal O_{\mathbb C^{\times}})_0$$
onto a compact torus (here and in what follows, $\mathcal O$ denotes the sheaf of holomorphic functions). The Lie algebra covering map
$$Hyp(S^1,\mathbb C)=H^0(S^1)^{*}\to H^1(\mathcal O)=H^{0,1}(\Sigma )\cong H^{1,0}(\Sigma
)^{*} $$
is the dual of the restriction map
$$H^{1,0}(\Sigma )\to \mathcal O(S^1):\theta\to\frac {c^{*}\theta}{dz}$$
where $z$ is the usual complex coordinate along $S^1$ (see Section 1 of \cite{BP0}).

Asymptotic invariance of Wiener measure implies that the pushforward of measures is asymptotically translation invariant. Hence in this reformulation, in this abelian context (and we restrict attention to the identity component of the loop group), the limit as $\beta\downarrow 0$ is the normalized Haar measure for the image of $P(c)$. This image is all of the identity component for the moduli space precisely when the image of $c$ is not a `straight line'; see the appendix to \cite{BP0}.
\end{remark}

Part (b) of the conjecture raises a critical completeness question:

\begin{question}Do functions as in (b) of Conjecture \ref{conj4} span a dense subspace of $L^2(\mu_0)$?
Below we will note there are similar smooth vectors associated to parabolic bundles, and perhaps this generalization is essential to obtain a dense subspace.
\end{question}

\subsection{Reconstructing $\mu_0$}

We now want to expand on this to propose a direct construction of $\mu_0$, in analogy with the Wiener measure construction above.

Suppose that $C$ is a finite collection of analytically parameterized embedded loops $c:S^1 \to \Sigma_c$. There is then a map
$$P(C):Hyp(S^1,G) \to \prod_{c\in C} Bun_G(\Sigma_c):g \to (P(c)(g)) $$

\begin{conjecture}\label{conj41}  Suppose that there does not exist a Riemann surface isomorphism of $\Sigma_{c_1}$ and $\Sigma_{c_2}$ which maps $Im(c_1)$ to $Im(c_2)$ for distinct $c_1,c_2\in C$. Then the pushforward $P(C)_*\mu_0$ is in the Lebesgue class for $\prod_{c\in C} Bun_G^0(\Sigma_c)$.
\end{conjecture}

In the cases which we have excluded in this conjecture, the map $Hyp(S^1,G) \to \prod_C Bun_G(\Sigma_c)$ is not surjective, hence $[\mu_0]$ will not project to the Lebesgue class. It is possible that a stronger hypothesis is needed, e.g. maybe it is necessary to require that the free homotopy classes of the $c_i$ are distinct.

Assuming the truth of this conjecture, the basic task is to compute the density $\delta_C(\rho_1,...,\rho_n)$ on $H^1(\Sigma,K)\times ...\times H^1(\Sigma,K)$, where we are identifying stable bundles with irreducible $K$ representations of the fundamental group, using Narasimhan-Seshadri-Ramanathan, and we use the product of
of the mapping class group invariant measures as background.

\begin{question}\label{ques41} Suppose that the surfaces $\Sigma_c$ are topologically distinct. Is $\delta_C=1$, implying that the distributions $c_*\mu_0$ are independent, $c\in C$? More generally if $C$ is partitioned into
subsets $C_1,..,C_n$ corresponding to topologically distinct surfaces, are the $P(C_i)*\mu_0$ independent?

\end{question}

In general $\delta_C$ will depend on some kind of interaction of the loops, but what is the nature of this interaction? For example a priori the density $\delta_C$ depends on how the loops are parameterized. If the measure $\mu_0$ is invariant with respect to reparameterizations, then one can simultaneously reparameterize all of them, but not necessarily individually. The simplest hypothesis is the following: If $C=\{c_1,...,c_n\}$, then we obtain
$n$ random $\rho_i\in Hom(\pi_1(\Sigma),K)$ (modulo conjugation by $K$ for each $i$, where $\rho_i$ corresponds to $P(c_i)(g)$,
$g\in Hyp(S^1,G)$ distributed according to $\mu_0$). Maybe the density only depends on the $n^2$ elements of $K$, $\rho_i([c_j])$, where $[c_j]$ denotes the homotopy class of $c_j$. This would be a stunning simplification. However I do not see some natural formula for the density emerging.

Suppose that $l$ is positive level. One can pose the analogous question for $\mu^{|\mathcal L|^{2c}}$, although
the basic intuition seems to be lacking for a measure having values in a line bundle.

\subsection{Parabolic Reductions}

There is an obvious, but technically demanding, generalization of these conjectures in which the $\Sigma_c$ are replaced by punctured surfaces with parabolic markings. This may be essential to know that we are actually obtaining a dense subspace of functions. It also leads to some consistency checks. We refer to \cite{TW} for a clear exposition of the background.

Fix a triangular decomposition for $\mathfrak g$. A parabolic subgroup is a Lie subgroup $P\subset G$ such that
$B^+\subset P$ (A theorem of Tits asserts that parabolic subgroups of the triple $(G,B^+,H)$ are in bijective
correspondence with subsets of the positive simple roots).

Assume that $\Sigma$ is a closed Riemann surface, and that $\{z_1,...,z_n\}$ is a finite set of points, and to each point $z_i$, there is
an associated parabolic subgroup $P_i$. If we are given an analytic embedding $c:S^1\to \Sigma\setminus\{z_i\}$,
then there are projections
\begin{equation}\label{projections}Hyp(S^1,G)\to Hyp(S^1,G)/H^0(\Sigma\setminus Im(c),z_1,...,z_n;G,P_1,...,P_n)
\end{equation}
$$\to Hyp(S^1,G)/H^0(\Sigma\setminus Im(c),z_1,...,z_{n-1};G,P_1,...,P_{n-1}) $$
$$\to ...\to Hyp(S^1,G)/H^0(\Sigma\setminus Im(c),G)=Bun_G(\Sigma)$$
The measure $\mu_0$ has to push forward to a coherent family of probability measures on these quotients,
and our aim is to identify these measures. This will incidentally generate some consistency checks.

Recall that given $(\Sigma,z_1,...,z_n)$ and associated
parabolic subgroups $P_i$, a parabolic bundle
is a G-bundle $P\to \Sigma$ together with, for each $i$,  a fixed isomorphism
$ P_{z_i}/P_i \to G/P_i$. Two parabolic bundles are isomorphic if there is an isomorphism of
holomorphic bundles, i.e. a $G$ -equivariant
holomorphic bijection $F: P \to P'$, such that for each $i$ there are commuting diagrams
$$\begin{matrix} P_{z_i}/P_i & & \\ &  \searrow &\\ \updownarrow & & G/P_i\\& \nearrow& \\
P'_{z_i}/P_i & &\end{matrix} $$

\begin{proposition} Using a hyperfunction as a (pair of) transition functions (as at the top of this section) induces
a natural identification
$$Hyp(S^1,G)/H^0(\Sigma\setminus Im(c),z_1,...,z_n;G,P_1,...,P_n) \to Bun_G(\Sigma,z_1,...,z_n;P_1,...,P_n)$$
the set of isomorphism classes of parabolic bundles with the given marked points.
\end{proposition}

In order to define stable, and semistable, for parabolic bundles,
it is necessary to introduce more structure.  Given
$P_i$ we assume that there is a fixed conjugacy class $C_i$
with the property that it is related to $P_i$ in the
following way.  There are bijective correspondences
$$K/conj\leftrightarrow T/\dot W \leftrightarrow \mathfrak t/W$$
where $\dot W$ is the Weyl group of $(K,T)$ and  $W=\dot W\propto\check{\dot T}$ is the affine Weyl group.
The last space is the closure of the fundamental positive alcove.
$C_i$ represents a point in these identified spaces, and we
additionally fix a representative $x_i$ for $C_i$ in $\mathfrak t$. We
require that the inclusion $K/C_K(x_i)\to G/P_i$ is an isomorphism.
Using this one can define stable and semistable for parabolic bundles;
see \cite{TW}. But we will bypass this.

Suppose that we are given a representation
$$g:\pi_1(\Sigma\setminus\{z_i\},z_0) \to K $$
with the property that for fixed closed loops $\gamma_i$ surrounding the points $z_i$,
$g(\gamma_i)\in C_i$. The inclusion
$$\Sigma\setminus\{z_i\},z_0) \to (\Sigma,z_0)$$
induces a map
$$\pi_1(\Sigma\setminus\{z_i\},z_0) \to \pi_1(\Sigma,z_0)$$
Using the homomorphism $g$, there is an induced holomorphic bundle
$$\widetilde{\Sigma\setminus\{z_i\}}\times_{\pi_1(\Sigma\setminus\{z_i\},z_0)} G $$
(the tilde indicates the universal covering) together with a reduction
$$\widetilde{\Sigma\setminus\{z_i\}}\times_{\pi_1(\Sigma\setminus\{z_i\},z_0)} K $$
that defines a unitary connection which has prescribed holonomy around the marked points.

\begin{proposition}The map
$$H^1(\Sigma,z_1,...,z_n;C_1,...C_n) \to Bun_G(\Sigma,z_1,...,z_n;P_1,...P_n)$$
induces a bijection of the set of irreducible
representations, up to conjugation, with the set of stable parabolic bundles.
\end{proposition}

If $genus(\Sigma )>0$ then there is a
unique Lebesgue class probability measure on
$H^1(\Sigma ,z_i;K,C_i)$ which is invariant with respect to the mapping class
group $MCG(\Sigma ,z_1,..,z_n)$.

\begin{conjecture} Suppose that $c$ is a analytic embedding of $S^1$ in $\Sigma\setminus\{z_i\}$. The pushforward  $P(c)_*\mu_0$ is absolutely continuous with respect to the unique mapping class group invariant probability measure on the set of stable bundles, $Bun_G^0(\Sigma,z_1,...,z_n;P_1,...P_n)$. The density is
invariant with respect to the subgroup of the mapping class group which fixes the homotopy class of $c$.
\end{conjecture}

One can now ask if it might be possible to reconstruct the measure $\mu_0$ from these pushforwards.
Unfortunately there is something missing. Even if one could prove these measures are coherent (which I have not done), relative to the maps (\ref{projections}), this would not give an independent method of constructing $\mu_0$, because it is necessary somehow vary the surface.

There is a relatively straightforward generalization of all of this to the measures with values in line bundles from the previous section.

\section{Computing Invariant Measures II: Root Subgroup Factorization}\label{calculate2}

\subsection{The $SU(2)$ Case}

We first consider $K=SU(2,\mathbb C)$ and $G=SL(2,\mathbb C)$. From a technical point of view, this turns out to be
a dramatic simplification.

A triangular factorization for $g\in LG$ is a multiplicative
factorization of the form
\begin{equation}\label{trifactorization}g=l\cdot m\cdot a\cdot u,\end{equation}
where
\[l=\left(\begin{array}{cc}
l_{11}&l_{12}\\
l_{21}&l_{22}\end{array} \right)\in H^0(\Delta^{*},G),\quad
l(\infty )=\left(\begin{array}{cc}
1&0\\
l_{21}(\infty )&1\end{array} \right),\] $l$ has appropriate
boundary values on $S^1$ (depending on the smoothness properties
of $g$), $m=\left(\begin{array}{cc}
m_0&0\\
0&m_0^{-1}\end{array} \right)$, $m_0\in S^1$,
$a=\left(\begin{array}{cc}
a_0&0\\
0&a_0^{-1}\end{array} \right)$, $a_0>0$,
\[u=\left(\begin{array}{cc}
u_{11}&u_{12}\\
u_{21}&u_{22}\end{array} \right)\in H^0(\Delta ,G),\quad
u(0)=\left(\begin{array}{cc}
1&u_{12}(0)\\
0&1\end{array} \right),\] and $u$ has appropriate boundary values
on $S^1$, where $\Delta$ ($\Delta^*$) is the open unit disk
centered at $z=0$ ($z=\infty$, respectively), and $H^0(U)$ denotes
holomorphic functions in a domain $U\subset \mathbb C$. The basic
fact is that for $g\in LK$ having a triangular factorization,
there is a second unique `root subgroup factorization'
\begin{equation}\label{coordinate}g(z)=k_1^{*}(z)\left(\begin{matrix} e^{\chi(z)}&0\\
0&e^{-\chi(z)}\end{matrix} \right)k_2(z),\quad \vert
z\vert=1,\end{equation} where
\begin{equation}\label{k1product}k_1(z)=\left(\begin{matrix}a_1(z)&b_1(z)\\-b_1^*(z)&a_1^*(z)\end{matrix}\right)=
\lim_{n\to\infty}\mathbf a(\eta_n)\left(\begin{matrix} 1&-\bar{\eta}_nz^n\\
\eta_nz^{-n}&1\end{matrix} \right)..\mathbf
a(\eta_0)\left(\begin{matrix} 1&-\bar{
\eta}_0\\
\eta_0&1\end{matrix} \right),\end{equation} $\mathbf{\chi}(z)
=\sum\mathbf{\chi}_jz^j$ is a $i\mathbb R$-valued Fourier series
(modulo $2\pi i\mathbb Z$),
\begin{equation}\label{k2product}k_2(z)=\left(\begin{matrix}d_2^*(z)&-c_2^*(z)\\c_2(z)&d_2(z)\end{matrix}\right)=
\lim_{n\to\infty}\mathbf a(\zeta_n)\left(\begin{matrix} 1&\zeta_nz^{-n}\\
-\bar{\zeta}_nz^n&1\end{matrix} \right)..\mathbf
a(\zeta_1)\left(\begin{matrix} 1&
\zeta_1z^{-1}\\
-\bar{\zeta}_1z&1\end{matrix} \right),\end{equation} $\mathbf
a(\cdot )=(1+\vert\cdot\vert^2)^{-1/2}$, and it is understood that
if $g\in C^{\infty}(S^1,K)$, then the coefficients are rapidly
decreasing, and similarly for other function spaces; conversely a root
subgroup factorization as in (\ref{coordinate}) implies that $g$ has a
triangular factorization (see \cite{Pi4}).

For $l>-1$ (minus half the dual Coxeter number, as in Conjecture \ref{conj1}),
let $\tilde{\mu}_l$ denote the product probability
measure
\begin{equation}\label{productmeasure1}\left(\prod_{i=0}^{\infty}\frac {1+(l+2)i}{\pi}\frac
{d\lambda (\eta_
i)}{(1+\vert\eta_i\vert^2)^{2+(l+2)i}}\right)\times \left(\prod_{
j=1}^{\infty}\frac
{2j(l+2)}{\pi}e^{-2j(l+2)\vert\mathbf{\chi}_j\vert^2}d\lambda
(\mathbf{\chi}_j)\right)$$
$$\times d\lambda
(e^{\chi_0})\times\left(\prod_{k=1}^{\infty}\frac
{(l+2)k-1}{\pi}\frac {d\lambda (\zeta_
k)}{(1+\vert\zeta_k\vert^2)^{(l+2)k}}\right),\end{equation} where
$d\lambda$ denotes Lebesgue measure for $\mathbb C$, or normalized
Haar measure for $S^1$.  The following is technically useful, because it reduces $0-1$
issues to questions about Gaussians.

\begin{lemma}\label{lemma1}In the sense of measures, $\tilde{\mu}_l$ is equivalent
to the Gaussian measure
\begin{equation}\label{Gaussianbackground}\left(\prod_{i=0}^{\infty}\frac
{2+(l+2)i}{\pi}e^{-(2+(l+2)i)\vert\eta_i\vert^2} d\lambda (\eta_
i)\right)\times \left(\prod_{ j=1}^{\infty}\frac
{2j(l+2)}{\pi}e^{-2j(l+2)\vert\mathbf{\chi}_j\vert^2}d\lambda
(\mathbf{\chi}_j)\right)$$
$$\times d\lambda
(e^{\chi_0})\times\left(\prod_{k=1}^{\infty}\frac
{(l+2)k}{\pi}e^{-(l+2)k\vert \zeta_k\vert^2}d\lambda (\zeta_
k)\right)\end{equation}
\end{lemma}

This follows from Kakutani's criterion
for equivalence of product measures.

Our basic heuristic claim is that for $l=0$, $\tilde{\mu}_l$ is a
coordinate expression for the invariant (ordinary) measure $\mu_0$ for $LK$ (more generally $\tilde{\mu}_l$
is a coordinate expression for the positive line bundle valued measure $\mu^{|\mathcal L|^{2l}}$). To give
rigorous meaning to this claim, it is necessary to consider a
completion of the loop group, as in the statement of Theorem
\ref{theorem1}. The basic difficulty is that the measure
$\tilde{\mu}_l$ is supported on sequences for which the sum $\chi$
and the products
(\ref{k1product}) and (\ref{k2product}) (marginally) fail to
converge (the behavior of the random sum
$\sum_{n>0}\chi_n z^n$ is analyzed in Chapter 13
of \cite{K}; because of Lemma \ref{lemma1}, the same qualitative analysis applies
to the products). Consequently it is {\bf not} possible to view $\tilde{\mu}_l$ as a
countably additive measure on any kind of pointwise defined, or
even measurable, loop space of $K$ or $G$.

Initially, following standard practice, we consider a relatively
thick completion, the formal completion $\mathbf LG$. A generic point in
$\mathbf LG$ has a unique formal triangular factorization
\begin{equation}\label{trifactorization2}g=l\cdot m \cdot a\cdot u \end{equation}
where $m\in T$ (the diagonal
torus in $SU(2)$),
$a\in A:=exp(\mathbb R\left(\begin{matrix} 1&0\\
0&-1\end{matrix} \right))$, $l\in \mathcal N^{-}$, the (profinite
nilpotent) group consisting of formal power series in $z^{-1}$,
$$l=\left(\begin{matrix} 1+\sum_{j=1}^{\infty}A_jz^{-j}&\sum_{j=1}^{\infty}B_jz^{
-j}\\
\sum_{j=0}^{\infty}C_jz^{-j}&1+\sum_{j=1}^{\infty}D_jz^{-j}&\end{matrix}
\right ),$$ with $det(l)=1$ (as a formal power series in
$z^{-1}$), and $u\in \mathcal N^{+}$, the group consisting of
formal power series in $z$,
$$u=\left(\begin{matrix} 1+\sum_{j=1}^{\infty}a_jz^j&\sum_{j=0}^{\infty}b_jz^j\\
\sum_{j=1}^{\infty}c_jz^j&1+\sum_{j=1}^{\infty}d_jz^j&\end{matrix}
\right),$$ with $det(u)=1$. The Fourier coefficients of the factors
in the triangular factorization of a generalized loop are well-defined
random variables (as is typically the case for fields that are
relevant to quantum field theory).

The formal completion $\mathbf LG$ is {\bf not} a group, but it is a standard
Borel space which is acted on
naturally from the left and right by the complex polynomial loop
group $L_{fin}G:=G(\mathbb C[z,z^{-1}])$.

The first claim is that the composite mapping
\begin{equation}\label{mapping}\{(\eta,\chi,\zeta)\} \to \mathbf
LG \end{equation}
$$(\eta,\chi,\zeta) \to g=k_1(\eta)^*\left(\begin{matrix}e^{\sum\chi_jz^j}&0\\0&e^{-\sum\chi_jz^j}\end{matrix}\right)k_2(\zeta)
\to l(g)\cdot m(g)\cdot a(g)\cdot u(g)$$ which is a priori only
defined for rapidly decreasing sequences, extends to a measurable
mapping which is defined $\tilde{\mu}_l$-almost surely. To be more precise, if $\eta$, $\chi$ and $\zeta$
are $l^2$ sequences, then the matrix products defining $g$ and the triangular factors
are deterministically defined as Lebesgue measurable functions of $z\in S^1$ (see section 2 of \cite{BP2}). But the measure (\ref{productmeasure1}) is
supported on sequences with a $l^2$ logarithmic divergence; the composite mapping
(jumping over $g$), where the `products' are now understood to be formal,
can be extended in an almost sure sense. As a consequence
$\tilde{\mu}_l$ can be pushed forward to a probability measure
$\mu_l$ on $\mathbf LG$, where $L_{fin}K \times
L_{fin}K$ acts. This is somewhat analogous to Ito's
uniformization of Brownian motion on nonlinear targets, using
stochastic differential equations.

In the case of $K=SU(n)$ the line bundle $\mathcal L\to C^{\infty}(S^1,G)$
can be realized using Toeplitz operators: the line bundle $\mathcal L^{-1}=Det(A)$
is the the pullback of a Fredholm determinant line bundle, where $A(g)$ is the Toeplitz operator
corresponding to $g$; see chapter 6 of \cite{PS}.

\begin{conjecture}\label{conj5} The measure $\mu_0$ is bi-invariant
with respect to $L_{fin}K$. More generally $\mu_l$
is bi-quasi-invariant with respect to $L_{fin}K$, and
the Radon-Nikodym derivative can be read off from the heuristic
expression
\begin{equation}\label{heuristic}d\mu_l(g)=\frac 1{\mathcal Z}det(A(g)^*A(g))^{l}d\mu_0
(g),\end{equation} where $A(g)$ is the (block) Toeplitz operator
associated to $g:S^1 \to K$. In more sophisticated terms, the
measure $\mu_l$ is a coordinate expression for a bi-invariant
measure with values in the positive line bundle $|\mathcal L|^{2l}$.

\end{conjecture}

\begin{remark}\label{conjremark} There are multiple ways to think about this conjecture. For example
suppose that $g\in C^0(S^1,K)$ is distributed according to
Wiener measure $\nu_{\frac 1T E}$. The following is known: (1)  $g$ has a triangular factorization $\nu_{\beta}$-almost surely, (2) the coefficients of the triangular factors, and also the variables $\eta_i,\chi_j$ and $\zeta_k$ (which can be viewed as functions defined on the top stratum of the formal completion) are tight with respect to the measures $\nu_{\beta}$ as $\beta\to 0$, and hence (3) for each $i$, the family of measures $(\eta_i)_*(\nu_{\beta})$ has finite limits as $\beta\to 0$ (i.e. the mass does not escape to infinity), and similarly for $\chi_j$ and $\zeta_k$. When the level $l=0$, the conjecture predicts the values of these limits. Some of these are known. For example it is known that both $(\eta_0)_*(\nu_{\frac 1T E})$ and $(\zeta_1)_*(\nu_{\frac 1T E})$ converge to the rotationally invariant distribution on a sphere, because these variables are equivariant with respect to appropriate (root subgroup) actions of $SU(2)$ and Wiener measure is asymptotically invariant. \end{remark}

What justifies this speculation that $\mu_l$ might factor in root subgroup coordinates?
The utility of root subgroup coordinates (\ref{coordinate}) is
manifested by the (Plancherel-esque) identities
\begin{equation}\label{Toeplitz0}det(A(g)^*A(g))=
\left(\prod_{i=0}^{\infty}\frac{1}{(1+\vert\eta_i\vert^2)^{i}}\right)\times
\left(\prod_{
j=1}^{\infty}e^{-2j\vert\mathbf{\chi}_j\vert^2}\right)\times
\left(\prod_{k=1}^{\infty}\frac
{1}{(1+\vert\zeta_k\vert^2)^{k}}\right)\end{equation}
\begin{equation}\label{Toeplitz1}det(A_1(g)^*A_1(g))=
\left(\prod_{i=0}^{\infty}\frac{1}{(1+\vert\eta_i\vert^2)^{i+1}}\right)\times
\left(\prod_{
j=1}^{\infty}e^{-2j\vert\mathbf{\chi}_j\vert^2}\right)\times
\left(\prod_{k=1}^{\infty}\frac
{1}{(1+\vert\zeta_k\vert^2)^{k-1}}\right)\end{equation}
(where $A_1$ is the shifted Toeplitz operator)
\begin{equation}\label{Toeplitzdiag}a_0(g)^2=\frac{det(A_1(g)^*A_1(g)) }{det(A(g)^*A(g))}
=\left(\prod_{i=0}^{\infty}\frac{1}{(1+\vert\eta_i\vert^2)}\right)\times
\left(\prod_{k=1}^{\infty}(1+\vert\zeta_k\vert^2)\right)\end{equation}
These formulas are valid for
$g\in W^{1/2}(S^1,K)$, and generalize venerable identities of Szego and Widom, see \cite{Pi4}.
This fits seamlessly with the continuity claim in Conjecture \ref{conj2}.

A corollary of this product expression is the following result,
which expresses the diagonal distribution for $\mu_l$ in terms of
an affine analogue of Harish-Chandra's $\mathbf c$-function (see
Section 4.4 of Part III of \cite{Pi1}).

\begin{conjecture}\label{HCformula}For $\lambda \in \mathbb R$,
$$\int a_0(g)^{-2\sqrt{-1}\lambda}d\mu_{l}(g)=
\frac {sin(\frac {\pi}{2+l})}{sin(\frac {\pi}{2+l}
(1-\sqrt{-1}\lambda))}$$
\end{conjecture}

The `$2$' is included in the exponent to match up with the general diagonal distribution conjecture below
(or because $\dot {\alpha}_1(h_{\dot{\alpha}_1})=2$).\\

\noindent{\bf Heuristic Proof of the Conjecture} In this (heuristic) proof we will write
$g=k_1^*e^{\chi}k_2$ as in root subgroup factorization. This could be dispensed with.
The main point is that (conjecturally) the root subgroup coordinates $\eta_i,\chi_j,\zeta_k$
are independent random variables relative to $\mu_l$ and $a_0$ factors in terms of these variables.

Suppose that $\lambda\in \mathbb R$. Then, using (\ref{Toeplitzdiag})
$$\int a_0(k_1^*e^{\chi h_1}k_2)^{-2\sqrt{-1}\lambda}d\mu_l$$
$$=\int a_0(k_1^*)^{-2\sqrt{-1}\lambda}a_0(k_2)^{-2\sqrt{-1}\lambda}d\mu_l$$
The important point here: there is no dependence on imaginary roots, i.e. the $\chi_j$.
In root subgroup coordinates this
$$=\int\prod_{i=0}^{\infty} (1+\vert\eta_i\vert^2)^{-\frac 12(-2\sqrt{-1}\lambda )}\prod_{k=1}^{\infty}
(1+\vert\zeta_k\vert^2)^{\frac 12(-\sqrt{-1}2\lambda )}$$
$$\prod_{i=0}^{\infty}\frac {1+(l+2)i}{\pi}\frac 1{(1+\vert\eta_i\vert^2)^{2+(l+2)i)}}d\lambda
(\eta_i)\frac {-1+(l+2)k}{\pi}\frac{1}{(1+\vert\zeta_k\vert^2)^{(l+2)k}}d\lambda (\zeta_k)$$
$$=\prod_{i=0}^{\infty}\frac {(l+2)i+1}{(l+2)i+(1-\sqrt {-1}\lambda )}\prod_{k=1}^{\infty}\frac {(l+
2)k-1}{(l+2)k-(1-\sqrt {-1}\lambda )}$$
$$=\frac 1{1-\sqrt {-1}\lambda}\prod_{i>0}\frac {1+\frac 1{(l+2)i}}{
1+\frac {1-\sqrt {-1}\lambda}{(l+2)i}}\prod_{k>0}\frac {1-\frac 1{
(l+2)k}}{1-\frac {1-\sqrt {-1}\lambda}{(l+2)k}}$$
$$=\frac {sin(\frac {\pi}{l+2})}{sin(\frac {\pi}{l+2}(1-\sqrt {-1}
\lambda )}$$
This `proves' that the conjecture follows from the product expression for root subgroup coordinates
and (\ref{Toeplitzdiag}).

\begin{remark}\label{aRVremark} In this calculation there are individual products that do not make sense; this reflects the
fact that $\mathbf a_1:=a_0(k_1)$ and $\mathbf a_2:a_0(k_2)$
are not individually well-defined random variables. But the product $\mathbf a_1\mathbf a_2$
is a well-defined random variable, so in the end the formula is valid. We are gliding over a number of subtle issues
of this sort. \end{remark}

At a heuristic level this formula can be viewed as an infinite dimensional example of
the Duistermaat-Heckman exact stationary phase formula (see
section 7 of \cite{Pi2}). Conjecturally root subgroup coordinates are
essentially action angle variables for the homogeneous Poisson structure
that is in the background.

\subsection{Root Subgroup Factorization: The General Case}

There is a generalization of root subgroup coordinates (see \cite{PP}) and Conjecture \ref{conj5} for a general
simply connected $K$ (with simple Lie algebra).

\subsubsection{The Short Version}\label{shortversion}
In heuristic terms, the first step is to choose
a reduced factorization for the (fictitious) `longest Weyl group element' for the affine Weyl group. One uses this choice to `order' the positive affine
roots for the affine Lie algebra $\widehat{ \mathfrak g(\mathbb C[z,z^{-1}])}$. The ordering has the form
$$\tau_1,\tau_2,...; \text{imaginary roots}; ... \tau'_0,...,\tau'_{-N} $$
where the $\tau_k$ is a listing of the real positive affine roots of the form $q\delta-\dot{\alpha}$, $q>0$ and $\dot{\alpha}$ is a positive root of $\mathfrak g$, the ordering of the imaginary roots is irrelevant (the corresponding root subgroup elements all commute), and the $\tau'_i$ list the other positive affine roots; see \cite{PP}. Let $\dot g$ denote the dual Coxeter number, and $\rho=\sum_{i=0}^{r}\Lambda_i$, the sum of the fundamental affine positive weights. The general form of the invariant measure $d\mu_0$ is
\begin{equation}\label{productmeasure2}\prod\frac{1}{\mathfrak Z} \frac{d\lambda(\eta_i)}{(1+|\eta_i|^2)^{1+\rho(h_{\tau'_i})}}  \prod \frac{1}{\mathfrak Z} exp(- \dot g j|\chi_j|^2  )d\lambda(\chi_j)  \prod\frac{1}{\mathfrak Z} \frac{d\lambda(\zeta_k)}{(1+|\zeta_k|^2)^{1+\rho(h_{\tau_k})}} \end{equation}
(times Haar measure for $T$). We will further parse this, and the corresponding expression for $\mu_l$, in
the Longer Version below.

\begin{remarks} (a) A note on possible notational confusion: $\chi_j\in i\mathfrak h_{\mathbb R}$; in the $SU(2)$ case we were writing this element as $\left(\begin{matrix}\chi_j&0\\0&-\chi_j\end{matrix}\right)$.
This explains the non-appearance of a $2$ in the Gaussian measure involving $\chi_j$.

(b) This formula is a Kac-Moody analogue of a formula known to be valid in finite dimensions.\end{remarks}

There is a similar product expression for the measure $d\mu_l$,
$$d\mu_l=|\sigma_0|^{2l}d\mu^{|\mathcal L|^{2l}}$$
where $\sigma_0$ is the matrix coefficient corresponding to $\Lambda_0$, the basic fundamental positive weight
(in the case of $SL(n,\mathbb C)$, $\sigma_0(g)=det(A)(\widehat{g})$, where $A(g)$ is the Toeplitz operator corresponding to $g:S^1 \to SL(n,\mathbb C)$, and the determinant is really a holomorphic function on the central extension, as in \cite{PS}). This product expression emerges because $\sigma_0$ (and other fundamental matrix coefficients) factor completely in terms of root subgroup coordinates. This reduces to our previous expression in the case $K=SU(2)$.

\subsubsection{The Detailed Version}

In this detailed version we will use the supplementary notation in
Subsection \ref{supplement}. Thus we will be replacing $K$ with $\dot K$.
The following is a synopsis of the main results in \cite{PP}.

The Weyl group $W$ for $(\widehat L\dot{\mathfrak g},\mathbb
Cd+\mathfrak h)$ acts by isometries of $(\mathbb R d+\mathfrak
h_{\mathbb R},\langle \cdot,\cdot \rangle)$. The action of $W$ on
$\mathbb R c$ is trivial. The affine plane $d+\dot{\mathfrak h}$
is $W$-stable, and this action identifies $W$ with the affine Weyl
group and its affine action (\ref{affineaction}) (see Chapter 5 of
\cite{PS}). In this realization
\begin{equation}\label{r0eqn}r_{\alpha_0}= \dot h_{\dot{\theta}}\circ r_{\dot{\theta}}
,\quad \text{and}\quad r_{\alpha_i}=r_{\dot{\alpha_i}},\quad
i>0.\end{equation}

\begin{definition}\label{minimal} (a) A sequence of simple reflections
$r_1,r_2,..$ in $W$ is called reduced if $w_n=r_nr_{n-1}..r_1$ is
a reduced expression for each $n$.

(b) A reduced sequence of simple reflections
$\{r_j\}$ is affine periodic if, in terms of the identification of
$W$ with the affine Weyl group, (1) there exists $l$ such that
$w_l \in \check{\dot T}$ and (2) $w_{s+l}= w_s \circ w_l$, for all
$s$. We will refer to $w_l^{-1}$ as the period ($l$ is the length
of the period).

\end{definition}

Given a reduced sequence of simple
reflections $\{r_j\}$, corresponding to simple positive roots
$\gamma_j$, the positive roots which are mapped to negative roots by $w_n$
are
$$\tau_j=w_{j-1}^{-1}\cdot\gamma_j=r_1..r_{j-1}\cdot\gamma_j,\quad j=1,..,n.$$

\begin{theorem}\label{periodic}(a) There exists an affine
periodic reduced sequence $\{r_j\}$ of simple reflections such
that as immediately above
\begin{equation}\label{flips}\{\tau_j:1\le j<\infty\}=\{q \delta -\dot{\alpha
}:\dot{\alpha }>0,q=1,2,..\},\end{equation} i.e. such that the
span of the corresponding root spaces is $\dot{\mathfrak
n}^{-}(z\mathbb C[z])$. The period can be chosen to be any point
in $C \cap \check{\dot T}$.

(b) Given a reduced sequence as in (a), and a reduced expression
for $\dot {w}_0=r_{-N}..r_0$ (where $\dot w_0$ is the longest
element of $\dot W$), the sequence
$$r_{-N},..,r_0,r_1,..$$
is another reduced sequence. The corresponding set of positive
roots mapped to negative roots is
$$\{q \delta +\dot{\alpha }:\dot{\alpha }>0,q=0,1,..\}
,$$ i.e. the span of the corresponding root spaces is $\dot
{\mathfrak n}^{+}(\mathbb C[z])$.

\end{theorem}

From now on we fix a reduced
sequence $\{r_j\}$ as in Theorem \ref{periodic}, and a reduced
expression $\dot w_0=r_{-N}..r_0$.
We set
$$i_{\tau_n}=\mathbf w_{n-1} i_{\gamma_n} \mathbf
w_{n-1}^{-1}, \qquad n=1,2,..$$
$$i_{\tau_{-N}^{\prime}}=i_{\gamma_{-N}},\quad
i_{\tau_{-(N-1)}^{\prime}}=\mathbf
r_{-N}i_{\gamma_{-(N-1)}}\mathbf r_{-N}^{-1},..,\quad
i_{\tau_0^{\prime}}=\dot{\mathbf w}_0 i_{\gamma_0} \dot{\mathbf
w}_0^{-1}
$$ and for $n>0$
$$i_{\tau_n^{\prime}}=\dot{\mathbf w}_0 \mathbf w_{n-1}
i_{\gamma_n} \mathbf w_{n-1}^{-1} \dot{\mathbf w}_0^{-1}.$$

Also for $\zeta\in \mathbb C$, let $\mathbf
a(\zeta)=(1+\vert\zeta\vert^2)^{-1/2}$ and
\begin{equation}\label{kfactor}k(\zeta)=\mathbf a(\zeta)\left (
\begin{matrix}1&-\bar{\zeta}\\\zeta&1
\end{matrix}\right)=\left ( \begin{matrix}1&0\\\zeta&1
\end{matrix}\right)\left ( \begin{matrix}\mathbf a(\zeta)&0\\0&\mathbf a(\zeta)^{-1}
\end{matrix}\right)\left ( \begin{matrix}1&-\bar{\zeta}\\0&1
\end{matrix}\right) \in SU(2).\end{equation}

\begin{theorem}\label{Ktheorem1smooth} Suppose that $\tilde k_1 \in \tilde L \dot{K}$
and $\Pi(\tilde k_1)=k_1$. The following are equivalent:

(I.1) $m(\tilde k_1)=1$, and for each complex irreducible
representation $V(\pi)$ for $\dot{G}$, with lowest weight vector
$\phi \in V(\pi)$, $\pi(k_1)^{-1}(\phi)$ has holomorphic extension
to $\Delta$, is nonzero at all $z\in \Delta$, and is a positive
multiple of $v$ at $z=0$.

(I.2) $\tilde k_1$ has a factorization of the form
$$\tilde k_1=\lim_{n\to\infty}i_{\tau'_n}(k(\eta_n))..i_{\tau'_{-N}}(k(\eta_{-N})),$$
for a rapidly decreasing sequence $(\eta_j)$.

(I.3) $\tilde k_1$ has triangular factorization of the form
$\tilde k_1=l_1a_1u_1$ where $l_1\in H^0(\Delta^*,\dot N^-)$ has
smooth boundary values.

Moreover, in the notation of (I.2),
$$a_1=\prod_{j=-N}^{\infty}\mathbf a(\eta_j)^{h_{\tau^{\prime}_j}}.$$

Similarly, the following are equivalent: for $\tilde k_2 \in
\tilde L \dot{K}$,

(II.1) $m(\tilde k_2)=1$; and for each complex irreducible
representation $V(\pi)$ for $\dot{G}$, with highest weight vector
$v \in V(\pi)$, $\pi(k_2)^{-1}(v)\in H^0(\Delta;V)$ has
holomorphic extension to $\Delta$, is nonzero at all $z\in
\Delta$, and is a positive multiple of $v$ at $z=0$.

(II.2) $\tilde k_2$ has a factorization of the form
$$\tilde k_2=\lim_{n\to\infty}i_{\tau_n}(k(\zeta_n))..i_{\tau_1}(k(\zeta_1))$$
for some rapidly decreasing sequence $(\zeta_j)$.

(II.3) $\tilde k_2$ has triangular factorization of the form
$\tilde k_2=l_2a_2u_2$, where $l_2 \in H^0(\Delta^*,\infty;\dot
N^+,1)$ has smooth boundary values.

Also, in the notation of (II.2),
\begin{equation}\label{product2}a_2=\prod_{j=1}^{\infty}\mathbf a(\zeta_j)^{h_{\tau_j}}.\end{equation}

\end{theorem}

\begin{theorem}\label{smooththeorem}Suppose $\tilde g \in \tilde L\dot K$ and $\Pi(\tilde g)=g$.

(a) The following are equivalent:

(i) $\tilde g$ has a triangular factorization $\tilde g=lmau$,
where $l$ and $u$ have $C^{\infty}$ boundary values.

(ii) $\tilde g$ has a factorization of the form
$$\tilde g=\tilde k_1^* exp(\chi)\tilde k_2$$ where $\chi \in \tilde L\dot{\mathfrak t}$,
and $\tilde k_1$ and $\tilde k_2$ are as in Theorem
\ref{Ktheorem1smooth}.

(b) In reference to part (a),
\begin{equation}\label{diagonalclaim}a(\tilde g)=a(g)=
a(k_1)a(exp(\chi))a(k_2),\quad
\Pi(a(g))=\Pi(a(k_1))\Pi(a(k_2))\end{equation} and
\begin{equation}\label{abeliandiagonal}a(exp(\chi))=\vert \sigma_0
\vert(exp(\chi))^{h_0}\prod_{j=1}^r \vert \sigma_0
\vert(exp(\chi))^{\check a_j h_j}.  \end{equation}
\end{theorem}

\begin{remarks} (a) As in the $\dot K=SU(2)$ case, we expect that if $\eta$, $\chi$ and $\zeta$
are $l^2$ sequences, then the matrix products defining $g$ and the triangular factors
are deterministically defined as Lebesgue measurable functions of $z\in S^1$.

(b) There is a technical challenge lurking here: In the $SU(2)$ case there is a known algorithm for finding
$\eta,\chi,\zeta$, given $g=g_-g_0g_+$. In higher rank cases any algorithm will depend on the choice of `ordering of the positive roots', i.e. the choice of reduced sequence in Theorem \ref{periodic}.

\end{remarks}

\subsubsection{Root Subgroup Factorization and Measures}
Now we want to discuss measures in connection with root subgroup factorization. Because the $\dot K=SU(2)$ case is far more explicit, there are a number of technical things which we understand for $\dot K=SU(2)$ and which currently elude us for general $\dot K$.

Recall that $d\mu_l$ is to have the official expression
$$d\mu_l=(\sigma_0\otimes \overline{\sigma_0})^{\otimes l}d\mu^{|\mathcal L|^{2l}}$$
and the heuristic expression
\begin{equation}\label{heuristic8}d\mu_l=\frac{1}{\mathfrak Z}\langle \sigma_0,\sigma_0\rangle^l d\mu_0\end{equation}

Consider the measure $\widetilde{\mu}_l$ on sequences $\eta,\chi,\zeta$ given by
\begin{equation}\label{productmeasure3}\prod_{i=-N}^{\infty}\frac{1}{\mathfrak Z} \frac{d\lambda(\eta_i)}{(1+|\eta_i|^2)^{1+\rho(h_{\tau'_i})+lq(\tau'_i)}}  \prod_{j=1}^{\infty} \frac{1}{\mathfrak Z} exp(- (l+\dot g)j|\chi_j|^2  )d\lambda(\chi_j) \end{equation}
 $$\times \prod_{k=1}^{\infty}\frac{1}{\mathfrak Z} \frac{d\lambda(\zeta_k)}{(1+|\zeta_k|^2)^{1+\rho(h_{\tau_k})+lq(h_{\tau_k})}} $$
(times Haar measure for $e^{\chi_0}\in \dot T$), where $h_{\tau}$ denotes the coroot corresponding to the real root $\tau$, $\tau=q(\tau)\delta+\dot {\alpha}$ (Note: $q$
is a function of the root $\tau$, but it will be convenient to occasionally write it as a function
of the coroot $h_{\tau}$), and $\rho:=\sum_{i=0}^r\Lambda_i$.

\begin{remark}Where does this formula come from? When $l=0$ this is the exact analogue of a corresponding formula in finite dimensions for $\dot K$. When the level is nonzero, we are adding in a known formula for $|\sigma_0|^{2l}$.
\end{remark}

To make this formula more explicit, we use the following formulas: Given a positive root $\dot \alpha$
\begin{equation}\label{formulas}h_{q\delta \pm\dot\alpha}=qc\pm h_{\dot \alpha}, \qquad \rho=\dot g\gamma+\dot \rho
\qquad \lambda(h_{\dot\alpha})=2\frac{\langle\lambda, \dot\alpha\rangle}{\langle\dot\alpha,\dot\alpha\rangle} \end{equation}
$$\rho(h_{q\delta\pm\dot \alpha})=(\dot g\gamma+\dot \rho)(qc\pm h_{\dot\alpha})=\dot gq+\dot \rho(h_{\dot\alpha})$$
Then (\ref{productmeasure3}) equals

\begin{equation}\label{productmeasure4}\prod_{i=-N}^{\infty}\frac{1}{\mathfrak Z} \frac{d\lambda(\eta_i)}{(1+|\eta_i|^2)^{1+(l+\dot g)q(h_{\tau'_i})+\dot\rho(h_{\dot \alpha})}}  \prod_{j=1}^{\infty} \frac{1}{\mathfrak Z} exp(- (l+\dot g)j|\chi_j|^2  )d\lambda(\chi_j) \end{equation}
$$ \prod_{k=1}^{\infty}\frac{1}{\mathfrak Z} \frac{d\lambda(\zeta_k)}{(1+|\zeta_k|^2)^{1+(l+\dot g)q(h_{\tau_k})-\dot \rho(h_{\dot\alpha})}} $$
(times Haar measure for $e^{\chi_0}\in \dot T$)
\begin{equation}\label{productmeasure5}=\prod_{\dot\alpha>0}\left(\prod_{q'=0}^{\infty}\frac{1}{\mathfrak Z} \frac{d\lambda(\eta_{q'\delta+\dot\alpha}}{(1+|\eta_{q'\delta+\dot{\alpha}}|^2)^{1+(l+\dot g)q'+2\frac{\langle\dot\rho,\dot\alpha\rangle}{\langle\dot\alpha,\dot\alpha\rangle}}} \prod_{q=1}^{\infty}\frac{1}{\mathfrak Z} \frac{d\lambda(\zeta_{q\delta-\dot\alpha})}{(1+|\zeta_{q\delta-\dot{\alpha}}|^2)^{1+(l+\dot g)q-2\frac{\langle\dot\rho,\dot\alpha\rangle}{\langle\dot\alpha,\dot\alpha\rangle}}}\right)\end{equation}
 $$\times \prod_{j=1}^{\infty} \frac{1}{\mathfrak Z} exp(- (l+\dot g)j|\chi_j|^2  )d\lambda(\chi_j) $$
(times Haar measure for $e^{\chi_0}\in \dot T$), where in this last expression we are using the roots
$q\delta-\dot{\alpha}$ to index the zeta variables, rather than $i$, and similarly for the eta variables.

\begin{conjecture}\label{conj7} For general $\dot K$
the composite mapping
\begin{equation}\label{mapping}\{(\eta,\chi,\zeta)\} \to \mathbf
LG \end{equation}
$$(\eta,\chi,\zeta) \to g=k_1(\eta)^*\left(\begin{matrix}e^{\sum\chi_jz^j}&0\\0&e^{-\sum\chi_jz^j}\end{matrix}\right)k_2(\zeta)
\to l(g)\cdot m(g)\cdot a(g)\cdot u(g)$$ which is a priori only
defined for rapidly decreasing sequences as in Theorem \ref{smooththeorem}, extends to a measurable
mapping which is defined $\tilde{\mu}_l$-almost surely.
\end{conjecture}

To clarify the notation, $\sum \chi_j$ is shorthand for $\sum_{j=-\infty}^{\infty}\chi_j$, where $\chi_{-j}=-\chi_j^*$.

As a consequence of Conjecture \ref{conj7},
$\tilde{\mu}_l$ can be pushed forward to a probability measure
$\mu_l$ on $\mathbf LG$, where $L_{fin}K \times
L_{fin}K$ acts.

\begin{conjecture}\label{conj8} The measure $\mu_0$ is bi-invariant
with respect to $L_{fin}K$. More generally $\mu_l$
is bi-quasi-invariant with respect to $L_{fin}K$, and
the Radon-Nikodym derivative can be read off from its heuristic
expression (\ref{heuristic8}). (In the case $\dot K=SU(n)$
\begin{equation}\label{heuristic9}d\mu_l(g)=\frac 1{\mathcal Z}det(A(g)^*A(g))^{l}d\mu_0
(g),\end{equation} where $A(g)$ is the (block) Toeplitz operator
associated to $g:S^1 \to \dot K$).

\end{conjecture}

As in the $SU(2)$ case, there are (at least) two possible strategies for proving this conjecture.
One is to prove that $\mu_l$ is invariant with respect to $\mathbf L\dot G \to \mathbf L \dot G:g\to g^*$ (this restricts to inversion on unitary loops). In the $SU(2)$ case, $g\to g^*$ can be written down explicitly, but this
does not seem to help in proving invariance. If we could push this through, this first strategy would work for any level $l$.

The second strategy only applies when $l=0$. The following is known: If $g\in C^0(S^1,\dot K)$ is distributed according to
Wiener measure $\nu_{\frac 1T E}$, then (1)  $g$ has a triangular factorization $\nu_{\frac 1T E}$-almost surely, (2) the coefficients (in a coordinate system) of the triangular factors are tight with respect to the measures $\nu_{\frac 1T E}$ as $T\uparrow \infty$. This together with asymptotic invariance guarantees the existence of an invariant measure.

\begin{conjecture} The variables $\eta_i,\chi_j$ and $\zeta_k$ can be expressed as algebraic functions (in fact
using only roots and rational expressions) in terms of coefficients of the triangular factors for $g$.
Consequently the variables $\eta_i,\chi_j$ and $\zeta_k$ are also tight with respect to the measures $\nu_{\beta}$ as $\beta\to 0$, and hence (3) for each $i$, the family of measures $(\eta_i)_*(\nu_{\frac 1T E})$ has finite limits as $\beta\to 0$ (i.e. the mass does not escape to infinity), and similarly for $\chi_j$ and $\zeta_k$.
\end{conjecture}

From this second point of view, conjecture \ref{conj7} is predicting these $\beta\downarrow 0$ limits. This strikes me as very plausible, but I could easily be suffering from a form of Stockholm syndrome.

\subsection{The Diagonal Distribution Conjecture}\label{Diagdistribution}

A corollary is the following diagonal distribution conjecture. The statement that follows is slightly different from
the way in which I have stated the conjecture in the past. The crucial point is the inclusion of the factor
 $\frac{2}{\langle\dot\alpha,\dot\alpha\rangle}$, which is simply unity in simply laced cases.

\begin{conjecture}
$$\int a(g)^{-\sqrt{-1}\lambda}d\mu_{l}(g)=\prod_{\dot\alpha >0}\frac {sin(\frac {\pi}{l+\dot {g}}\frac{2}{\langle\dot\alpha,\dot\alpha\rangle}\langle 2\dot \rho
,\dot\alpha\rangle )}{sin(\frac {\pi}{l+\dot {g}}\frac{2}{\langle\dot\alpha,\dot\alpha\rangle}\langle 2\dot\rho -\sqrt{-1}\lambda
,\alpha\rangle )} $$
where $\dot {g}$ is the dual Coxeter number, $2\dot\rho$ is the sum of the
positive roots, $\lambda\in \mathfrak h_{\mathbb R}^{*}$, and $\langle,\cdot,\cdot\rangle$
is any positive multiple of the Killing form.
\end{conjecture}

A standing assumption in these notes is that $\langle,\cdot,\cdot\rangle$ is normalized so that the length squared
of a long root is $2$. The point we are making is that, in our revised conjecture, that assumption is not needed.
(Of course this is a conjecture, so I should be careful to not be so dogmatic on this point).\\

\noindent{\bf Heuristic proof of the Conjecture}
$$\int a(k_1^*e^{\chi }k_2)^{-\sqrt{-1}\lambda}d\mu_l=\int a(k_1^*)^{-\sqrt{-1}\lambda}a(k_2)^{-\sqrt{-1}\lambda}d\mu_l$$
It follows from (\ref{productmeasure5}) that this equals
equals
\begin{equation}=\prod_{\dot \alpha>0}\left(\prod_{q\ge 0}
\frac{1+\frac{1}{(l+\dot g)q}\frac{2}{\langle\dot\alpha,\dot\alpha\rangle}\langle 2\dot\delta,\dot\alpha\rangle}{1+\frac{1}{(l+\dot g)q}\frac{2}{\langle\dot\alpha,\dot\alpha\rangle}\langle 2\dot\delta-\sqrt{-1}\lambda,\dot\alpha\rangle}
\prod_{q> 0}\frac{1+\frac{1}{(l+\dot g)q}\frac{2}{\langle\dot\alpha,\dot\alpha\rangle}\langle 2\dot\delta,\dot\alpha\rangle}{1-\frac{1}{(l+\dot g)q}\frac{2}{\langle\dot\alpha,\dot\alpha\rangle}\langle 2\dot\delta-\sqrt{-1}\lambda,\dot\alpha\rangle}\right)
\end{equation}

The standard product expansion for $sine$ now (at least heuristically) implies the Conjecture!

For $l=0$ this formula should be
compared with the known formula of Harish-Chandra,
$$\int_K a(g)^{-\sqrt{-1}\lambda}d\lambda(g)=
\mathbf c(\rho -\sqrt{-1}\lambda )=\prod_{\alpha >0}\frac {\langle\rho ,\alpha
\rangle}{\langle\rho -\sqrt{-1}\lambda ,\alpha\rangle}$$
When we incorporate the level $l$, the generalization of the conjecture is
$$\int a^{-i\lambda}d\mu_{l}=\prod_{
\alpha >0}\frac {sin(\frac {\pi}{2(\dot {g}+l)}\langle\rho ,\alpha
\rangle )}{sin(\frac {\pi}{2(\dot {g}+l)}\langle\rho -i\lambda ,\alpha
\rangle )}$$
As $l\to\infty$, we recover Harish-Chandra's formula. This is consistent with the standard
intuition that $l\to\infty$ is a `classical limit'. The original heuristic derivation of these conjectures
is in Part III of \cite{Pi1}. There is a stationary phase interpretation in \cite{Pi2}. And the
formulas follow directly from the factorization of the measures in terms of root subgroup coordinates (and the usual product expression for the sine function); the upshot is that the factorization of the measures has a much deeper significance.

A triangular factorization for $g\in Hyp(S^1,G)$ implies a Riemann-Hilbert factorization
$$g=g_{-}\cdot g_0\cdot g_{+}$$
The diagonal distribution conjecture determines the distribution
for $g_0$. This is most neatly expressed in terms of the Harish-Chandra
transform
$$\mathcal H(\mu_l)(\lambda)=\prod_{\alpha}\Gamma(1+\frac{i\pi}{l+\dot g}\frac{\langle\lambda,\dot\alpha\rangle}{\langle\dot\alpha,\dot\alpha\rangle})$$
(this follows from 4.4.27 of Part II of \cite{Pi1}).

\subsection{ Missing Formulas}\label{missingformulas}

Let's return to root subgroup factorization and the $SU(2)$ case. For a sufficiently regular
loop $g:S^1 \to SU(2)$, the invertibility of $A(g)$ and $A_1(g)$ is equivalent to a
factorization
\begin{equation}\label{coordinate2}g(z)=\left(\begin{matrix}a_1^*(z)&-b_1(z)\\b_1^*(z)&a_1(z)\end{matrix}\right)\left(\begin{matrix} e^{\chi(z)}&0\\
0&e^{-\chi(z)}\end{matrix} \right)\left(\begin{matrix}d_2^*(z)&-c_2^*(z)\\c_2(z)&d_2(z)\end{matrix}\right)
\end{equation}
where in particular $c_2,d_2$ are holomorphic in $\Delta$ and do not simultaneously vanish (and are subject to some mild normalizations which we ignore). One can say more: $k_2$ has a triangular factorization
\begin{equation}\label{k2formula}k_2=\left(\begin{matrix}d_2^*&-c_2^*\\c_2&d_2\end{matrix}\right)=
\left(\begin{matrix} 1&x^{*}\\
0&1\end{matrix} \right)\left(\begin{matrix} a_2&0\\
0&a_2^{-1}\end{matrix} \right)\left(\begin{matrix} \alpha_2&\beta_2\\
\gamma_2&\delta_2\end{matrix} \right)\end{equation}
where $x=x(z)$ is holomorphic in $\Delta$ (with a similar expression for $k_1$).

Now suppose that $\eta,\chi,\zeta$ are distributed according to $\mu_l$. As we already remarked,
$k_2$ is not well-defined on the circle. It is useful to rewrite the triangular factorization of $k_2$ as
\begin{equation}\label{k2formula2}k_2=\left(\begin{matrix} \mathbf a_2&0\\
0&\mathbf a_2^{-1}\end{matrix} \right)\left(\begin{matrix} 1&X^{*}\\
0&1\end{matrix} \right)\left(\begin{matrix} \alpha_2&\beta_2\\
\gamma_2&\delta_2\end{matrix} \right) ,\end{equation} where
$X=\mathbf a_2^{-2}x$. It turns out that $x(z)$ is not a well-defined random function, but $X(z)$ is.
It also easy to see that ($\mathbf a_1$ and) $\mathbf a_2$ are not well-defined random variables,
but the product $\mathbf a_1\mathbf a_2$ is.

Most intriguing, $\mathbb P(c_2,d_2):\Delta\to \mathbb P^1$ is a well-defined random holomorphic function, i.e.
$$c_2/d_2=\gamma_2/\delta_2=(g_+)_{21}/(g_+)_{22}$$
is a well-defined random meromorphic function in the disk. This function has a Taylor series
$\sum_{n=1}^{\infty}\xi_nz^n$
where $\xi_n$ is the sum of terms
$$(-1)^r(-\overline \zeta_{i_{0}})\left(\zeta_{j_1}(-\overline \zeta_{i_1})\right)...\left(\zeta_{j_r}(-\overline \zeta_{i_r})\right) $$
where $j_s<i_s$ and $j_s\le i_{s-1}$ for $s=1,..,r$, and $\sum_{s=1}^{r+1} i_s -\sum_{s=1}^r j_s=n$; in particular
$$\xi_n=(-\overline{\zeta}_n)\prod_{s=1}^{n-1}(1+|\zeta_s|^2)+polynomial(\zeta_s,\overline{\zeta}_s,s<n)$$
For example
$$c_2/d_2=(-\overline{\zeta}_1) z+(-\overline{\zeta}_2)(1+|\zeta_1|^2)z^2+\left((-\overline{\zeta}_3)(1+|\zeta_1|^2)(1+|\zeta_2|^2)+
(-\zeta_1\overline{\zeta}_2^2)(1+|\zeta_1|^2)\right)z^3$$
$$+((-\overline{\zeta}_4)(1+|\zeta_1|^2)(1+|\zeta_2|^2)(1+|\zeta_3|^2)
+(1+|\zeta_1|^2)(\zeta_2\overline{\zeta}_3^2(1+|\zeta_2|^2)$$
$$+2\zeta_1\overline{\zeta}_2\overline{\zeta}_3(1+|\zeta_2|^2)
+\overline{\zeta}_1^2\overline{\zeta}_2^3) )z^4+...$$
How to go back and forth from the distribution for $\zeta$ and the distribution for $\mathbb P(c_2,d_2)$
has not been resolved.

The natural generalization of $\mathbb P(c_2,d_2)$ to a higher rank group involves a holomorphic map into
the flag space of $\dot G$ in place of $\mathbb P^1$. It is far more complicated to recover the $\zeta$ variables
because this depends on a choice of ordering of roots. In any event I do not have heuristic formulas that might lead to reasonable conjectures for these distributions (or, in the simplest $SU(2)$ case, the point processes that would describe their zeroes and poles).

Now let's return to Riemann-Hilbert factorization, $g=g_-g_0g_+$. We have a conjecture for the $g_0$ distribution.
There is a heuristic expression for the distribution for $\theta_+:=g_+^{-1}\partial g_+\in H^1(\Delta,\mathfrak g)$
(which I refer to as `time ordered exponential coordinates'):
\begin{equation}\label{Wformula}(\theta_{+})_{*}\mu_l=\frac 1{\mathcal Z}det(1+W^{
*}W)^{-\dot {g}-l}dm(theta_{+})\end{equation}
$$=\lim_{n\to\infty}\frac 1{\mathcal Z_n}det(1+W^{
*}W)^{-\dot {g}-l}dm(P_n\theta_{+})$$
where $W=W(g_{+})=A(g_+^{-1})B(g_{+})$ (the graph operator, following the notation in
\cite{PS}), $g_{+}$ corresponds to $\theta_{+}$, and $P_n$ projects $
\theta_{+}$ to its
first $n$ coefficients (so that it is an orthogonal
projection for $H^1(\Delta,\mathfrak g)$. The heuristic expression (\ref{Wformula}) is an analogue of yet another formula of Harish-Chandra from finite dimensions.  The first heuristic expression is manifestly
$PSU(1,1)$ (conformally) invariant, which is consistent with the uniqueness conjectures. Unfortunately
I do not know how to turn this heuristic expression for the $\theta_+$ distribution into a precisely stated conjecture.

\begin{remark}Understanding the meaning of (\ref{Wformula}) is important because there are analogous formulas
in other contexts, such as unitarizing measures for the Virasoro group (sometimes called Malliavin measures) and the Kontsevich-Suhov generalization of Werner's measure on self-avoiding loops on Riemann surfaces. In Section \ref{YM}
there is a discussion of an important analogous measure that is related to $YM_3$ (the existence of this analogous measure is an open question).
\end{remark}

\section{Harmonic Analysis}\label{representations}

\subsection{The Loop Group Action on $L^2(\mu_0)$}\label{loopgroupaction}

Let $\nu_{\beta}$ denote Wiener measure on $C^0(S^1,K)$ with inverse temperature
$\beta$. This family of measures interpolates between Haar measure $\delta_k$ for $K$ and $\mu_0$ for $LK$:
$$\delta_K=\nu_{\infty} \stackrel{\beta\uparrow \infty}{\leftarrow} \nu_{\beta} \stackrel{\beta\downarrow 0}{\rightarrow}\mu_0 $$

Let $L_{\beta}$ ($R_{\beta}$), denote the left (the right, respectively) regular representation of $L_{fin}K$ on half-densities for the measure
class of $\nu_{\beta}$, $0\le \beta\le \infty$.

For $\beta=\infty$, the finite dimensional case, the Peter-Weyl theorem implies that the Von Neumann algebras $L_{\beta}''$ and $R_{\beta}''$ are commutants, each is a sum (over positive weights) of matrix algebras, and the intersection is equivalent to (a completion of) the convolution algebra of central functions on $K$.

In infinite dimensions the situation appears to be rather different:

\begin{conjecture} (a) Suppose that $0<\beta<\infty$.  Then $L_{\beta}''$ and $R_{\beta}''$ are commutants,
each is a factor of type III, and the intersection is trivial, i.e. the action of $L_{fin}K \times L_{fin}K$
is irreducible.

(b) When $\beta=0$, i.e. for $\mu_0$, the same statements hold for the complement of the invariant subspace of constants.
\end{conjecture}

Part (a) is stated in \cite{AHTV}, although this paper appears to be incomplete.

I think the expectation is the factors in (a) are of type $III_1$. The obvious strategy is to first prove that the constant $1$ is a cyclic vector for the left and right representations, in order to concretely realize the modular automorphism group. This appears to be open; see section 3 of \cite{ADGV} for the case of paths (as opposed to loops).

In the case $\beta=0$, i.e. for $\mu_0$, there does not appear to be a convenient choice of cyclic vector
(in the complement of the invariant vector $1$). However there does appear to be an obvious candidate for the modular group (which is well-defined up to inner automorphisms):

\begin{conjecture}Suppose that $\beta=0$. In this case the modular group is realized by the action
of the standard hyperbolic one parameter subgroup of $PSU(1,1)$ acting on $Hyp(S^1,G)$.
\end{conjecture}

In general we can consider the (conjectural) unitary action
$$W^{1/2}(S^1,K)\times \mathcal H^{1/2}([\mu_l])\times W^{1/2}(S^1,K) \to \mathcal H^{1/2}([\mu_l])$$
for $l>-1$. Presumably this is irreducible for $l\ne 0$.

If this is correct, the basic conclusion is that it is essentially not possible to decompose the loop group actions
on half-densities. As we will argue below, the more productive thing to look at is the action of $Diff(S^1)$.

\subsection{Holomorphic Actions}

Suppose that the level $l$ is a positive integer. In this case the previous discussion can be modified
by considering the action on $\mathcal L$ as opposed to the action on $|\mathcal L|$.

There are natural holomorphic actions of the Kac-Moody extension $\widehat{L_{fin}G}$ on the holomorphic line bundle $\mathcal L \to \mathbf LG$ covering
the action of $L_{fin}G$ on $\mathbf L G$ from the left and the right. Although it is a long story,
Kac and Peterson proved an algebraic version of the holomorphic Peter-Weyl theorem, and of the Borel-Weil theorem.
Their work can be expressed in terms of the action
of $\widehat{L_{fin}G}\times\widehat{L_{fin}G}$ acting on holomorphic sections of
$L^{*\otimes l}\to \mathbf L G$; see section 1.7 of Part I of \cite{Pi1} (which applies to all symmetrizable Kac-Moody groups). However, as we will see in the next section, it is important to have an analytic version.

For this reason consider the natural holomorphic actions of the Kac-Moody extension
$\widehat{C^{\omega}(S^1,G)}$ on the holomorphic line bundle $\mathcal L \to Hyp(S^1,G)$ covering
the actions of $C^{\omega}(S^1,G)$ on $Hyp(S^1,G)$ from the left and the right. Assuming the truth of our conjectures
regarding the measure $\mu^{|\mathcal L|^{2l}}$, there are now unitary representations $L_l$ and $R_l$ for $\widehat{C^{\infty}(S^1,K)}$ acting on sections of $\mathcal L^{*\otimes l}$ which are square integrable relative to $\mu^{|\mathcal L|^{2l}}$.   The subspace of holomorphic square integrable sections is an invariant subspace, which we provisionally denote by $L^2H^0_l(\widehat{Hyp(S^1,G)})$. Unfortunately, whereas the space of all square integrable sections is complete, because the base $Hyp(S^1,G)$ is quite thick, it will not be the case that the subspace of holomorphic sections is complete. The conjectural way to resolve this is to identify a core subspace such that
each element of the completion of $L^2H^0_l(\widehat{Hyp(S^1,G)})$ will be holomorphic on this core (this is modeled on how I approached the same issue for measures on infinite rank Grassmannians in \cite{Pi1}). Let $\Phi$ denote the finite set of integrable highest weight representations of level $l$ for the affine Lie algebra $\widehat{\mathfrak g(\mathbb C[z,z^{-1}])}$. Each of these representations can be globalized and unitarized, as in \cite{PS}, using purely algebraic methods (following Garland, \cite{Garland}).

The following has the general shape of the result we are looking for.

\begin{conjecture}\label{Peter-Weyl} (a) The algebraic Peter-Weyl isomorphism of Kac and Peterson extends to an
equivariant Hilbert space isomorphism
$$\bigoplus_{\pi\in\Phi} H(\pi)\otimes H(\pi)^* \to \overline{L^2H^0_l(\widehat{Hyp(S^1,G)}} $$
where the overline indicates the Hilbert space completion.

(b) Given $g\in L^{\infty}W^{1/2}(S^1,G)$, there exists a continuous evaluation map
$$L^2H^0_l(\widehat{Hyp(S^1,G)}\to \mathcal L^{*\otimes l}|_g:s \to s(g)$$
so that the completion consists of square integrable sections which are holomorphic
on the subspace $L^{\infty}W^{1/2}(S^1,G)$.
\end{conjecture}

This statement is flawed because in (b) we need to be able to say precisely how to estimate the value of
a holomorphic section at $g\in L^{\infty}W^{1/2}(S^1,G)$ in terms of its $L^2$ norm. Note that it is essential that $g\in L^{\infty}W^{1/2}(S^1,G)$ as opposed to $g\in W^{1/2}(S^1,G)$, because in the absence of the boundedness condition (which guarantees the associated Toeplitz operator is bounded and Fredholm), it is not the case that $g\in Hyp(S^1,G)$.

It may be that the complement of $L^2H^0_l(\widehat{Hyp(S^1,G)})$ in the space of all square integrable sections is similar to the $\mu_0$ situation: the left and right actions are commutants, the combined action is irreducible,
and the left and right Von Neumann algebras are type $III_1$ factors.

\subsection{Reparametrization Actions}

Assuming the truth of Conjecture \ref{conj2}, there is an action of
$K\times C^{\omega}Homeo(S^1)\times K$ on $H^{1/2}([\mu_0])$ (half-densities).
Of course we can identify  $H^{1/2}([\mu_0])$ with $L^2(\mu_0)$.

\begin{conjecture}\label{Virdecomposition} Assuming the truth of Conjecture \ref{conj2} and an affirmative answer to Question \ref{ques41},
for the action of $ K\times C^{\omega}Homeo(S^1)\times K$, $L^2(\mu_0)$ can be regarded as a completion of the tensor
product
$$L^2([\mu_0])=\otimes_{g\ge 0}L^2([\mu_0])_g $$
where $L^2([\mu_0])_g $ is generated by functions of the form $P(c)^*f$, $c:S^1 \to\Sigma$, $f\in C^{\infty}(Bun_G^0(\Sigma))$, and $g=genus(\Sigma)$ (see Subsection \ref{modulispace} for the notation).

\end{conjecture}

Fix the genus. There is a decomposition
$$L^2([\mu_0])_g=\bigoplus_{\lambda ,\nu}\lambda\otimes W(\lambda ,\nu
)_g\otimes\nu^{*},$$
where $\lambda$ and $\nu$ denote irreducible representations of $K$. This seems to exhaust
the obvious discrete ways of decomposing the representation.

\begin{remark} The natural functions on $H^1(\Sigma,K)$ involve holonomy around nontrivial loops. These are very far from compactly supported. They are giving us generalized functions, although they are not defined everywhere and not smooth on all of $Bun_G$.

\end{remark}

\begin{question}

How does one decompose the action of $C^{\omega}Homeo(S^1)$ on $W(\lambda, \nu)_g$?
\end{question}

More generally for any positive integral level $l$, there will be a similar
decomposition for $L^2$ sections of $\mathcal L^{*\otimes l}$.

\begin{conjecture}\label{Virdecomposition2}
$$L^2\Omega_l^0(\widehat{Hyp})=\otimes_{g\ge 0}L^2\Omega_l^0(\widehat{Hyp})_g $$
where $\Omega_l(\widehat{Hyp})_g  $ is generated by sections of the form $P(c)^*s$, $c:S^1 \to\Sigma$, $s\in \Omega_l(\widehat{Bun_G^0(\Sigma)})$ has compact support or is otherwise reasonably well-behaved, and $g=genus(\Sigma)$.

\end{conjecture}

In the holomorphic
sector, the action of $\widehat{C^{\omega}Homeo(S^1)}$ is the diagonal action
as a subgroup of $\widehat{C^{\omega}Homeo(S^1)}\times\widehat{C^{\omega}Homeo(S^1)}$
where these factors are acting on the tensor product of a highest weight representation and its dual.

To fix ideas, we will consider the holomorphic sector and suppose that $G=SL(2,\mathbb C)$ and $l=1$.  In
this case there are two irreducible highest weight representations of level $1$.
Hence as a representation of $\widehat {LG}\times\widehat {LG}$
$$H^0_{l=1}=H(0)\otimes H(0)^{*}\oplus H(\frac 12)\otimes H(\frac
12)^{*}$$
It is known that the decompositions of $H(0)$ and $H(1/2)$ with respect to
$SU(2,\mathbb C)\times \widehat {Diff(S^1)}$ are multiplicity free, and more precisely,
$$H(0)=\bigoplus_{n=0}^{\infty}V(2n+1)\otimes L(c=1,h=(\frac n2)^
2)$$
and
$$H(1/2)=\bigoplus_{n=1}^{\infty}V(2n)\otimes L(c=1,h=(\frac n2)^
2)$$
where $V(N)$ is the unique irreducible representation of
$SU(2,\mathbb C)$ of dimension $N$.  Thus $H(0)\otimes H(0)^*$ equals
$$\bigoplus_{n, m\ge 0}V(2n+1)\otimes L(1,(\frac n
2)^2)\otimes L(1,(\frac m2)^2)^{*}\otimes V(2m+1)^{*}$$
There is a similar decomposition for $H(1/2)\otimes H(1/2)^*$ with $V(2n)$ in place of $V(2n+1)$.
The structure of a representation of the form $L\otimes L^{*}$ for
$Diff(S^1)$ is not known; it is possible that it might be
irreducible.

Given a closed Riemann surface of genus $g$ and a real analytic embedding $c:S^1 \to \Sigma$, there is an inclusion (of a conformal block)
induced by pullback
$$P(c)^*H^0_{l=1}\left(\widehat{ Bun_G(\Sigma)}\right) \subset H^0_{l=1}$$

\begin{question}Suppose that $\Sigma_1$ and $\Sigma_2$ have distinct genuses and $c:S^1 \to\Sigma_i$ are real analytic embeddings. are the subspaces (generated by conformal blocks)
$$P(c_1)^*H^0_{l=1}\left(\widehat{ Bun_G(\Sigma_1)}\right)\perp P(c_2)^*H^0_{l=1}\left(\widehat{ Bun_G(\Sigma_2)}\right)$$
\end{question}

When the level is zero, as in Conjecture \ref{Virdecomposition}, there is some probabilistic intuition
which suggests the corresponding subspaces are perpendicular. But when we are thinking about bundle valued measures, this seems less clear. But let's assume perpendicularity.

On the one hand there is a decomposition
\begin{equation}\label{Virdecomphol}H^0_{l=1}=\left(\bigoplus_{n,m\ge 0}V(2n+1)\otimes L(1,(\frac n
2)^2)\otimes L(1,(\frac m2)^2)^{*}\otimes V(2m+1)^{*} \right)\bigoplus\end{equation}
$$\left(\bigoplus_{n,m\ge 1}V(2n)\otimes L(1,(\frac n
2)^2)\otimes L(1,(\frac m2)^2)^{*}\otimes V(2m)^{*}\right) $$
On the other hand there is a decomposition
$$H^0_{l=1}=\bigoplus_{g\ge 0}\bigoplus_{N,M\ge 1} V(N)\otimes  W(N,M) \otimes V(M)^*$$
How are these related?
Suppose that the genus $g=0$. In this case $W(N,M)$ vanishes unless $N=M=1$, and
$$W(1,1)\subset L(c=1,h=0)\otimes L(c=1,h=0)^* $$
These might actually be equal. In any event we can see how this genus zero piece fits into
the decomposition \ref{Virdecomphol}. How about higher genus?

\subsection{Locality}\label{locality}

The Haar measure
\begin{equation}\label{localmeas}\prod_{v\in S^1}d\lambda_K(g_v) \text{ for } \prod_{v\in S^1} K \end{equation}
is ultralocal in the sense that if $S^1=E\sqcup E^c$, then the measure splits
as a product. Here the structure of the circle is essentially irrelevant.

Wiener measure is heuristically of the form
$$d\nu_{\beta}=\frac{1}{\mathfrak Z}e^{-\beta E(g)}\prod_{v\in S^1}d\lambda_K(g_v)$$
where $E(g)=\frac{1}{2}\int_{S^1}\langle dg\wedge *dg\rangle$ is the standard energy
(see \cite{Driver2} and references for a thorough justification of this point of view, for mathematicians).
This is local, in the sense
that (1) it is possible to define Wiener measure on intervals with Dirichlet boundary condition,
and (2) if $S^1=I\sqcup I^c$, where $I$ is an interval, then it is possible to express $d\nu_{\beta}$
as a convex combination of local Wiener type measures as in (1).

Heat kernel measures $\nu^{(s)}_t$ are not local in any sense that I can imagine. It might be possible
to formulate heat kernel measures for an interval with appropriate boundary conditions, but it is not at all
clear how to split up $\nu^{(s)}_t$ as in the case of Wiener measures.

If $s=1$, then $[\nu_t^{(1)}]=[\nu_{\beta}]$, $t=1/\beta$. Hence from a representation theoretic point of view (as above), it does not matter whether we think about $\nu_t^{(1)}$ or $\nu_{\beta}$. But from a heuristic point of view it does not seem possible to express $\nu^{(s)}_t$ in terms of a local functional against the ultralocal background (\ref{localmeas}). The upshot seems to be that heat kernel measures are nonlocal.

In the next section (see Subsubsection \ref{flat torus}) we will see that it is important to identify locality characteristics for $\mu_0$.
The situation here could be similar to what Jones and Wassermann
discovered for loop groups acting on highest weight representations; see \cite{Wassermann}. Loops supported in $I$ and loops supported in $I^c$ commute, hence their actions on $L^2(\mu_0)$ commute (A major complication is that we have not proven that $C^{\infty}$ loops - as opposed to $C^{\omega}$ loops - fix $\mu_0$. We will put this aside). In the case of highest weight representations (which in some respects is more complicated, because of the central extension), the corresponding Von Neumann algebras are commutants. Is this true for the natural representation associated to $\mu_0$ (one should perhaps first investigate this for Wiener measure $\nu_{\frac 1T E}$ and let $T\uparrow\infty$)? In the next section we will toy with the idea of thinking of $\mu_0$ as having a kind of tautological density relative to the ultralocal Haar measure for $\prod_{S^1}K$, namely the characteristic function for its support (which we also think of as a surrogate for the missing unitary form for $Hyp(S^1,G)$). But this is clearly too vague to be useful. We clearly need to understand $L^2(\mu_0)$ from the Jones-Wassermann point of view, to get started.

\section{ Generalizations}\label{Generalizations}

\subsection{Symmetric Space Target}

Suppose that $X$ is a simply connected compact symmetric space
with a fixed basepoint.  From this we obtain a diagram of groups,
\begin{equation}\label{diagram_of_groups}
\begin{matrix}& & G & &\\
 & \nearrow& & \nwarrow &\\
 G_0  & & & & U  \\
 & \nwarrow& & \nearrow &\\
& & K  && \end{matrix}
\end{equation}
where $U$ is the universal covering of the identity component of
the isometry group of $X$, $X\simeq U/K$ (using the basepoint), $G$ is the
complexification of $U$, and $X_0=G_0/K$ is the noncompact type
symmetric space dual to $X$; and a diagram of equivariant totally
geodesic (Cartan) embeddings of symmetric spaces:
\begin{equation}\label{CartanEmbeddings}
\begin{matrix}
U/K & \rightarrow & U & &  \\
\downarrow & & \downarrow & &\\
G/G_0 & \rightarrow & G & \leftarrow &G/U  \\
 & &\uparrow & & \uparrow\\
& & G_0 &\leftarrow & G_0/K  \end{matrix}
\end{equation}

Without loss of generality one can suppose that $X$ is irreducible (i.e. cannot be factored as a product). 
There are two types: Type I spaces, meaning that $\mathfrak u$, the Lie algebra of $U$ is simple,
and type II spaces, meaning that $X=K$ is a group, in which case $U=K\times K$.
 
\begin{example}\label{symmetricspaces} Below we will be interested in the Type I example

\begin{equation}\label{CartanEmbedding1}
\begin{matrix}
SU(n)/SO(n) & \rightarrow & SU(n) & &  \\
\downarrow & & \downarrow & &\\
SL(n,\mathbb C)/SL(n,\mathbb R) & \rightarrow & SL(n,\mathbb C) & \leftarrow &SL(n,\mathbb C)/SU(n)  \\
 & &\uparrow & & \uparrow\\
& & SL(n,\mathbb R) &\leftarrow & SL(n,\mathbb R)/SO(n)  \end{matrix}
\end{equation}

This is embedded in the `group case',

\begin{equation}\label{CartanEmbedding2}
\begin{matrix}
SU(n) & \rightarrow & SU(n)\times SU(n) & &  \\
\downarrow & & \downarrow & &\\
SL(n,\mathbb C) & \rightarrow & SL(n,\mathbb C)\times SL(n,\mathbb C) & \leftarrow &SL(n,\mathbb C)/SU(n)\times SL(n,\mathbb C)/SU(n)  \\
 & &\uparrow & & \uparrow\\
& & SL(n,\mathbb C) &\leftarrow & SL(n,\mathbb C)/SU(n)  \end{matrix}
\end{equation}

These two examples can be generalized in the following way. Fix a complex group $G$ (e.g. $SL(n,\mathbb C)$, as above). Let $G_0$ be the
normal real form (e.g. $SL(n,\mathbb R)$), and let $U$ and $K$ be maximal compact subgroups of $G$ and $G_0$,
respectively (e.g. $SU(n)$ and $SO(n)$, respectively).

We have focused exclusively on the group case in these notes. We will mention why we are interested in the first set of Type I examples in the next subsection.
\end{example}

There is a prolongation of the diagram  (\ref{CartanEmbeddings}) to loop spaces. The formal completion of the loop space of $G/G_0$ is defined in terms of the inclusion
$$\mathbf L(G/G_0):=\{[g_1,g_2]\in \mathbf LG: [\Theta(g_2^*),\Theta(g_1^*)]=[g_1,g_2]\}$$
where $\Theta$ is the complex linear extension to $G$ of the Cartan involution which fixes $K$ inside $U$ (see \cite{Pi2}). The hyperfunction completion is similarly defined.
The inclusions $U/K \subset G/G_0$ and 
$$C^0(S^1,U/K) \subset Hyp(S^1,G/G_0)\subset \mathbf L G$$ 
are homotopy equivalences, i.e. these completions faithfully remember the original topologies, as in the group cases.

We now discuss $L_{fin}U$ invariant measures on these completions. There is an existence result (see \cite{Pi2}):

\begin{theorem} For the natural action
$$L_{fin}U \times \mathbf L(G/G_0) \to \mathbf L(G/G_0)$$
there exists an invariant probability measure.
\end{theorem}

\begin{conjecture} The invariant measure is unique.
\end{conjecture}

As in the group case, uniqueness, and proof that the measure is supported on the hyperfunction completion, would imply
invariance with respect to $C^{\omega}Homeo(S^1)$. There should be an analogue of monotonicity for Wiener measure
on $C^0(S^1,U/K)$.

In the group case we formulated a conjecture which purports to characterize invariant measure classes
having appropriate symmetry. There is almost certainly a generalization of this to the symmetric space context, using
Toeplitz determinants. 

It is instructive to reflect on what we might hope to compute in this more general context:\\

1. A generic point in $\mathbf L(G/G_0)$ has Riemann-Hilbert factorization
$g=g_- g_0 g_+$, where $g_+:\Delta \to G$ and $g_-:\Delta^* \to G$ are holomorphic,
$g_0\in G/G_0 \subset G$ and $g_-=g_+^{*\Theta}$. The analogue of linear Riemann-Hilbert coordinates
for $Hyp(S^1,G/G_0)$ is the pair $(g_0,\theta_+:=g_+^{-1}\partial g_+)\in G/G_0 \times H^1(\Delta,\mathfrak g)$.
There is a conjecture for the distribution
 of $g_0$, see 3. below. It is not obvious how to compute the distribution
for $\theta_+$. As noted previously, I have so far failed
to find a reasonable conjecture even in the group case. \\

2. Given a real analytic embedding $c:S^1\to \Sigma_c$, one can consider the composition
$$Hyp(S^1,G/G_0) \to Hyp(S^1,G) \stackrel{P(c)}{\rightarrow} Bun_G(\Sigma_c)$$
For a generic $c$ this composition is probably surjective, and I have no guess for the pushforward
of the invariant measure. However suppose that $c$ is a parameterization of the waist for a Riemann surface with reflection symmetry, i.e. $\Sigma_c$ is a double $\Sigma^*\circ \Sigma$. In this case the image of
the composition, in terms of the parameterization $Bun_G^0(\Sigma_c)\sim Hom_{irred}(\pi_1,U)/U$, is identified
with a submanifold of homomorphisms $\rho$ which satisfy the reality condition $\rho\circ R_*=\Theta\circ \rho$, where $R$ denotes reflection symmetry. This submanifold has a natural probability measure, akin to
the normalized symplectic volume of $Hom_{irred}(\pi_1,U)/U$, which conjecturally can be characterized in terms of its invariance with respect to the subgroup of the mapping class group compatible with reflection symmetry. Conjecturally this is the image of the unitarily invariant measure on $Hyp(S^1,G/G_0)$. This restriction on $c$ seemingly precludes
the possibility of constructing the invariant measure in analogy with the construction of Wiener measure that we outlined in the group case.\\

3. Although it is more involved to formulate than in the group case,
there is a diagonal distribution conjecture for $\mu_0$ which determines the distribution
for the zero mode $g_0$ (see \cite{Pi2}). As we previously pointed out, in the group case this is a consequence of the existence of root subgroup coordinates. There does exist a homogeneous Poisson structure on $L(U/K)$. Consequently one can ask whether it is possible to formulate some analogue of root subgroup coordinates that
would serve as action-angle variables. The answer (even for the finite dimensional symmetric space $U/K$) is apparently negative. Is there a substitute, or is this telling us
that something is not solvable in the Type I case?

\subsection{Homogeneous Space Targets}

There has been recent interest in sigma models with compact homogeneous space targets, see the review \cite{ABK}, or \cite{OSS}. A generic flag manifold is of the form $U/T$, where $T$ is a maximal torus in $U$. In general (beyond the symmetric space case) there is not an essentially unique invariant metric, which introduces a family of parameters. The approach advocated in \cite{ABK} and references is to consider a differential geometric (not necessarily isometric) embedding of $U/T$ into a product of symmetric spaces.  This approach does suggest an extrinsic way to realize a kind of hyperfunction completion of the loop space of $U/T$, as a submanifold of the hyperfunction completions of the symmetric space factors. I have not made any progress in understanding how to formulate an existence result for an invariant measure that would have some intrinsic meaning. One obstacle is to understand the intrinsic meaning of the zero mode from this point of view (in the symmetric space context above, the zero mode is $g_0\in G/G_0$, for which there is a conjecture for the distribution).

In \cite{Witten} there is a speculative discussion concerning 2D sigma models with moduli space targets.

\subsection{Symmetrizable Kac-Moody Groups}\label{symmetrizable}

As evident from the title of \cite{Pi1}, I once believed that the proper framework for this work was
symmetrizable Kac-Moody Lie algebras. I am now skeptical. The following discussion might help to put these notes
into perspective.

Given a symmetrizable generalized Cartan matrix $A$, there is a formal completion $\mathbf{G}(A)$ of the Kac-Peterson group $G(A)$. The inclusion $G(A)\subset \mathbf{G}(A)$  is a homotopy equivalence, and the formal completion is a
natural algebro-geometric framework for the Kac-Peterson version of the Peter-Weyl theorem (see \cite{Pi1}). In affine and indefinite cases, this completion is not a group and there does not exist a unitary form. In affine cases (ignoring the automorphism group $exp(\mathbb C L_0)$), this completion is a $\mathbb C^{\times}$ bundle over a completion $\mathbf L \dot G$ (or a twisted version of this) in the sense of Section \ref{background}.

\begin{question} Suppose that $A$ is a generalized symmetrizable Cartan matrix.
Does there exist an invariant probability measure for the action
$$K(A)\times \mathbf{G}(A) \times K(A) \to\mathbf{G}(A) $$
\end{question}

If $A$ is of finite type, then the answer to the question is yes, and there are many bi-invariant measures which are classified using the Harish-Chandra transform. For example there is the Haar measure of $K(A)$, the unique bi-invariant measure on the unitary form. At the other extreme, when we consider Riemann-Hilbert factorization for $g\in \mathbf L G(A)$, $g=g_-g_0g_+$, the $\mu_0$ distribution for $g_0$ is a bi-invariant measure which we conjecture is absolutely continuous with respect to Haar measure for $G(A)$, hence has a support which is very spread out (and of course it is very far from ergodic).

\begin{question}\label{complexRSF} Haar measure for $K(A)$ can be realized as a product measure using root subgroup coordinates, and Haar measure for $G(A)$ can be realized as a product measure using a complex analogue of root subgroup coordinates (see \cite{Pi4.5}). It is not known (to me) if there are $K(A)$ bi-invariant measures on $G(A)$ which are absolutely continuous with respect to Haar measure for $G(A)$ and can be realized in a simple way using root subgroup factorization. This might be at least heuristically relevant below.
\end{question}

Suppose that $A$ is of (for simplicity untwisted) affine type, and $\dot G$ is the corresponding finite dimensional complex group. In this case Conjecture \ref{conj1} asserts that there is a unique $K(A)$ bi-invariant probability measure $\mu_0$ on the completion $\mathbf L \dot G$, and we have emphasized that the support of $\mu_0$ is a kind of surrogate for a unitary real form for $\mathbf L \dot G$. A $K(A)$ bi-invariant probability measure $\widetilde{\mu}_0$ on $\mathbf G(A)$ would, according to Conjecture \ref{conj1}, necessarily project to $\mu_0$. I do not know how to prove the non-existence of this lift. However the existence of such a lift is inconsistent with basically everything we know about invariant measures. For example in order to integrate matrix coefficients, we have to fix a level and consider a bundle-valued measure on $\mathbf L\dot G$, and these bundle-valued measures have disjoint support, which is not consistent with the existence of one lift.

As we have emphasized in these notes, there are multiple ways in which $\mu_0$ can conjecturally be realized:
using root subgroup coordinates, using projections to moduli spaces of $G$-bundles on Riemann surfaces, using
time ordered exponential coordinates (i.e. $\theta_+=g_+^{-1}\partial g_+$ as a coordinate), and using holomorphic maps to flag spaces (we did not present concrete proposals in the latter two cases). There are lifts of these structures to the appropriate central extensions, but the corresponding lifted formulas for measures simply do not make any sense. For example a key point in Section \ref{calculate2} is that for triangular factorization for
a loop in $SL(2,\mathbb C)$,
$$g=l (ma)^{\dot h_1} u$$
the scalar $ma$ is a (meromorphic) quotient $ma=\frac{\sigma_1(\widehat g)}{\sigma_0(\widehat g)}$. The conjectural distribution for the quotient makes sense, but the numerator and denominator are simply not well-defined random variables.

I now want to explain why the existence of a unitarily bi-invariant measure in indefinite cases seems unlikely, or at least is definitely inconsistent with Conjecture \ref{conj1}.

A preliminary comment: It is not clear whether there are interesting analogues of root subgroup coordinates, and so on, in indefinite cases. The existence of root subgroup coordinates is related to presently intractable questions about imaginary roots for $\mathfrak g(A)$, such as their multiplicities. To produce root subgroup coordinates it is necessary to `order' the roots in some generalized sense, as in the affine case, see (\ref{shortversion}). In the affine case the ordering of imaginary roots is unimportant because the root vectors commute. This is not true in general in the indefinite cases. It is possible that commutativity of imaginary root vectors is a characterization of the finite and affine cases.

A central indefinite example is the maximal hyperbolic Kac-Moody algebra $E_{10}$. In this case one can exploit the inclusion of $E_9 \subset E_{10}$
in a way which is analogous to the inclusion $\dot{\mathfrak g} \subset L\widehat{\dot{\mathfrak g}}$ (see \cite{KMW}, following ideas of Feingold and Frenkel). $E_9$ is
the untwisted affine algebra corresponding to $E_8$,
\begin{equation}\label{extension}\mathfrak g(E_9)=\mathbb CL_0 \propto \widehat{L\dot{ \mathfrak g}},\text{ where } \dot {\mathfrak g}=\mathfrak g(E_8) \end{equation}
The $E_{10}$ adjoint action of the center of $E_9$ induces an `affine grading' (by $\mathbb Z$) for $\mathfrak g(E_9)$ (see \cite{KMW}),
and in turn this produces an analogue of the linear Riemann-Hilbert decomposition for loop algebras,
$$\mathfrak g(E_{10})= \mathfrak g(E_{10})_-\oplus \mathfrak g(E_{9})\oplus\mathfrak g(E_{10})_+ $$
For the formal group completion there is a corresponding analogue of (generic) Riemann-Hilbert factorization for loop groups,
$$g=g_-g_0g_+$$
where now (and this is the crucial point) $g_0$ is in the formal completion for the Kac-Moody group corresponding to
(\ref{extension}). According to the Uniqueness Conjecture \ref{conj1}, the $g_0$ distribution would have to be an invariant probability measure which pushes down to the measure $\mu_0$ on $\bf{L} \mathfrak{g}(E_8)$. As we have argued in the affine case above, the existence of such a lift seems unlikely.\\

Conclusions: The truth of Conjecture \ref{conj1} implies a negative answer to the question of whether there (always) exists a unitarily bi-invariant measure in indefinite cases. It is at least conceivable that one could use the complex version of root subgroup factorization (see \cite{BP1}) to produce a counterexample to Conjecture \ref{conj1}, and in turn it might be possible to use root subgroup factorization to produce an example of a unitarily bi-invariant measure in some indefinite case.\\

There has been considerable speculation in the physics literature about a connection between  $E_{10}$ and M Theory.
For example in \cite{BGH} it is conjectured that (one aspect of) M Theory is related (in a necessarily heuristic way)
to a Laplace type operator on the `non-compact type symmetric space' $G_0(E_{10})/K$, where $G_0(E_{10})$ denotes the group associated to the normal real form of $E_{10}$, as in Example \ref{symmetricspaces}. The proposals in these notes do not apply to the `non-compact type symmetric spaces $G_0/K$ in Example \ref{symmetricspaces}. It is commonplace to instead consider a double coset space $\Gamma\backslash G_0/K$, but it is not clear if we have anything interesting to say about this. In any event one wonders if the `dual symmetric space', $U(E_{10})/K(E_{10})$,
or more simply $U(E_{10})$, the group corresponding to the unitary real form of $E_{10}$, might also be relevant.

This suggests the following questions. In our musings about the chiral model with compact simply connected target $K$, we emphasized the natural role of the completions
$$L_{fin}G \subset C^{0}(S^1,G) \subset Hyp(S^1,G) \subset \mathbf L G $$
For $E_{10}$ we only have analogues of the first group (essentially the Kac-Peterson group) and the last space
$$G(E_{10}) \subset \mathbf G (E_{10}) $$
A first question is whether there exists a Lie group completion of the Kac-Peterson group $G(E_{10})$
analogous to $C^0(S^1,G)$, or $C^{\infty}(S^1,G)$. I suspect the answer to this question is `no'; see \cite{Pi5} for some evidence. A second question is whether there exists some analogue of $Hyp(S^1,G)$, which we have argued
is a kind of universal version of $Bun_G$ for Riemann surfaces.

This second question is clearly intertwined with the long standing puzzle of how to think about the transition
from $E_9$, the affine case, to $E_{10}$.

\section{ Potential Applications to Sigma Models}\label{sigmamodel}

In this section we will discuss the potential relevance of the measure $\mu_0$, and its conjectural structure, to the 2D sigma model with target $K$, i.e. the principal chiral model. It is useful to broaden the discussion to include a level $l$ and the Wess-Zumino-Witten term. For simplicity we will mainly focus on the case $K=SU(2)$. 

If $\Sigma$ is a closed Riemann surface, we would like to make sense of the heuristic Feynman measure on maps (or fields)  $g:\Sigma\to K$,
\begin{equation}\label{sigmaaction}exp(-\int_{\Sigma}\left(\frac{ 1}{2T}\langle g^{-1}dg\wedge *g^{-1}dg\rangle+2\pi i l\Gamma(g)\right))\prod_{v\in\Sigma} d\lambda(g_v)\end{equation}
where $\Gamma$ is the (multi-valued) WZW term, $T>0$ is a dimensionless parameter, the multi-valuedness of $\Gamma$ forces $l$ to be integral, and $\prod_{v\in\Sigma} d\lambda(g_v)$
denotes the background Haar measure for the compact group $\prod_{\Sigma}K$. At a heuristic level, this Feynman measure is conformally invariant. At a rigorous level, it may
simply not exist, or involve some exotic interpretation. In any event the ultimate goal is to construct a quantum field theory satisfying the axioms in (for example) \cite{KS}.

If $l=0$, then this is the sigma model with target $K$, i.e. the principal chiral model, denoted $CM=CM_0$ (or sometimes $PCM$). It is not known how to mathematically construct the quantum chiral model in finite volume. The action is local, and the model should be a kind of continuum central limit for a class of statistical mechanical models. Renormalization group heuristics suggest that conformal invariance is broken at the quantum level, and the dimensionless parameter $T$ transmutes into
a mass parameter $M$. In $\mathbb R^{1,1}$ there is a remarkable Yangian symmetry (or to put it another way, the scattering is assumed to be elastic), and consequently there is a conjectural description of the scattering theory, see e.g. \cite{Z1}, \cite{ORW} and \cite{Bernard}. The chiral model is also believed to be asymptotically free. This suggests that
for the Minkowski type space $RS^1\times \mathbb R$, if the radius $R$ is small (or maybe just finite), then in some sense to be made precise, the theory should approximately behave like a free massless theory (with $\dot r$ elementary particles, where $\dot r$ is the rank of $K$). Below we will speculate that the conjectured factorization of $\mu_0$ in root subgroup coordinates might lead to an explanation for this.

If $l\in \mathbb N$, then the situation is murkier. The action is not local (the $WZW$ functional is not local), and to my knowledge there does not exist a tangible connection with statistical mechanics. Nonetheless renormalization ideas are relevant. The special value $T=1/l$ corresponds to a renormalization group fixed point, the $WZW_l$ conformally invariant model. One possible interpretation of \cite{Z2} is that there should be a one parameter family of massless integrable theories which interpolates between a massless theory $CM_l$ and the conformally invariant
$WZW_l$ model. The parameter $T$ presumably transmutes into the flow parameter, the nature of which is unclear. I am confused by the interpretation in \cite{Bernard1}, which seems to imply that $CM_l$ is conformally invariant, and which seems to suggest that inversion invariance is broken, yielding a left right Yangian symmetry, analogous
to the left right symmetry for the WZW model (the recent preprint \cite{PW} is also relevant to $CM_l$).

The $WZW_l$ model, as a conformal field theory involving closed strings only, has apparently been constructed in the sense of Segal (see \cite{Kirillov1} and \cite{Kirillov2}), which is a major achievement. Construction of the corresponding boundary conformal field theory is apparently open. One also wonders if there is some
more elementary approach which might shed some light on the flow from $CM_l$ to $WZW_l$.

In this section we consider the ordinary sigma model ($l=0$), initially with target $X$, a Riemannian manifold.
The first subsection, on the classical theory, is included to motivate our point of view, nothing more.

We will then discuss some possible conjectures about the quantum theory, suggested by the structure of $\mu_0$.

\subsection{The Classical Euclidean Perspective}

The classical fields for the sigma model are maps $x:\Sigma\to X$, where $\Sigma$ is a space time and $X$ (the target)
is a Riemannian manifold. The action is given by
$$S: W^1(\Sigma,X)\to \mathbb R:x \to \frac 12\int_{\Sigma}\langle dx\wedge *dx\rangle$$
where the derivative $dx$ is regarded as a one form on $\Sigma$ with values in the tangent bundle of the target
and $*=*_{\Sigma}$ is the star operator. When $dim(\Sigma)=2$, this action depends only on the conformal
structure of $\Sigma$, because the star operator is conformally invariant on one forms in dimension 2.

The traditional Hamiltonian point of view is developed for example in \cite{Bernard} or section 2 of \cite{TU}.
From the Hamiltonian point of view one considers a Minkowskian type space time $\Sigma=\mathbb R^{1,1}$ or $\Sigma:=(S^1\times \mathbb R,dt^2-d\theta^2)$.  The classical solutions are so called wave maps $\Sigma \to X$, and the initial value problem is known to be globally well-posed if the target $X$ is compact.

Suppose that $X=K$. If $\Sigma=\mathbb R^{1,1}$, then the wave map system is integrable in the sense that there exist a zero curvature representation of the equations and a classical Yangian symmetry (see e.g. \cite{Bernard}). If $\Sigma=S^1\times \mathbb R$, with nontrivial topology, the Yangian symmetry does not exist, and it seems unlikely that the system is completely integrable in the sense of the existence of action-angle variables (on the complement of some `small subset').

We will adopt a Euclidean perspective which is more akin to the point of view of \cite{KS}. In this view we think of the classical two dimensional sigma model as a functor from
Segal's category of compact Riemann surfaces, where the objects are compact oriented 1-manifolds, and the morphisms are compact Riemann surfaces, to (an infinite dimensional version of) the symplectic category of Guillemin/Sternberg and Weinstein, where the objects are symplectic manifolds and the morphisms are Lagrangian submanifolds (see \cite{NP} for a tutorial on this point of view). More precisely a compact oriented 1-manifold $S$ maps to the cotangent bundle of the configuration space, $W^{1/2}(S,X)$ (closed strings in $X$ parameterized by $S$), and a compact Riemann surface $\Sigma$ with boundary $S$ maps to the Lagrangian submanifold of $T^*W^{1/2}(S,X)$ defined by
$$W^{1}Harm(\Sigma,X)\subset T^*W^{1/2}(S,X):x \to *dx|_S$$
where $W^1Harm(\Sigma,X)$ is the space of harmonic maps from $\Sigma$ to $X$. Note $*dx|_S$ remembers the boundary values and outward pointing normal derivative of the harmonic map $x$, i.e. the Dirichlet and Neumann data associated to the map. If $X$ is linear, then this Lagrangian submanifold is a graph, because the Neumann data is a function of the Dirichlet data. In general the geometry of these Lagrangian submanifolds is highly complex, reflecting the nonlinearity of the harmonic map equations.

\begin{proposition}\label{sewing}Consider a composition $\Sigma_2\circ \Sigma_1$ in Segal's category. Then
$$Harm(\Sigma_2,X) \circ Harm(\Sigma_1,X) = Harm(\Sigma_2\circ\Sigma_1,X)$$\end{proposition}

\begin{remarks} (a) On the left hand side of this proposition, $\circ$ denotes composition in the infinite dimensional generalization of the symplectic category, while on the right hand side $\circ$ denotes composition in Segal's category of Riemann surfaces.

(b) When all of the abstraction is unwound, this simply says that given harmonic maps $x_1:\Sigma_1\to X$ and $x_2:\Sigma_2\to X$, to put them together to obtain a harmonic map $\Sigma_2\circ \Sigma_1 \to X$, it is necessary and sufficient that the values of the maps agree along the common boundary and that the normal derivatives match up.

(c) Roughly as in the Minkowskian setting, when $\Sigma=\mathbb C\mathbb P^1$ is simply connected, harmonic maps into $K$ are in some sense known (or at least expressible in terms of a holomorphic map into the basic homogeneous space for $LK$ and a Gram-Schmidt process); for non-simply connected surfaces a `classification' is elusive (see e.g. \cite{Uhlenbeck1}, \cite{Guest} and \cite{Hitchin}).
\end{remarks}

\subsection{On the Hamiltonian for $CM_0$}

Our main goal in this subsection is to formulate a possible conjecture for the $CM_0$ Hamiltonian. This will involve the introduction of three hypotheses. At least in my view, the first two hypotheses are quite plausible. The third hypothesis is probably too simplistic. But I have not been able to rule it out, and it might be useful as an approximation.

For the classical chiral sigma with target $X$, in the Euclidean framework,
a compact one manifold $S$ is mapped to the phase space, the cotangent bundle of $W^{1/2}(S,X)$. For the quantum sigma model, in the Euclidean framework, $S$ should naively map to a quantization of the cotangent bundle, e.g. to the Hilbert space of half-densities associated to the configuration space $W^{1/2}(S,X)$. In general this is meaningless.
The basic problem is that to define a space of half-densities, we need a measure class, and interesting measure classes tend to live on completions, or a thickening, of $W^{1/2}(S,X)$. From a physics perspective, because of the uncertainty principle, one cannot expect that the time zero quantum fields will have values in the possibly curved space $X$. A  solution to this problem, at least in principle, is to embed $X$ into $\mathbb R^n$, use `$\mathbb R^n$-valued' generalized functions defined on $S$, and use a potential to coax fields to have some affinity for having values near $X$. In practice a renormalization group process is necessary, which modifies the potential at each scale, and what one ends up with is invariably obscure.

Now suppose that $X=K$. In this case $Hyp(S,G)$ is a natural thickening of $TW^{1/2}(S,K)\sim W^{1/2}(S,G)$
which is homotopically faithful and equivariant with respect to real analytic $K$ valued loops acting from the left and right. There is a natural measure class associated to $W^{1/2}(S,K)$, namely the measure class of $\mu_0$ (assuming $S$ is a simple circle; otherwise we interpret this as the product measure for the loop groups corresponding to the different components of $S$). As we have stressed previously, a basic theme is that the support
of the measure class of $\mu_0$ is a kind of surrogate for the classical configuration space, $W^{1/2}(S,K)$.

\subsubsection{First Hypothesis}  $S$ maps to the Hilbert space of half-densities of the measure class, $\mathcal H^{1/2}([\mu_0])$.  In other words we are substituting the measure class of $\mu_0$ for the configuration space. We can and will identify the state space with $L^2(\mu_0)$ whenever this is convenient. As we will see below, the fact that there is a canonical measure representing the measure class $[\mu_0]$ is important.

\subsubsection{Comments}
Any two separable infinite dimensional Hilbert spaces are isomorphic. The question is whether this specific realization of the quantum state space has some natural meaning. 

We will initially focus on the flat Euclidean cylinder $RS^1\times \mathbb R$. The nature of the dependence on the radius $R$ is important, but we will temporarily assume $R=1$. There will be an associated one parameter semigroup of contractions, which we can interpret as a homogeneous Markov process (if we analytically continue to the Minkowski point of view, we obtain a one parameter group of isometries, and the generator, the Hamiltonian $H=H(1)$, has positive energy).

\begin{remark}\label{remark10} As an aside, I will briefly mention some standard heuristics - these are probably of limited utility
for this problem involving dimensional transmutation.

For general Riemannian target $X$, on the cylinder the action is given by
\begin{equation}\label{1.1}S(x:S^1\times \mathbb R \to X)=\frac
12\int_{\Sigma}\{\vert\frac {\partial x}{\partial t}\vert^2+\vert\frac {
\partial x}{\partial\theta}\vert^2\}d\theta dt\end{equation}
The time zero fields constitute the loop space
$Map(S^1,X)$. The tangent
space to the loop space at $x$ is naturally identified with $\Omega^0(x^{*}
TX)$,
the space of vector fields along the loop $x$.  There is a
$W^0$ Riemannian metric on this tangent space, given by
\begin{equation}\label{1.2}\langle v,w\rangle_x=\int_{S^1}\langle v(\theta ),w(\theta )\rangle_{
x(\theta )}d\theta \end{equation}
where $v(\theta ),w(\theta )\in TX\vert_{x(\theta )}$, and $\langle
\cdot ,\cdot\rangle_{x(\theta )}$ denotes the inner
product (Riemannian metric) for $X$ at the point $x(\theta )$.  In
this way we can view $Map(S^1,X)$ as a Riemannian manifold, which we denote by $W^0(S^1,X)$.

In the second expression in (\ref{1.1}) for the action, the first term is
the usual kinetic energy for a path in the Riemannian
manifold $W^0(S^1,X)$, and the second term represents a
potential energy term, corresponding to the energy
function on the finite energy loop space $W^1(S^1,X)$,
\begin{equation}\label{1.3}E(x:S^1\to X)=\frac 12\int_{S^1}\langle dx\wedge *dx\rangle =\frac
12\int\vert\frac {\partial x}{\partial\theta}\vert^2d\theta \end{equation}
Note that the Riemannian metric (\ref{1.2}) and $E$ depend upon
the radius of $S^1$.

From this we heuristically deduce
that the quantum Hamiltonian for the sigma model is of
the form
$$H=\Delta_{W^0} +E$$
where $\Delta_{W^0}$ is the Laplacian for the Riemannian manifold
$W^0(S^1,X)$, and $E$ is viewed as a (extremely singular)
multiplication operator.

A puzzle is that this point of view does not at all suggest why $W^{1/2}(S^1,X)$
is the natural analytic configuration space.
\end{remark}

\subsubsection{Second Hypothesis}

As above, our first hypothesis is that the state space is $\mathcal H^{1/2}([\mu_0])$, which we will identify with $L^2(d\mu_0)$. Our second hypothesis is that the ground state is $\mu_0^{1/2}\in \mathcal H^{1/2}([\mu_0])$, or in other words the characteristic function of the support of $\mu_0$ in $L^2(d\mu_0)$.  \\

\subsubsection{Comments} Heuristically, if we consider $H_0=\Delta_{W^0} $ (the Laplacian from Remark \ref{remark10},
then we would classically be considering geodesics on the product group  $\prod_{S^1}K$, and the ground state would be the Haar measure $\prod_{S^1}d\lambda_K$ (or more precisely its square root). In this case the ordering of the points of space is completely irrelevant.

For the chiral model the energy function for the loop group is added as a potential (see the heuristics above). This potential is attempting to maintain the ordering
of the points of space $S^1$. Our second hypothesis is essentially asserting that the ground state is now shaped more like the characteristic function for `the support of $\mu_0$', our quantum surrogate for $W^{1/2}(S^1,K)$. In particular the restriction of a field to
space is marginally distributional in nature; we can only make sense of such a field in terms of, for example, its root subgroup (or nonlinear Fourier series) coefficients.

In combination with Segal's axioms, this second hypothesis has a surprising consequence. Segal's axioms mimic properties of the path integral, and when we assume the vacuum has a measure theoretic realization, it basically means that the path integral actually exists and defines a stochastic process. We will discuss this below.

There is no finite dimensional justification for this second hypothesis. If one considers a generalized Laplace type operator $H=\Delta+U$ on a compact Riemannian manifold, the large $T$ limit of $\frac{1}{\mathfrak Z}exp(-\frac 1T U)dV$ is simply the Riemannian volume, which is the ground state for $\Delta$ and completely forgets the potential. We will see something similar for $YM_3$ (which is subcritical). The reader will have to be patient.    

\subsubsection{Third Hypothesis}

Our ambition now is to use what we have learned (or at least conjectured) about the measure $\mu_0$ to make an educated guess at the dynamics for the chiral model. For a free field (essentially the chiral model with target $\mathbb R$ in place of $K$), the Feynman measure and the vacuum are Gaussian, and from this and Fourier series, one can conclude
that the quantum field is an assembly of harmonic oscillators. This is true independent of the dimension of space, and whether space is compact or otherwise. For the nonabelian $K$ valued chiral model dimension and compactness matter a great deal, and we are considering the critical case.

We are now focusing on the chiral model restricted to $S^1\times \mathbb R$, i.e. space is compact and one dimensional ($R=1$). The analogue of the Fourier transform for a loop in $K$ is root subgroup factorization. As it happens the vacuum in our second hypothesis factors in root subgroup coordinates. This should tell us something.\\

{\bf Our third hypothesis is that the infinitesimal generator $H$ also factors in root subgroup coordinates}. At first sight this must seem nuts. We will first spell this out in a precise way, then discuss why this might actually be possible. Even if this is naive, it might be useful.\\

Given a $g\in W^{1/2}(S^1,K)$ having a triangular factorization, there is a unique root subgroup factorization (or multiplicative Fourier series)
\begin{equation}\label{RSF}g(z)=\prod_{i\ge 0}^{\rightarrow}\mathbf a(\eta_i)\left(\begin{matrix} 1&\overline{\eta}_iz^i\\
-\eta_iz^{-i}&1\end{matrix} \right)\left(\begin{matrix} e^{\chi(z)}&0\\
0&e^{-\chi(z)}\end{matrix} \right)\prod_{k\ge 1}^{\leftarrow}\mathbf a(\zeta_k)\left(\begin{matrix} 1&\zeta_kz^{-k}\\
-\overline{\zeta}_kz^k&1\end{matrix} \right),\quad \vert
z\vert=1,\end{equation}
where $\chi(z)=\sum\mathbf{\chi}_jz^j$ is a $i\mathbb R$-valued Fourier series
(modulo $2\pi i\mathbb Z$).

\begin{remark}\label{leaking} It is useful to recall that the $\eta_i$ and $\zeta_j$ variables should be viewed as
something similar to affine coordinates for spheres. This is because the limit points for
$$\mathbf a(\eta_i)\left(\begin{matrix} 1&\overline{\eta}_iz^i\\
-\eta_iz^{-i}&1\end{matrix} \right) \text{  where  } \mathbf a(\eta_i)=\frac{1}{(1+|\eta_i|^2)^{1/2}} $$
are of the form
$$\left(\begin{matrix} 0&z^i\\
-z^{-i}&0\end{matrix} \right)\left(\begin{matrix} \lambda&0\\
&\lambda^{-1}\end{matrix} \right)  \text{  where  } \lambda\in S^1$$
These infinite points represent a Weyl group element for a root subgroup corresponding to a real root. One could think of this totality of points as a kind of skyscraper sheaf over a sphere. But in this measure theoretic context, we will just think in terms of a sphere.
\end{remark}

Although $W^{1/2}$ has measure zero with respect to $\mu_0$, the $\eta_i$, $\chi_j$ and $\zeta_k$
can be interpreted as random variables with respect to $\mu_0$, and conjecturally
$\mu_0$ factors:
\begin{equation}\label{productmeasure7}d\mu_0=\left(\prod_{i=0}^{\infty}\frac {1+2i}{\pi}\frac
{d\lambda (\eta_
i)}{(1+\vert\eta_i\vert^2)^{2+2i}}\right)\times \left(\prod_{
j=1}^{\infty}\frac
{4j}{\pi}e^{-4j\vert\mathbf{\chi}_j\vert^2}d\lambda
(\mathbf{\chi}_j)\right)$$
$$\times d\lambda
(e^{\chi_0})\times\left(\prod_{k=1}^{\infty}\frac
{2k-1}{\pi}\frac {d\lambda (\zeta_
k)}{(1+\vert\zeta_k\vert^2)^{2k}}\right),\end{equation}

\begin{remark}\label{leaking2} Note that the densities for $\eta_i$ and $\zeta_k$ have relatively large variances (compared to a Gaussian such as $\chi_j$). Hence there is a considerable amount of mass which is migrating from the real root directions to the imaginary root directions, per Remark \ref{leaking}.
\end{remark}

This form of the vacuum suggests a form for the Hamiltonian. First, the Gaussian density for $\chi$ suggests that $\chi$ is a massless free field, with the zero mode compactified. 
This is striking, because the chiral model is asymptotically free, and more precisely at short distances (e.g. when the radius $R$ is small), the field should behave like a massless free field with $rank(K)$ elementary particles.

The most naive way to think about the spectrum and partition function for the free field is the following. For the harmonic oscillator $-\frac 12 (\frac{d}{dq})^2+\frac{\omega^2}{2} q^2$ ($q\in \mathbb R$), the ($L^2$ normalized) ground state is $(\frac{\omega}{\pi})^{1/2}e^{-\frac{\omega}{2}q^2}$ and the spectrum is $\{\frac{\omega}{2}+n\omega:n=0,1,...\}$. Since we will have infinitely many modes, we need to subtract the ground state energy, so that the spectrum is $\{n\omega:n=0,...\}$. If $q$ denotes the real or imaginary part of $\chi_j$, then we take $\omega_j=4j$, $H_{Re(\chi_j)}=-\frac 12 (\frac{d}{dq})^2+\frac{\omega_j^2}{2} q^2-\frac{\omega_j}{2}$, and the partition function is
$$trace(e^{-tH_{Re(\chi_j)}})=\sum_{n=0}^{\infty}e^{-t\omega_n}=\frac{1}{1-e^{-4jt}} $$
We should emphasize that this is contingent on subtracting the ground state energy, which potentially leads to problems when we consider all the modes at once (because then we are subtracting infinity).

Naively, the partition function for the $\chi$ Hamiltonian is the product
$$trace(e^{-tH_{\chi}})=(\prod_{j=1}^{\infty}\prod_{n=0}^{\infty}\frac{1}{1-e^{-4jt}})^2=\frac{1}{\phi(e^{-4t})^2}$$
If we remember that we subtracted $\sum_{n=1}^{\infty}n$, then we should correct this using a zeta function regularization and write
$$trace(e^{-tH_{\chi}})=\frac{1}{\eta(4\tau)^2} \text{ where } \tau=\frac{it}{2\pi}$$
One can arrive at this same formula by using the standard zeta regularization of the determinant for the Laplace operator on a torus (see sections 1 and 2 of \cite{Gawedzki2}), or for path integral heuristics, see page 341 of \cite{DMS}. At least in my view, the justification for using
zeta function regularization is that it is consistent with Segal's axioms for quantum field theory (this is one point of \cite{Pi2.5}).

In the general case of $\dot K$, using the same reasoning, the partition function is, in the first form,
$$trace(e^{-tH_{\chi}})=\frac{1}{\phi(e^{-2\dot g t})^{2r}}$$
where $r$ is the rank.\\

Now consider (in the $SU(2)$ case) one of the modes corresponding to a real root $\tau$, say corresponding to the parameter $\eta_i$. The corresponding measure has a density proportional to
$$\frac{1}{(1+|\eta_i|^2)^{2+2i}}$$

\begin{remark} \label{geomremark1} Recall from Remark \ref{leaking} that one should actually think of $\eta_i$ as an affine parameter for a sphere. Let $\kappa$ denote the canonical bundle (cotangent bundle) for $\mathbb P^1$. For the bundle $\kappa^{-1/2}$ on $\mathbb P^1$ (which can be identified with the dual of the tautological bundle) the norm of the canonical section is
$$\frac{1}{(1+|\eta_i|^2)}$$
So there is a geometric interpretation lurking here.
\end{remark}

Our basic observation is that this density determines a kind of spherical harmonic oscillator (see \cite{HP} for a more formal introduction).

Consider the rotationally invariant metric on $\mathbb P^1$ with unit Gaussian curvature. In the standard affine coordinate $z=x+iy=re^{i\theta}$, the metric is
$$ds^2=\frac{4}{(1+z\overline z)^2}(dx^2+dy^2)$$
Let $\Delta_{S^2}$ denote the corresponding nonnegative Laplace operator.

\begin{definition} \label{sphosc}Given $\omega\ge 0$ (in the standard affine coordinate $z=re^{i\theta}$)
$$L_{\omega}=\Delta_{S^2}+\omega^2 r^2=-\frac 14(1+r^2)^2\left((\frac{\partial}{\partial r})^2+\frac 1 r\frac{\partial}{\partial r}+\frac{1}{r^2}(\frac{\partial}{\partial \theta})^2\right)+\omega^2r^2$$
The corresponding quadratic form is
$$Q_{\omega}(f)=\int_{\mathbb P^1}df\wedge*df+\omega^2r^2f^2 *1$$
\end{definition}

\begin{remark}\label{sphosc2} Beyond our third hypothesis that the Hamiltonian of the sigma model is diagonalized by root subgroup factorization, what motivates/justifies this definition? In a nutshell, we want $L{\omega}$ to be a $S^1$ invariant self-adjoint second order
differential operator having principal symbol determined by the rotationally invariant metric for the 2-sphere
and having the ground state in part (a) of the following theorem. Imagine adding lower order terms to $L_{\omega}$. The corresponding quadratic form, in terms of the radial parameter $r$ (we can ignore $\theta$ because of rotational invariance), would (because of self-adjointness) be of the form
$$\int_{r=0}^{\infty} b(r)(f'(r)g(r)+f(r)g'(r))\frac{r}{(1+r^2)^2} dr$$
$$=\int_{r=0}^{\infty} (b(r)f'(r)g(r)\frac{r}{(1+r^2)^2}-\frac{\partial }{\partial r}(b(r)f(r)\frac{r}{(1+r^2)^2}) dr$$
This calculation shows that for the differential operator, the derivative terms cancel, and this only adds zeroth order terms. The fact that we know the lowest energy state (as we claim in part (a) below) implies that
$L_{\omega}$ is determined up to a constant. When we apply this to the sigma model, these constants will be determined by the qft consideration that the ground state energy must be zero.

\end{remark}

The following is from \cite{HP}.

\begin{theorem}\label{spectrum}

(a) The ground state for $L_{\omega}$ is
$$(1+r^2)^{-\omega}$$
with corresponding eigenvalue $\omega$.

(b) The spectrum is 
$$\lambda_{m,n}:=(m+\frac{n+1+\sqrt{n^2+4\omega^2}}{2})^2-\omega^2-\frac 14$$
$m,n\ge 0$, counted with multiplicity 1 when $n=0$ and multiplicity 2 otherwise (We are not claiming these are all distinct; see (c)).
The partition function is
$$tr(e^{-tH})=(\sum_{m,n=0}\mathbf m(m,n)e^{-t\lambda_{m,n}}$$
where $\mathbf m(m,n)=1$ if $n=0$ and $=2$ otherwise.

(c) If $\omega=0$, then $\lambda_{m,n}=(m+n)(m+n+1)$. In this case the spectrum is $N(N+1)$ with multiplicity $2N+1$, $N=0,1,...$. The partition function
$$tr(e^{-tH})=\sum_{N=0}^{\infty}(2N+1)e^{-tN(N+1)}$$

\end{theorem}

\begin{remark}\label{abstractarg}  This is a continuation of Remark \ref{geomremark1}. The ground state calculation can be cast
in more abstract terms. Suppose that $L$ is a holomorphic hermitian line bundle with canonical connection $\nabla$, where $\nabla^{(0,1)}=\overline \partial$. There is an associated nonnegative Laplace operator
$$\Delta  $$
If $s$
is a holomorphic section, then
$$\partial (s,s)=\theta (s,s)$$
(see page 73 of Griffiths and Harris) and
$$\overline {\partial}\partial (s,s)= \overline \partial (\theta) (s,s)-\theta \wedge \overline \theta (s,s)$$
Thus
$$\overline {\partial}\partial (s,s)+\theta \wedge \overline \theta (s,s)=\Theta (s,s)$$
where $\Theta$ is the curvature, a (1,1) form.
As we noted in Remark \ref{geomremark1}, in our case the line bundle is the kth power of the dual of the canonical bundle (or the 2kth power
of the generating positive line bundle on $\mathbb P^1$, the dual of the tautological bundle).

\end{remark}

\subsubsection{The $CM_0$ Hamiltonian for $K=SU(2)$}

Recall that we are initially assuming that the radius of the circle is $R=1$, hence we will write the Hamiltonian
as $H(1)$. According to our third hypothesis, $H(1)$ factors in terms of the coordinates $\eta_0,...,\eta_i,...$, $\chi_1,...,\chi_j,...$,
$\zeta_1,...,\zeta_k,...$. We are considering the possibility that $\chi_j$ is a standard linear harmonic oscillator. Consider
$\zeta_k$. We are considering the possibility that this is a spherical harmonic oscillator with $\omega=2(k-1)$.
Since there are infinitely many modes, we need to fix the spherical Hamiltonian so that the ground state energy vanishes. We therefore take
$$H_{\zeta_k}=L_{2(k-1)}-2(k-1) \text{  and  } H_{\eta_i}=L_{2i}-2i$$

Putting everything together (with $l=0$ in what follows, but included for later reference)
\begin{equation}\label{SigmaHamiltonian}H(R=1):=\sum_{i=0}^{\infty}(\Delta^{S^2}_{\eta_i}+((2+l)i)^2|\eta_i|^2-(2+l)i)+
\end{equation}
$$
\sum_{j=1}^{\infty}(\Delta^{\mathbb R^2}_{\chi_j}+(2(2+l))^2|\chi_j|^2-2(2+l)j   )+
\sum_{k=1}^{\infty}(\Delta^{S^2}_{\zeta_k}+((2+l)(k-1))^2|\eta_i|^2-(2+l)(k-1))$$

\subsubsection{The $CM_0$ Partition Function for $K=SU(2)$}

The conjectural full partition function is an infinite product. We have already evaluated the free massless part of the product.

If $\omega=2i$, let $\lambda_{i,m,n}$ denote the corresponding eigenvalues
in Theorem \ref{spectrum} for $L_{\omega}$.
The full partition function is
$$trace(e^{-tH(1)})=\frac{\left(e^{t}\prod_{k=1}^{\infty}\sum_{m,n=0}\mathbf m(m,n)
e^{-t(\lambda_{k,m,n}-2k)}\right)^2}{\eta(e^{-4t})^2}$$

\begin{lemma} Suppose $q<1$. Then the product
$$=\prod_{k=1}^{\infty}\left((\sum_{m,n=0}\mathbf m(m,n)
q^{m^2+m+m(n+\sqrt{n^2+16k^2})+(n+1)\frac{n+\sqrt{n^2+16k^2}}{2}-2k}\right)$$
converges and is nonzero.
\end{lemma}

\begin{proof} We must show that
$$\sum_{k=0}^{\infty}\left((\sum_{m>0,n=0}+2\sum_{m\ge 0,n\ge 1})
q^{m^2+m+m(n+\sqrt{n^2+16k^2})+(n+1)\frac{n+\sqrt{n^2+16k^2}}{2}-2k}\right)$$
converges. This can be split into two sums. The first sum is
$$\sum_{k=0}^{\infty}\sum_{m=0}^{\infty}
q^{m^2+m+4mk}$$
$$=\le \sum_{m=1}^{\infty}q^{m^2+m}(1-q^{4m})^{-1}$$
Since $(1-q^{4m})$ converges to $1$ as $m\uparrow \infty$, this sum is convergent.
The second sum is $2$ times
$$\sum_{k=0}^{\infty}\sum_{m\ge 0,n\ge 1}
q^{m^2+m+m(n+\sqrt{n^2+16k^2})+(n+1)\frac{n+\sqrt{n^2+16k^2}}{2}-2k}$$
$$\le \sum_{k=0}^{\infty}\sum_{m\ge 0,n\ge 1}
q^{m^2+m+m(n+4k)+(n+1)(n+2k)-2k}$$

$$\le \sum_{k=0}^{\infty}\sum_{m\ge 0,n\ge 1}
q^{m^2+m+m(n+4k)+(n+1)n+2nk}$$
$$=\sum_{m\ge 0,n\ge 1} \sum_{k=0}^{\infty}
q^{m^2+m+m(n+4k)+(n+1)n+2nk}$$
$$=\sum_{m\ge 0,n\ge 1}
q^{m^2+m+n^2+n+mn}(1-q^{4m+2k})^{-1}$$
Since $(1-q^{4m+2k})$ tends to $1$ as $m,n\uparrow \infty$, this will converge
iff
$$=\sum_{m\ge 0,n\ge 1}
q^{m^2+m+n^2+n+mn}$$
converges. This does converge, and this implies the lemma.

\end{proof}

To this point we have not discovered anything of special mathematical interest about this expression
for the partition function. Perhaps this is what we should expect, because $CM_0$ is not conformally invariant. In any event we now
have a possible formula for the partition function. 

\begin{question} Can this be numerically tested?
\end{question}

\subsubsection{The Feynman Measure for $S^1\times \mathbb R$}\label{Feynmanmeasure}

By abstract nonsense there is a path integral, or Feynman measure on the path space
$C^0((-\infty,\infty),Hyp(S^1,G))$. Given any time $t_1$, the projection of the measure by the evaluation map $g\to g(t_1)$ is the vacuum $d\mu_0$. Given times $t_1<t_2<...<t_n$ the corresponding projected measure is
$$\prod_{j=2}^n\langle g_{j-1}|e^{-tH}|g_j\rangle\prod_{k=1}^n d\mu_0(g_k) $$
where $\langle g_{1}|e^{-tH}|g_2\rangle$ denotes the kernel (It might be possible to evaluate this kernel exactly, similar to the Mehler kernel for the standard harmonic oscillator).

Just as we can define (unnormalized) Wiener measure on $C^0(S^1,K)$ using evaluation projections
$$(eval_V)_*(w_t)=\prod_E p_{tl(e)}(g_{\partial e})\prod_V d\lambda_K(g_v) $$
we can, modulo some technical issues, define a Feynman type measure $W_t$ on $C^0(S^1,Hyp(S^1,G))$ using evaluation projections
$$(eval_V)_*(W_t)=\prod_E P_{tl(e)}(g_{\partial e})\prod_V d\mu_0(g_v) $$
where now
$$P_{t}(g_{\partial e})=\langle g_{1}|e^{-tH}|g_2\rangle$$ denotes the kernel for $e^{-tH}$.

The total integral of the Wiener measure is $p_{2\pi t}(1)$. What is the analogue for $W_t$? It is $trace(e^{-2\pi t H})$.

A mystery is how to think about this `$SU(2)$ quantum field' in a coordinate independent way. This should be compared with a 
free scalar field theory. For a free scalar field, the field is almost surely a continuous function of a transverse parameter (e.g. the time parameter $t$ above)
with values in distributions in spacial directions. We are speculating the same is true for the sigma model. The free scalar field can be more invariantly
constructed (as a limit) by using a conformal loop ensemble.
 
\subsubsection{The Flat Torus, and the Dependence of $H$ on the Radius $R$}\label{flat torus}

Consider the flat torus with projections
$$\begin{matrix} & & R_1S^1 \times R_2S^1 & & \\
& \swarrow& & \searrow & \\
R_1S^1 & & & R_2 S^1 \end{matrix} $$

For each of these projections, there is a disintegration formula for a candidate for the Feynman measure
corresponding to this flat torus. By calculating the total measure, using the previous subsections, we obtain
\begin{equation}\label{consistency}\mathfrak Z(R_1S^1 \times R_2S^1)=trace(e^{-2\pi R_1 H(R_2)})=trace(e^{-2\pi R_2 H(R_1)})\end{equation}

In conformal field theory, at least according to page 337 of \cite{DMS}, $H(R)=\frac 1R H(1)$. This is definitely true
for a massless free field in two dimensions) This would imply the consistency of (\ref{consistency}).

The chiral model is expected to be asymptotically free, vaguely meaning that as $R\downarrow 0$, the $rank(K)$ elementary particles should be approximately free massless and independent (In the case of 3D Yang-Mills, there is a technical formulation of what this might mean in \cite{CC}). 

Superficially it might seem that we actually have $dim(K)$ independent free fields, one for each positive root
of $\mathfrak g$ and $r=rank(K)$ for the torus. But this is misleading. The apparent field that corresponds
to a given positive root of $\mathfrak g$ is not really an independent quantum field. There is a fundamental difference for the vacua for the real roots versus the imaginary roots. 

The imaginary roots correspond to a free massless field, which is an assembly of linear harmonic oscillators.
The ordinary Fourier integral expansion can be derived from Fourier series: if $f$ has compact support, for sufficiently large
$L$, there is an orthonormal Hilbert space/Fourier series expansion
$$f(x)=\sum_{-\infty}^{\infty}\langle f,\frac{1}{\sqrt{2L}}e^{2\pi in x/2L}\rangle\frac{1}{\sqrt{2L}} e^{2\pi in x/2L} $$
$$=\sum_{-\infty}^{\infty}\int f(y)e^{-2\pi i yn/2L} \frac{dy}{2L}e^{2\pi i xn/2L} $$
$$=\sum_{-\infty}^{\infty}\widehat f(\frac{n}{2L}){2L}e^{2\pi i xn/2L} $$
This is a Riemann sum which converges
to
$$\int_{-\infty}^{\infty}\widehat f(p)exp(2\pi ix\cdot p)dp $$
From this one can easily derive the infinite volume expansion of the free field from the finite volume expansion.

This is not so for the real roots. Root subgroup factorization involves products. The fundamental problem: It is not clear how to think about the infinite volume limit, which naively should
involve a continuum product. In the limit as $R\uparrow \infty$ we have to recover the Yangian symmetry and the scattering solution. I have not made any progress on this.

\subsection{Coupling with Gravity and Sewing}

Of our three hypotheses in this section, only the first is relevant to the construction of a model in the sense of Segal. Thus given a compact 1-manifold $S=S^{(1)}$, the corresponding Hilbert space is $\mathcal H^{1/2}([\mu_0])$ (where if $S^{(1)}$ has multiple components, then we are associating a copy of $\mu_0$ to each component).
The fundamental problems are to associate a vector $\mathcal Z\in \mathcal H^{1/2}([\mu_0])$ to a morphism $\Sigma$ with boundary $S$, and to prove the Segal sewing axioms are satisfied. The standard heuristic prescription (as in \cite{Pi2.5}) for obtaining this vector has to be supplemented with the renormalization group. 
Our basic hypothesis is simply that the state will have the form
$\rho_{\Sigma} d\mu_0^{1/2}$. Can we come up with a laundry list of constraints which will determine these densities, i.e. does knowing the measure class help?
I am stuck at this point.

\subsection{ On the Hamiltonian for the Chiral Model $CM_l$, $l=1,2,...$, $K=SU(2)$ }

This is completely parallel to the case $l=0$. We will introduce parallel hypotheses, without much discussion.

Various aspects of the classical theory have been considered by many authors; for a sampling, see e.g. \cite{AM}, \cite{Bernard}, \cite{Bernard1}, \cite{Uhlenbeck2}. The relevance of this to the Hamiltonian is that we need to consider sections of a line bundle, indexed by the level $l$, over the phase space, rather than simply functions.

\subsubsection{First Hypothesis} Space $S$ maps to the Hilbert space of square integrable sections
of $\Omega^0_l(\widehat{Hyp(S^1,G)})$. Note that the Hilbert space for $WZW_l$ is the subspace of holomorphic
functions, and in turn this contains the `primary fields':
$$ \Omega^0_l(\widehat{Hyp(S^1,G)})\supset H^0_l(\widehat{Hyp(S^1,G)}) \supset \oplus_{level(\Lambda)=l}\mathbb C \sigma_{\Lambda}$$

For simplicity we will consider the case of $K=SU(2)$.

\subsubsection{Second Hypothesis} The ground state is $\sigma_0^l$. Recall that
$$d\mu_l:= \sigma_0^l \otimes \overline{\sigma_0^l} d\mu^{|\mathcal L|^{2l}} $$
We think of this groundstate as a square root.

\subsubsection{Third Hypothesis}

As in the case $l=0$, the measure $\mu_l$, and also the square root $\sigma_0^l$, factors in root subgroup coordinates. The one complication is that for $\sigma_0$ we have to use root subgroup coordinates for the central extension.

Our third hypothesis is that the infinitesimal generator $H$ also factors in root subgroup coordinates.

 \begin{remark} It would make absolutely no sense to introduce a hypothesis like this for the $WZW_l$ model.
We know what the infinitesimal generator is in the WZW case: $H=L_0+\overline {L_0}$, which is only defined
 relative to the left right factorization of the Hilbert space for the WZW model:
$$ L^2H^0_l(\widehat{Hyp(S^1,G)}) =\oplus_{level(\Lambda)=l}H(\Lambda)\otimes \overline{H(\Lambda)}$$
In some sense we are trying to understand if there is some kind of flow from root subgroup coordinates
to the left-right factorization for the WZW model. 
\end{remark}

As before the form of the vacuum suggests a form for the Hamiltonian. The density for $\chi$ (and asymptotic freedom) suggests that $\chi$ is a massless free field, with the zero mode compactified.

Now consider (in the $SU(2)$ case) one of the modes corresponding to a real root $\tau$, say corresponding to the parameter $\eta_i$. The corresponding measure has a density proportional to
$$\frac{1}{(1+|\eta_i|^2)^{2+(2+l)i}}$$

Fix a parameter $\omega\ge 0$. We now consider what we will call
the spherical harmonic oscillator
$$L_{\omega}=\Delta+\omega^2 r^2=\frac 14(1+r^2)^2((\frac{\partial}{\partial r})^2+\frac 1 r\frac{\partial}{\partial r}+r^{-2}(\frac{\partial}{\partial \theta})^2)+\omega^2r^2$$
where $z=re^{i\theta}$ and now
$\Delta$ denotes the Laplacian acting on sections of $\mathcal T^{*\otimes (2+l)i}$, the $(2+l)$ power of the dual of the tautological line bundle on $\mathbb P^1$.

We can calculate the spectrum exactly as before. The upshot is $\dot g=2$, the dual Coxeter number, is replaced
by $\dot g+l$ in the formulas.\\

As in the case $l=0$, the partition function will not be modular invariant. This is not consistent with \cite{Bernard}.

\section{Sewing Rules for WZW Modular Functors}\label{sewing}

It is well-known that at least heuristically the sewing rules for the $WZW_l$ modular
functor (the holomorphic part of the theory) should be a consequence of a holomorphic level $l$ Peter-Weyl theorem,
as in Conjecture \ref{Peter-Weyl}. The point of this section is to explain that if Conjectures \ref{conj3}
and and \ref{Peter-Weyl} are true, then the sewing rules can be proved using Peter-Weyl.
Whether the measure-theoretic point of view we are advocating is useful for proving Segal's axioms
for the full theory will be discussed in the next section.

In this section we will use, without comment, the
general definition of a modular functor, and
representation of a modular functor (or weakly
conformal field theory), which are in Section 5 of \cite{Segal}.  We
will begin by recalling Segal's construction of the level $l$
WZW modular functor and its representation.  We will
then discuss the sewing property.

\subsection{ Construction of the chiral $WZW_l$ theory}

Let $\Phi_l$ denote the finite set of dominant integral functionals $\Lambda$
of level $l$, or what is the same thing, the set of integrable highest weight
representations of the Kac-Moody group $\widehat {L_{fin}G}$ of level $
l$. A dominant integral functional $\Lambda$ of level $l$ is determined
by a dominant integral functional $\dot \Lambda$ on $\dot{\mathfrak h}_{\mathbb R}$
 satisfying $\dot\Lambda(h_{\dot\theta}\le l$); see (\ref{domfunct}).
These are the labels of the $WZW_l$ modular functor (or as a boundary conformal field theory).

Suppose that $\Lambda\in\Phi_l$.  We will realize the dual Hilbert
space representation, $H(\Lambda )^{*}$, using the Borel-Weil
theorem, in the following way:
\begin{equation}\label{7.1.1}L^2H^0(\mathcal L^{*}_{\Lambda}\to \mathcal F_{hyp})\subset H(\Lambda
)^{*}\subset H^0(\mathcal L_{\Lambda}^{*}\to \mathcal F_{an})\end{equation}
We will realize $H(\Lambda )$ analogously, using the right coset
flag space.

Suppose that we are given a collection, $\amalg C$, of labeled
oriented circles, together with a positive parameterization,
\begin{equation}p_C:S^1\to C, \label{7.1.2}\end{equation}
for each circle $C$ in the collection.  To this data (or
object) we associate the Hilbert space
\begin{equation}H(\coprod C)=\bigotimes_CH(\Lambda_C)^{\epsilon_C}, \label{7.1.3}\end{equation}
where $\Lambda_C$ is the label for the component $C$, and $\epsilon_
C$ is
vacuous if $C$ is positively parameterized and the dual
otherwise.

As in (\ref{7.1.1}) we can also associate to this object a
Borel-Weil realization of the Hilbert space,
$$L^2H^0(\mathcal L^{*}\to\prod \mathcal F_{C,hyp})\subset H(\coprod C
)\subset H^0(\mathcal L^{*}\to\prod \mathcal F_{C,an}). \label{7.1.4}$$
Here $\mathcal F_{C,an}$ is the right, resp.  left, coset flag space if
$C$ is positively, resp.  negatively, parameterized, and the
line bundle over this product is the product of the line
bundles corresponding to the factors.

The final space in (\ref{7.1.4}) is acted upon by the oriented
Baer product of the extensions corresponding to the
components, i.e.  the extension
\begin{equation}0\to \mathbb C^{*}\to\tilde {G}(H^0(\amalg C))\to G(H^0(\amalg
C))\to 0,\label{ 7.1.5}\end{equation}
where
\begin{equation}0\to ker\chi\to\prod\tilde {G}(H^0(C))\to\tilde {G}(H^0(\coprod
C))\to 0, \label{7.1.6}\end{equation}
and
\begin{equation}\chi :\prod \mathbb C^{*}\to \mathbb C^{*}:(\lambda_C)\to\prod\lambda_
C^{\epsilon_C}. \label{7.1.7}\end{equation}
We will denote the projection in (\ref{7.1.6}) by
\begin{equation}(\tilde {g}_C)\to [\tilde {g}_C].\label{ 7.1.8}\end{equation}
The center $\mathbb C^{*}$ acts on (\ref{7.1.4}) by the character $\lambda
\to\lambda^{-l}$.

Let $\Sigma$ denote a set of data of the
following type:  (1) a compact Riemann surface, each
connected component of which has a nonempty boundary,
(2) real analytic parameterizations of each of the
boundary components, and (3) to each boundary
component an assignment of a label from $\Phi_l$.

By Segal reciprocity, there is a canonical global
cross-section
\begin{equation}\begin{matrix} \quad&\quad&\tilde {G}(H^0(\partial\Sigma ))\\
\quad&\nearrow&\downarrow\\
G(H^0(\Sigma ))&\rightarrow&G(H^0(\partial\Sigma ))\end{matrix}  \label{7.1.9}\end{equation}
(because the induced Lie algebra cocycle vanishes on
$\mathfrak g(H^0(\Sigma ))$, by Cauchy's theorem, and the group $G
(H^0(\Sigma ))$ is
simply connected).  There is a continuous extension of
this cross-section to $H^0(\Sigma^0,G)$, which maps into the
multicomponent analogue for the groups $\tilde {G}(H^0(S^1_{\pm}
))$.

\begin{lemma}\label{7.1.10} The natural inclusions
$$L^2H^0(\mathcal L^{*\otimes l}\to \mathcal F_{\partial\Sigma ,hyp})^{H^0(\Sigma
,G)}\to H(\partial\Sigma )^{H^0(\Sigma ,G)}\to H^0(\mathcal L^{
*\otimes l}\to \mathcal F_{\partial\Sigma ,an})^{H^0(\Sigma ,G)}$$
are
isomorphisms.\end{lemma}

{\bf Idea of Proof} Suppose that $\sigma$ is an $H^0(\Sigma ,G)$-invariant
section of
\begin{equation}\mathcal L^{*\otimes l}\to \mathcal F_{\partial\Sigma ,an} \label{7.1.11}\end{equation}
The action of $H^0(\Sigma ,G)$ on $\mathcal L^{*}\to \mathcal F_{\partial
\Sigma ,hyp}$ extends continuously
to a holomorphic action by $H^0(\Sigma^0,G)$.  Since
\begin{equation}\mathcal F_{\partial\Sigma ,an}/H^0(\Sigma ,G)\cong \mathcal F_{\partial
\Sigma ,hyp}/H^0(\Sigma^0,G), \label{7.1.12}\end{equation}
it follows that $\sigma$ extends to a holomorphic section of $\mathcal L^{
*}$
over the hyperfunction flag space.  It is automatically
$L^2$ because the integral is performed over a compact
space by Conjecture \ref{conj3}.

Returning to the definition of the level k modular
functor, we associate to the morphism $\Sigma$ the space in
(\ref{7.1.7}), which we denote by $E_l(\Sigma )$.

\subsection{The sewing property}

Suppose that $\Sigma$ has two distinguished boundary
components $C_{in}$ and $C_{out}$ which are negatively and
positively parameterized, respectively.  Suppose that $C_{in}$
and $C_{out}$ have the same label $\Lambda$.  To indicate this, we
will write $\Sigma =\Sigma_{\Lambda}$.  Let $\check{\Sigma}$ denote the morphism obtained
by sewing along these components (we assume that each
component of $\check{\Sigma}$ has nonempty boundary).  Then
\begin{equation}\aligned
H(\partial\Sigma )&=H(\Lambda )^{*}\otimes H(\Lambda )\otimes H(\partial
\check{\Sigma })\endaligned
 \label{7.2.1}\end{equation}

\begin{lemma}
$$E_l(\Sigma_{\Lambda})\subset Domain(trace).$$
\end{lemma}

There is a natural map
\begin{equation}H^0(\Sigma ,G)\leftarrow H^0(\check{\Sigma },G) \label{7.2.3}\end{equation}
This together with the lemma will imply that there is a
natural map induced by trace,
\begin{equation}E_l(\Sigma_{\Lambda})\to E_l(\check{\Sigma }). \label{7.2.4}\end{equation}
We will verify the lemma in the course of proving the
following

\begin{proposition}The natural map
$$\bigoplus_{\Lambda\in\Phi}E_l(\Sigma_{\Lambda})\to E_l(\check{\Sigma }
)$$
is an isomorphism of vector
spaces.\end{proposition}

\noindent{\bf Idea of Proof}  Given a pre-Hilbert space $V$, we will
denote the completion by $V^{complete}$.  We claim that
\begin{equation}\aligned
\bigoplus_{\Lambda}E_l(\Sigma_{\Lambda})&=((\bigoplus_{\Lambda}H(
\Lambda )^{*}\otimes H(\Lambda ))\otimes H(\partial\check{\Sigma }
))^{H^0(\Sigma ,G)}\\
&=(L^2H^0(\tilde {L}_{hyp}^{*\otimes l}\to Hyp(S^1,G))^{complete}
\otimes H(\partial\check{\Sigma }))^{H^0(\Sigma ,G)}\\
&=(L^2H^0(\tilde {L}_{hyp}^{*\otimes l}\otimes H(\partial\check{
\Sigma })\to Hyp(S^1,G)))^{H^0(\Sigma ,G)}\\
&=(L^2H^0(\tilde {L}_{hyp}^{*\otimes l}\otimes H^0(\mathcal L^{
*}\to \mathcal F_{\partial\check{\Sigma },hyp})\to Hyp(S^1,G)))^{H^0
(\Sigma ,G)}\endaligned
. \label{7.2.6}\end{equation}
The second equality follows from the level $l$ Peter-Weyl
theorem.  In the third, resp.  fourth, line, the notation
means that we are considering sections having values in
$H(\partial\check{\Sigma })$, resp.  $H^0(\mathcal L^{*}\to \mathcal F_{
\partial\check{\Sigma },hyp})$.  The third and fourth
equalities follow from (\ref{7.1.10}).

If we view a holomorphic section of
\begin{equation}\tilde {L}_{hyp}^{*\otimes l}\otimes H^0(\mathcal L^{*}\to \mathcal F_{
\partial\check{\Sigma },hyp})\to Hyp(S^1,G) \label{7.2.7}\end{equation}
as a section-valued function on $\tilde {L}_{hyp}G$, then the
contraction map in (\ref{7.2.1}) is given simply by
\begin{equation}F\to F(1), \label{7.2.8}\end{equation}
where $1\in\tilde {L}_{an}G\subset\tilde {L}_{hyp}G$.  The action of $
H^0(\Sigma ,G)$ on such
functions extends to $H^0(\Sigma^0,G)$; for $g\in H^0(\Sigma^
0,G)$
\begin{equation}\label{7.2.9}(g\cdot F)(\tilde {g}_{hyp})=\tilde {g}_{\partial\check{\Sigma}}
\cdot F(\tilde {g}_{-}^{-1}\cdot\tilde {g}_{hyp}\cdot\tilde {g}_{
+}) \end{equation}
where $g_{\pm}\in H^0(S^1_{\pm})$ are the restrictions of $g$ to collars
adjacent to $C_{in}$ and $C_{out}$, respectively, and $g_{\partial
\check{\Sigma}}$ denotes
the analogous restriction for $\partial\check{\Sigma}$.  We also have chosen
$\tilde {g}_{\pm}\in\tilde {H^0}(S^1_{\pm})$ and $\tilde {g}_{
\partial\check{\Sigma}}$, so that
\begin{equation}[\tilde {g}_{\pm},\tilde {g}_{\partial\check{\Sigma}}]=g \label{7.2.10}\end{equation}
with respect to Segal reciprocity.

Because each component of $\check{\Sigma}$ has nonempty boundary,
hence is a Stein manifold, it follows that the map
\begin{equation}H^0(\Sigma^0,G)\to Hyp(S^1,G):g\to [g^{-1}_{-},g_{+}] \label{7.2.11}\end{equation}
is surjective.  Setting $\tilde {g}_{hyp}=\lambda\in \mathbb C^{\times}$ in (\ref{7.1.9}), we now
see that we can reconstruct $F$ in (\ref{7.2.9}) from $F(1)$.  Thus
the map in the proposition is certainly injective.

It remains to show that $F$ is $L^2$, if we construct $F$
using (\ref{7.2.9}) from an invariant $F(1)$.  Because of the
invariance properties that $F$ inherits, $F$ is the pullback
of a section over a compact space, so that this is
automatic.

\section{Comparing $CM_l/WZW_l$ and $YM_3/YM_3+CS$  }\label{YM}

It is an open problem is to construct a non-free three dimensional quantum field theory in the sense of Segal, in particular for $YM_3$, or $YM_3+$Chern-Simons.
The physicists Karabali, Nair and collaborators have pointed out that, at least from a Hamiltonian point of view,
there is a useful analogy with chiral models (see\cite{KKN1}, \cite{KKN2}, \cite{KNY},
\cite{LMY},  \cite{LMY2}, \cite{KN}), and in fact there is a potential direct connection with the sigma model having target $G/K$.

Suppose that $\Sigma^{(3)}$ is a closed oriented Riemannian three manifold (a Euclidean space-time). The associated heuristic Feynman measure
for $YM_3$+Chern-Simons is
$$exp(-(\frac 1T \int\langle F_A\wedge *F_A\rangle+2\pi i l CS(A))\mathcal D[A]$$
where $F_A$ denotes the curvature for a $K$ connection $A$, $l$ is a nonnegative integer, $CS(A)$ denotes
the (properly normalized, multi-valued) Chern-Simons term, and $\mathcal D[A]$ denotes a heuristic Riemannian volume form
on gauge equivalence classes. This is similar in form to the heuristic Feynman measure (\ref{sigmaaction}) for the sigma model energy coupled to the WZW term.

Although the heuristic Feynman measures for $YM_3$ and the 2d chiral model $CM$ are similar, the theories are fundamentally different at a technical level. On the one hand $YM_3$ is subcritical, whereas the 2d chiral model is critical. As a consequence the renormalization group considerations for $YM_3$ are far simpler than for the 2d chiral model, e.g. for $YM_3$ the coupling constant $T$ is finite, whereas for the chiral model
the coupling constant $T \to 0$ on small scales. On the other hand dealing with local gauge invariance is a challenge, the classical $YM_3$ model is not integrable in any sense, and there is not a special tuning of the parameters for $YM_3+$Chern-Simons which yields something similar to the solvable conformally invariant $WZW_l$ model.

Our present focus is on the Hamiltonian, with an emphasis on mathematical considerations. From this limited perspective, the two theories are mathematically similar, because the configuration spaces $C^{\infty}(S^1,K)$ and $\Omega^1(S^2,\mathfrak k)/\Omega^0(S^2,K)$ are essentially homotopic. My purpose is to discuss a measure-theoretic correspondence, involving invariant measures for loop groups and a family of measures on a completion of $\Omega^1(S^2,\mathfrak k)/\Omega^0(S^2,K)$, which is suggested by the work of Karabali and Nair. Recently Guillarmou, Kupiainen, and Rhodes (\cite{GKR}) have explained how to make sense of these measures. Karabali et al use this family to suggest a realization for the Hilbert spaces of the $YM/YM+CS$ theories, and they develop a procedure (a little obscure to me) for calculating the ground state.

The crux of the $CM/YM_3$ analogy is the following. For the chiral model the configuration space is the critical loop group,
and the Toeplitz operator defines a map
$$W^{1/2}(S^1,K) \to \text{Fredholm Ops}: g \to A(g) $$
In the $YM_3$ case, after choosing a complex structure for the 2-sphere, the chiral Dirac operator (which, using the conformal structure for the 2-sphere, is identified with a $\overline{\partial}$ operator) defines a map
$$W^0\Omega^1(S^{(2)},\mathfrak k) \to \text{Fredholm Ops}: A \to \overline{\partial}_a, \qquad A=a-a^*, \qquad a\in \Omega^{0,1}(S^2,\mathfrak g) $$
In the $YM_3$ case one has to think of $\overline{\partial}_a$ as acting on a square root of the canonical line bundle, in order for the index to vanish. In both cases
one has to choose a representation for $K$, which we will largely suppress.

In both cases there is a canonical section, $det A$ and $det \overline{\partial} $, respectively, for the pullbacks of the
Fredholm determinant bundle. In both cases these bundles are equipped with unitary structures such that
$$|det A|^2(g)=det(A(g)A(g^{-1})) \text{ and } |det \overline{\partial}_a|^2=det_{\zeta}(\overline{\partial}_a \overline{\partial}_a^*)$$

In the chiral case $det A$ (given an appropriate choice of representation) is
the vacuum for the WZW model, and we claimed in Section \ref{sigmamodel} that its norm is an ingredient in thinking about the vacuum for the bare sigma model. 
In the $YM$ case $det \overline{\partial} $ is probably not the vacuum for $YM_3+CS$; $det \overline{\partial} $ depends only on 
conformal structure, whereas the vacuum for $YM_3$ presumably depends on conformal structure plus area form, and could be smoother (not necessarily simpler) than $det \overline{\partial} $ (One should keep an open mind; this is an important technical question which has to be resolved). There are formidable technical differences. The Toeplitz operator is a zeroth order pseudo-differential operator, whereas the d-bar operator is first order. For the norms, $det(A(g)A(g^{-1})$ is an ordinary Fredholm determinant, where $det_{\zeta}(\overline{\partial} \overline{\partial}^*)$ is a zeta determinant.

In the following table I have listed some mathematical correspondences which are relevant in comparing the two models, from the Hamiltonian point of view. In the table $S^{(2)}$ denotes a compact surface, $\theta_+=g_+^{-1}\partial g_+$ when a loop
has a Riemann Hilbert factorization $g=g_-g_0g_-$, $\Omega:S^2 \to G/K$ is the pre-potential described below

$$\begin{matrix} \text{Sigma Model} &\text{Yang-Mills}\\
\\
\frac{1}{T} E+2\pi i l WZW & \frac{1}{T} YM_3+2\pi i l CS\\
\\
W^{1/2}(S^1,K)&\mathcal C:=W^{1/2}\Omega^1(S^{(2)},\mathfrak k)/W^1(S^{(2)},K)\\
\\
\Omega_l^0(\widehat{Hyp})\supset H_l^0(\widehat{Hyp})\subset \oplus \mathbb C \sigma_j& \Omega_l^0(\widehat{\mathcal C})\supset \text{Non-existent?} \supset \text{Chern-Simons conformal block}\\
\\
det(A) & det(\overline{\partial})\\
\\
\text{choice of } S^1-metric,\quad  E(g)&\text{choice of }\Sigma -area-form,\quad YM_2\\
\\
Rot(S^1)&SDiff(\Sigma )\\

\\
\text{Wiener measure,}\quad\nu_{\frac 1T\mathcal E}&\text{ Yang-Mills
measure,}\quad\nu_{\frac 1T YM_2}\\
\\

\theta_+ & prepotential \quad \Omega \\
\\
\frac{1}{\mathfrak z}det(1+W(\theta_+)W(\theta_+)^*)^{-(\dot g+l)}D\theta_+& \frac{1}{\mathfrak z}exp(-(\dot g+l)(E(\Omega)+2\pi i\Gamma(\Omega)))d\lambda(\Omega)\\
\\
\Delta_{W^0(S^1,K)}+E &  \Delta_{W^{0}\mathcal C}+YM_2

\end{matrix}
$$

For the heuristic form of the Hamiltonian for $CM$, $\Delta_{W^0(S^1,K)}+E$, see Remark \ref{remark10}. For the heuristic form of the Hamiltonian for $YM_3$, $\Delta_{W^{0}\mathcal C}+YM_2$, where $W^{0}\mathcal C$ denotes the weak Riemannian metric on the space
of gauge orbits $\mathcal C$, see \cite{Mitter}. In both cases the natural degree of smoothness for classical fields is $W^{1/2}$, which is not at all visible from the forms of the Hamiltonians. This is a puzzle for me. But in the subcritical case of $YM_3$, Karabali, Nair et al do seem to know how to use this kind formal expression to actually compute.

To spell out our interest in the comparison, suppose for simplicity that the level $l=0$. 

Recall that in the context of loop groups, Wiener measure has the following small and large temperature limits
\begin{equation}\label{limits2}d\lambda_K \stackrel{T\downarrow 0}{\leftarrow} \nu_{\frac 1T E}\stackrel{T\uparrow \infty}{\rightarrow} \{ \begin{matrix} \prod_{S^1}d\lambda_K& \text{ wrt } BC(\prod_{S^1} K)\\ \mu_0& \text{ wrt } BC(Hyp(S^1,G)) \end{matrix}\end{equation}

One way to think about (normalized) Wiener measure is in terms of its finite point distributions, i.e. given a `triangulation' for $S^1$, with vertices $V\subset S^1$, for
$$evaluation_V: C^0(S^1,K) \to \prod_V K,  \qquad  eval_*(\nu_{\frac 1T E})=\frac{1}{p_{T2\pi}(1)}\prod_{Edges}p_{Tl(e)}(g_{\partial e})\prod_V d\lambda_K(g_v)$$
These projections are coherent and define a finitely additive measure. It is then necessary to prove there is a completion (For this and other purposes, following Ito, it is often useful to think of the completion as a finite codimensional conditioning of Wiener measure on path space, which is the image of a Gaussian measure via use of a stochastic
differential equation; following Feynman, this is natural from the point of view of quantum mechanics). When we think about $\mu_0$, we are thinking in terms 
of the coefficients of the Riemann-Hilbert factorization of a loop, which is a kind of momentum space point of view.

What is the analogue involving $\nu_{\frac 1T YM_2}$? I do not presently know how to elegantly formulate a $YM_2$ analogue of (\ref{limits2}), but a rough formulation is
\begin{equation}\label{limits3}d\lambda_{H^1(S^{(2)},K)} \stackrel{T\downarrow 0}{\leftarrow} \nu_{\frac 1T YM_2}\stackrel{T\uparrow \infty}{\rightarrow} \{ \begin{matrix} \text{Ashtekar measure}& \text{ wrt } BC(\{\text{holonomy functors}\}) \\ \widetilde{\mu_0}& \text{ wrt } BC(\bigsqcup_{\rho\in H^1(S^{(2)},K)}G\backslash Map(\widetilde{S^{(2)}},G/K)^{\rho})  \end{matrix}\end{equation}
where the first measure is the normalized Goldman volume form on moduli space (this limit is known, as independently discovered by Forman, King,
Sengupta and no doubt others), for Ashtekar measure see \cite{Ashtekar}, and the measure $\widetilde {\mu_0}$ will turn out to be (the finite part of) the Feynman measure studied in \cite{GKR} (I have written `Map' in (\ref{limits3}); tentatively one can think of $C^0$).

\begin{remark} Remember that $YM_3$ is subcritical, hence there is no reason to believe this procedure should have anything to do with computing the ground state for $YM_3$.
What follows may be a purely mathematical exercise, but it difficult to rule anything out.
\end{remark}

If $S^{(2)}$ is equipped with a conformal structure, then there is a classical differential geometric diagram 
(where the degree of smoothness of forms is initially irrelevant)
\begin{equation}\label{diagram1}\begin{matrix}  \Omega^1\otimes \mathfrak k &\cong & \Omega^{0,1}\otimes \mathfrak g\\
\downarrow & &\downarrow\\
\frac{\Omega^1\otimes \mathfrak k}{\Omega(S^{(2)},K)} & \cong&  \frac{\Omega^{0,1}\otimes \mathfrak g}{\Omega(S^{(2)},K)} \\  \uparrow & &\downarrow\\ 
 H^1(S^{(2)},K)\supset Bun_G^0(S^{(2)}) & & Bun_G(S^{(2)})\cong \frac{\Omega^{0,1}\otimes \mathfrak g}{\Omega(S^{(2)},G)} \end{matrix} \end{equation}
where in the top line $A\in \Omega^1(S^{(2)},\mathfrak k)$, is written as $A=a-a^*$, where $a\in \Omega^{0,1}(S^{(2)},\mathfrak g)$.

In two dimensions, given an area form, there is a finitely additive `$YM_2$ measure' on $\Omega^1(S^{(2)},\mathfrak k)$, as we will recall below. To obtain a countably additive measure, it is necessary to project to gauge equivalence classes and complete the $YM_2$ measure on some thickening of the quotient space 
$$\frac{\Omega^1\otimes \mathfrak k}{\Omega(S^{(2)},K)} $$
It is not clear that there is one completion which serves all purposes. We will describe two realizations, one 
involving holonomy and the other (the Karabali and Nair approach) involving the pre-potential coordinate $\Omega$ in the table above. 

The holonomy realization for $ \nu_{\frac 1T YM_2}$, which I will dangerously oversimplify, is similar to that for Wiener measure, although it is more complex because of gauge invariance (the basic ideas go back to Migdal). Suppose that $S^{(2)}$ is a compact oriented surface (a space from the $YM_3$ point of view), equipped with an area form. In this context it is useful to think of a (generalized) connection as a functor from the category of paths in $S^{(2)}$ to morphisms of a (trivial) $K$-bundle: given a path $e$,
the corresponding morphism is parallel translation from the initial to final fiber of the bundle, $P_{e_0} \to P_{e_1}$. Given a tiling (e.g. a triangulation) $\mathcal T$ of $\Sigma$ with oriented edges, let $V$, $
E$, and $F$ 
denote the sets of vertices, oriented edges, and faces, 
respectively.  There is a natural projection 
$$\pi_{\mathcal T}:\begin{matrix}\parallel\text{transport}\\ \text{functors}\end{matrix}\to\prod_{e\in E}Mor(P_{e_o},P_{e_1}):g\to (g_
e)_{e\in E}$$
where $g_e$ denotes parallel translation from the initial fiber to 
the initial fiber along the oriented edge $e$. 
Define a measure on the space $\prod_{e\in E} Mor(P_{e_0},P_{e_1})$ by 
$$d\nu_{\mathcal T}=\prod_{f\in F}H_{T Area(f)}(g_{\partial f})\prod_{e
\in E}dg_e$$
where $g_{\partial f}$ is the ``holonomy'' around the boundary of $
f$, i.e. $g_{\partial f}$ 
represents the conjugacy class of the morphism
$$g_{e_n}^{\epsilon_n}\circ ...\circ g_{e_1}^{\epsilon_1}\in Mor(
P_{basepoint})$$
where
$$\partial f=\pm (\epsilon_1e_1+...+\epsilon_ne_n)$$
(Since the holonomy is a conjugacy class, we can view it as a 
conjugacy class in $K$ (which is non-canonically isomorphic to 
$Mor(P_q)$, for each $q$). As in the case of Wiener measure, these distributions are naturally coherent with respect to 
refinement of the tiling.

The upshot is that there is a finite finitely additive $SDiff(S^{(2)}$ and gauge invariant measure on a space of parallel translation functors (generalized connections). 
Unfortunately it is not countably additive because of gauge invariance. 

To obtain a countably additive measure, it is necessary to project to gauge equivalence classes. What I expect to be true, but for which I cannot find a direct treatment in the literature (maybe this follows from \cite{TLevy}), is that the (normalized) finitely additive $YM_2$ measure on parallel translation functors descends to a $SDiff$ invariant measure which can be completed to a countably additive probability measure on gauge equivalence classes, holonomy functors (defined on an appropriate class of loops). The $SDiff$ invariance conjecturally characterizes this family of measures $\nu_{\frac 1T YM_2}$, where I have inserted a $T$ parameter:

\begin{equation}\begin{matrix} SDiff \times \{\begin{matrix}\parallel\text{transport}\\ \text{functors}\end{matrix}\}& \supset& \Omega^1\otimes \mathfrak k \\
\downarrow & &\downarrow \\
SDiff \times \{\begin{matrix}\text{holonomy}\\ \text{functors}\end{matrix}\} &  \supset&\frac{\Omega^1\otimes \mathfrak k}{\Omega(S^{(2)},K)}  \\ \uparrow & & \uparrow \\ H^1(S^{(2)},K) & &  H^1(S^{(2)},K)  \end{matrix}\end{equation}

\begin{remark}It is more conventional to express the measures $\nu_{\frac 1T YM_2}$ in terms of iterates of Wiener measure, see Wikipedia, Two-dimensional Yang-Mills Theory, for
a nice overview, or \cite{Sengupta}. 
\end{remark} 

We now pivot to the Karabali and Nair point of view, which exploits the right hand side of (\ref{diagram1}), which should perhaps be referred to as a `conformal gauge'. To begin suppose that $S^{(2)}=S^2$, the standard 2-sphere
with its standard complex structure and area form. The introduction of complex structure breaks the $SDiff$ symmetry.

Given $A\in \Omega^1(S^2,\mathfrak k)$, $A=a-a^*$, where $a\in \Omega^{0,1}(S^2,\mathfrak g)$. Just as a loop $g:S^1 \to G$ has 
a factorization $g=g_-g_0g_+$ iff $A(g)$ is invertible,
$a$ is complex gauge equivalent to 0, i.e. $a=-\overline{\partial}(h)h^{-1}$ for some $h:S^2\to G$ (which is unique up to multiplication on the right
by a constant $h_0\in G$), iff $\overline{\partial}_a$ is invertible (acting
on an appropriate associated vector bundle tensored with $\kappa^{1/2}$, where $\kappa$ is the canonical bundle). The
prepotential is $\Omega=h^*h$ (which is not quite uniquely determined). It is often convenient to regard $\Omega$ as a map $\Omega:S^2\to G/K$, modulo
the action of $G$ on (this intentionally ambiguously defined) space of maps. 
This coordinate for gauge equivalence classes of $K$-connections is analogous to the linear Riemann-Hilbert (or time ordered exponential)
coordinate $\theta_+$ for $LK\subset Hyp(S^1,G)$.

In the case of simply connected $S^2$ (with the unique complex structure and standard area form) we are now enlarging (\ref{diagram1}) to
$$\begin{matrix} \Omega^1\otimes \mathfrak k &\cong & \Omega^{0,1}\otimes \mathfrak g & & \\
\downarrow & &\downarrow & & \\
\frac{\Omega^1\otimes \mathfrak k}{\Omega(S^{(2)},K)} & \cong&  \frac{\Omega^{0,1}\otimes \mathfrak g}{\Omega(S^{(2)},K)} & \supset &  G\backslash Map(S^2,G/K)\\ 
\uparrow & &\downarrow & &\downarrow  \\ 
H^1(S^2,K)=point & &   Bun_G(S^2) & \supset & Bun_G^0(S^{(2)})=point \end{matrix} $$

The normalized 
$YM_2$ probability measure has a heuristic (local) expression in the pre-potential coordinate. To see this recall
$A-a-a^*$, $a=h\cdot 0=-(\overline{\partial} h)h^{-1}$, hence
\begin{equation}\label{key identity}h^{-1}(a-a^{*})h+h^{-1}dh=\Omega^{-1}\partial\Omega \end{equation}
(This also follows abstractly from the fact that 
$d+A$ is the coordinate expression for the unique holomorphic 
unitary connection in the bundle $\Sigma\times V\to\Sigma$, where the unitary 
structure is constant, and the holomorphic structure is gotten 
by declaring $\vec{\epsilon }h$ to be a holomorphic frame, where $
\vec{\epsilon}$ denotes a 
constant frame; in the holomorphic frame $\vec{\epsilon }h$ the connection is 
given by 
$$\partial +\Omega^{-1}\partial\Omega +\bar{\partial }$$
The identity (\ref{key identity}) implies 
$$h^{-1}F_A h=\bar{\partial }(\Omega^{-1}\partial\Omega )$$
This together with invariance of the Killing form implies
\begin{equation}\label{key identity 2}\langle F_A\wedge *F_A\rangle =\langle\bar{\partial }(\Omega^{-
1}(\partial\Omega )\wedge *\bar{\partial }(\Omega^{-1}(\partial\Omega 
)\rangle \end{equation}

\begin{remark} In a local holomorphic coordinate $z=x+iy$
$$\bar{\partial }(\Omega^{-1}\partial\Omega )=(\Omega^{-1}\Omega_z )_{\bar{z}}d\bar{z}\wedge dz$$
$$=\frac 14 \left( (\Omega^{-1}(\Omega_x-i\Omega_y))_x+i( (\Omega^{-1}(\Omega_x-i\Omega_y))_y\right)$$
$$=\frac 14 \left( (\Omega^{-1}\Omega_x)_x-i(\Omega^{-1}\Omega_y)_x+i(\Omega^{-1}\Omega_x)_y+(\Omega^{-1}\Omega_y)_y\right)$$
$$\frac 14 \left( (\Omega^{-1}\Omega_x)_x+(\Omega^{-1}\Omega_y)_y+i[\Omega^{-1}\Omega_x,\Omega^{-1}\Omega_y]\right) $$

\end{remark}

This now implies that the normalized $YM_2$ measure has the heuristic expression
$$\nu_{\frac 1T YM_2}=\frac{1}{\mathfrak Z}exp(-\frac{1}{2T}\langle\overline{\partial }
(\Omega^{-1}(\partial\Omega ))\wedge *\overline{\partial }(\Omega^{-1}
(\partial\Omega ))\rangle) det_{\zeta}(\overline{\partial }\overline{\partial }^*)^{\dot g} \mathcal D\Omega $$
$$=\frac{1}{\mathfrak Z}exp(-\frac{1}{2T}\langle\overline{\partial }
(\Omega^{-1}(\partial\Omega ))\wedge *\overline{\partial }(\Omega^{-1}
(\partial\Omega ))\rangle) exp(-\frac {\dot g}{2\pi}(
E(\Omega )+2\pi iWZW(\Omega ))\mathcal D\Omega $$
where $\mathcal D\Omega$ denotes the fictitious invariant measure for $Map(S^{2},G/K)$, $\dot g$ is the
dual Coxeter number ($\dot g=N$ for $K=SU(N)$) and
the last line uses the Polyakov-Wiegman anomaly formula for the zeta determinant. 

Recall that there are diffeomorphisms
$$ KAN^+ \leftrightarrow G \text{  and  } N^+A \leftrightarrow G/K$$
In terms of the latter correspondence, usually referred to as horocycle coordinates for $G/K$ 
$$\Omega=qq^*, \text{ where } q=na=ne^{\phi} $$
Long ago someone (possibly Gawedzki and Kupiainen) made the critical observation that
$$S_{WZW}(qq^*)=E(\Omega )+2\pi iWZW(\Omega )=\frac{1}{2\pi i}\int_{S^{(2)}}\langle \partial\phi\wedge \overline{\partial}\phi+e^{-2\rho(\phi)}  n^{-1}\partial n\wedge n^{-1}\overline{\partial}n\rangle $$
where $\rho$ is the sum of the positive complex roots. 

Using this GKR give a rigorous meaning to the measure that heuristically is
$$\lim_{T\uparrow \infty}\frac{1}{\mathfrak Z}exp(-\frac{1}{2T}\langle\overline{\partial }
(\Omega^{-1}(\partial\Omega ))\wedge *\overline{\partial }(\Omega^{-1}
(\partial\Omega ))\rangle) exp(-\frac {\dot g}{2\pi}(
E(\Omega )+2\pi iWZW(\Omega ))\mathcal D\Omega $$
see section 1.4 of \cite{GKR} (note that for GKR there is a zero mode, which means their measure is infinite; we are moding out
 by $G$, hence we obtain a finite measure, as claimed by Karabali and Nair). This is conformally invariant. It is natural to ask if there is some alternate
 way of thinking about $\widetilde{\mu_0}$ which leads to a characterization.

For a general Riemannian surface $S^{(2)}$ the analogue is 
$$\begin{matrix} \Omega^1\otimes \mathfrak k &\cong & \Omega^{0,1}\otimes \mathfrak g & & \\
\downarrow & &\downarrow & & \\
\frac{\Omega^1\otimes \mathfrak k}{\Omega(S^{(2)},K)} & \cong&  \frac{\Omega^{0,1}\otimes \mathfrak g}{\Omega(S^{(2)},K)} & \supset &  \bigsqcup_{[\rho\in H^1(S^{(2)},K)]} G\backslash Map(\widetilde{S^{(2)}},G/K)^{\rho}) \\ 
\uparrow & &\downarrow & & \\ 
H^1(S^{(2)},K) & &  Bun_G(S^{(2)}) &\supset & Bun_G^0(S^{(2)}) \end{matrix} $$

This leads to many questions which I cannot answer, even at the heuristic level of these notes. The main takeaway is that the limit as $T\uparrow \infty$
does exist, as I had hoped, but it is probably not the ground state for $YM_3$, because it is conformally invariant and too rough by half a degree. 

\subsection{Return to Physics}

Suppose $K$ is a simply connected compact Lie group with simple Lie algebra and normalized $Ad$-invariant inner product $\langle \cdot\cdot\rangle$ (which
induces an $Ad^*$ invariant inner product on the dual). 

Consider the Hamiltonian system $(T^*K\simeq K\times \mathfrak k, \langle\cdot,\cdot\rangle)$. The classical solutions are geodesics. 
In dimension $0+1$ the dynamics for the quantum version is determined by Brownian motion with generator $\Delta_K$. This induces Wiener measure
on $C^0(S^1,K)$, denoted $\nu_{\frac 1T E}$, having the heuristic expression
$$\nu_{\frac 1T E}=\frac {1}{\mathfrak Z}e^{-\frac 1T E(g)}\prod_{S^1} d\lambda_K(g_v)$$

In $1+1$ dimensions the Hamiltonian of the 2d chiral model with target $K$
has the heuristic form
$$\Delta_{W_0}+E_{W^1}$$
where $E$ is the usual energy function on $W^1(S^1,K)$, see Remark \ref{remark10}. We have conjectured that to obtain the Hilbert space (state space) 
and ground state for the chiral model
we consider the large $T$ limit
$$ \delta_K \overline{T\downarrow 0}{\leftarrow}  \nu_{\frac 1T E} \overline{T\uparrow 0}{\rightarrow} \mu_0 $$
From the structure of $\mu_0$ and the existence of root subgroup factorization, we posited a formula for the chiral Hamiltonian and the corresponding
Brownian type motion on $C^0(\mathbb R,Hyp(S^1,G))$; we have not found a means to check this hypothesis.

Next we considered $YM_3$. The Hamiltonian has the heuristic form
$$\Delta_{W_{1/2}}+YM_2$$
The $YM_2$ measure is well-defined at least in `holonomy coordinates'. Karabali and Nair have asserted (in a more direct way) that the Hilbert space 
can be found using the large $T$ limit
$$d\lambda_{H^1(S^{(2)},K)} \stackrel{T\downarrow 0}{\leftarrow} \nu_{\frac 1T YM_2} \stackrel{T\uparrow \infty}{\rightarrow} \widetilde{\mu_0}$$
using a conformally invariant gauge. Guillarmou, Kupiainen, and Rhodes have shown how to give a rigorous meaning to the measure $\widetilde{\mu_0}$.
This is probably not the ground state for $YM_3$; the support of the measure $\widetilde{\mu_0}$ seems to be too rough (by half of a degree) to be the vacuum for $YM_3$ (furthermore
there was no justification for considering a $T\uparrow \infty$ limit in this subcritical context). Karabali and Nair
propose an expansion for a density that should represent the vacuum. I do not see how to put this into geometrically comprehensible framework.

Does this lead to any insight about $YM_4$? This is critical like the chiral model. The configuration spaces for the chiral model and $YM_3$ (when space is
the 2-sphere) are homotopic, and have interesting algebraic topology. This may be the ultimate explanation for why $\nu_{\frac 1T E}$ and $\nu_{\frac 1T YM_2}$ have interesting large $T$ limits (in conformally invariant gauges).

For $YM_4$ the simplest space, $S^{3}$, is a group, $SU(2)$, just as $S^1$ is a group in the chiral case. The Hamiltonian is now heuristically
$$\Delta_{W_{0}}+YM_3$$
(which once again obscures the relevance of $W^{1/2}$).
Recent constructive field theoretic progress on the problem of constructing the $YM_3$ measure has been made by \cite{CC} (with an emphasis on asymptotic freedom) and \cite{CCHS}) (who also consider the coupling with a Higgs field).
It is irresistible to wonder about a large $T$ limit. The conventional wisdom is that
there does not exist a conformally invariant gauge choice. Instead one must add in an additional scalar field, to understand how conformal invariance
is broken for the quantum theory.

\end{document}